\documentclass[12pt]{article}
\usepackage{amsmath} 
\input{epsf} 
\def\ltap{\raisebox{-.6ex}{\rlap{$\,\sim\,$}} \raisebox{.4ex}{$\,<\,$}} 
\def\gtap{\raisebox{-.6ex}{\rlap{$\,\sim\,$}} \raisebox{.4ex}{$\,>\,$}}

\newcommand\as{\alpha_{\mathrm{S}}} 
\newcommand\f[2]{\frac{#1}{#2}}

\def\la{\lambda} 
 
\def\beq{\begin{equation}} 
\def\eeq{\end{equation}} 
\def\beeq{\begin{eqnarray}} 
\def\eeeq{\end{eqnarray}} 
\def\bom#1{{\mbox{\boldmath $#1$}}} 
\def\to{\rightarrow}
\def\ito{\leftarrow} 
\def\nn{\nonumber} 
 
\def\qt{q_T} 

\def\ms{${\overline {\rm MS}}$} 
\def\msbar{{\overline {\rm MS}}} 
\def\asp{\f{\as}{\pi}} 
\def\b0{b_0}

\def\tL{{\widetilde L}}
\def\ep{\epsilon}

\begin{document} 

\begin{titlepage}
\renewcommand{\thefootnote}{\fnsymbol{footnote}}
\begin{flushright}
LPSC 05--63 \\
hep-ph/0508068
     \end{flushright}
\par \vspace{10mm}

\begin{center}
{\Large \bf
Transverse-momentum resummation and\\
\vskip .3cm
the spectrum of the Higgs boson at the LHC}
\end{center}
\par \vspace{2mm}
\begin{center}
{\bf Giuseppe Bozzi${}^{(a)}$,
Stefano Catani${}^{(b)}$,}\\ 
\vskip .2cm
{\bf Daniel de Florian${}^{(c)}$
and
Massimiliano Grazzini${}^{(b)}$}\\

\vspace{5mm}

${}^{(a)}$Laboratoire de Physique Subatomique et de Cosmologie,\\
Universit\'e Joseph Fourier/CNRS--IN2P3, F-38026 
Grenoble, France \\

${}^{(b)}$INFN, Sezione di Firenze and Dipartimento di Fisica,
Universit\`a di Firenze,\\ 
I-50019 Sesto Fiorentino, Florence, Italy\\

${}^{(c)}$Departamento de F\'\i sica, FCEYN, Universidad de Buenos Aires,\\
(1428) Pabell\'on 1 Ciudad Universitaria, Capital Federal, Argentina\\

\vspace{5mm}

\end{center}

\par \vspace{2mm}
\begin{center} {\large \bf Abstract} \end{center}
\begin{quote}
\pretolerance 10000

We consider the transverse-momentum ($q_T$) distribution of  
generic high-mass systems (lepton pairs, vector bosons, Higgs particles, ....)
produced in hadron collisions.
At small $q_T$,
we concentrate on the all-order resummation of the
logarithmically-enhanced 
contributions in QCD perturbation theory.
We elaborate on the $b$-space resummation formalism
and introduce some novel features:
the large logarithmic contributions
are systematically exponentiated in a process-independent
form and, after integration over $q_T$, they are constrained by
perturbative unitarity to give a vanishing contribution to the total
cross section.
At intermediate and large $q_T$, resummation is consistently
combined with fixed-order perturbative results, to obtain
predictions with uniform theoretical accuracy over the entire range of
transverse momenta.
The formalism is applied to Standard Model Higgs boson production at LHC 
energies. We
combine the most advanced perturbative information available
at present for this process: resummation up to
next-to-next-to-leading logarithmic accuracy and 
fixed-order perturbation theory up to next-to-leading order.
The results show a high stability with respect to perturbative QCD 
uncertainties.

\end{quote}

\vspace*{\fill}
\begin{flushleft}
LPSC 05--63  \\ August 2005

\end{flushleft}
\end{titlepage}

\setcounter{footnote}{1}
\renewcommand{\thefootnote}{\fnsymbol{footnote}}

\section{Introduction}
\label{sec:intro}

This paper is devoted to study the transverse-momentum ($\qt$) spectrum
of high-mass systems produced by hard-scattering of partons 
in hadron--hadron collisions.
In Ref.~[\ref{Bozzi:2003jy}] we presented some quantitative results on the
$\qt$ spectrum of the Standard Model (SM) Higgs boson, produced
via the gluon fusion mechanism, at LHC energies.
The formalism used in Ref.~[\ref{Bozzi:2003jy}]
is quite general and applies to 
the transverse-momentum distribution of generic high-mass systems 
(lepton pairs, vector bosons, Higgs particles, ..) 
produced in hadron collisions.
The purpose of the present paper is twofold. Owing to its general
applicability, we find it useful to first describe and discuss the formalism 
with quite some details. We then perform a more
systematic phenomenological analysis of the $q_T$ distribution of
the Higgs boson at the LHC.

In this introductory section, rather than illustrating the resummation
formalism in general terms, we mainly consider the explicit case of the $q_T$
spectrum of the Higgs boson. This also serves for underlying some general
features of the formalism in concrete, rather than abstract, terms. 

Within the SM of electroweak interactions, the Higgs boson
[\ref{Gunion:1989we}] is responsible 
for the mechanism of the electroweak symmetry breaking,
but this particle has so far eluded experimental
discovery. Direct searches at LEP have established a lower bound
of  $114.4$~GeV [\ref{Barate:2003sz}] on 
the mass $M_H$ of the SM Higgs boson,   
whereas SM fits of electroweak precision
data lead to the upper limit 
$M_H < 260$~GeV at $95\%$ CL [\ref{Group:2004qh}].
The next search for Higgs boson(s) will be carried out at hadron colliders,
namely, the Fermilab Tevatron [\ref{Carena:2000yx}, \ref{Yao:2004jv}] 
and the CERN LHC [\ref{atlascms}, \ref{lhcupdate}].

The main production mechanism of the SM Higgs boson $H$
at hadron colliders is the gluon
fusion process $gg \to H$, through a heavy-quark (mainly, top-quark) loop.
When combined with the decay channels $H \to \gamma \gamma$,
$H \to WW$ and $H \to ZZ$, this production mechanism is one of 
the most important for Higgs boson searches and studies over the entire
mass range, 100~GeV$\ltap M_H \ltap$1~TeV, to be investigated at the LHC
[\ref{atlascms}]. To fully exploit the physics potential of the gluon fusion
process, it is relevant to provide reliable theoretical predictions for the
corresponding total cross section and for the associated distributions, 
such as, for instance, the Higgs $\qt$ distribution.
The dominant source of theoretical uncertainties on these quantities is
the effect of QCD radiative corrections, which, therefore, 
have to be carefully investigated.

The total cross section for Higgs boson production by gluon fusion
has been computed in QCD perturbation theory at the leading order 
(LO), ${\cal O}(\as^2)$, at the next-to-leading order
(NLO) [\ref{Dawson:1991zj}, \ref{Spira:1995rr}] 
and at the next-to-next-to-leading order (NNLO)
[\ref{Harlander:2000mg}--\ref{NNLOtotal}]
in the QCD coupling $\as$. The NNLO computation of the 
rapidity distribution of the Higgs boson 
has recently been completed [\ref{Anastasiou:2004xq}].
A key point of this theoretical activity
is that the origin of the dominant perturbative 
contributions to the total cross section has been identified 
and understood:
the bulk of the radiative corrections is due to virtual and soft-gluon
terms [\ref{Catani:2001ic}]. 
This point has a twofold relevance.
On one side, it explains the observation [\ref{Kramer:1996iq}] 
of the validity of the large-$M_t$ approximation 
($M_t$ being the mass of the top quark) in the calculation at the NLO, and,
therefore, it justifies the use of the same approximation at and beyond 
the NNLO. On the other side, it allows to estimate higher-order QCD
contributions by supplementing the NNLO calculation with an all-order
resummation of the logarithmically-enhanced terms due to multiple
soft-gluon emission [\ref{Catani:2003zt}].
Having these terms under control allows us to reliably predict the value 
of the cross section and, more importantly, to reduce the associated 
perturbative 
uncertainty at the level of about $\pm 10\%$~[\ref{Catani:2003zt}].

When studying the $q_T$ distribution of the Higgs boson in QCD perturbation
theory, it is convenient to start by considering separately
the large-$q_T$ and small-$q_T$ regions.

The large-$q_T$ region is identified by
the condition $q_T \sim M_H$.
In this region, the perturbative series is controlled by a small expansion
parameter, $\as(M_H^2)$, and calculations based on the truncation of the series
at a fixed order in $\as$ are theoretically justified.
SM Higgs boson production at large $q_T$ via gluon fusion
has to be accompanied by the radiation
of at least one recoiling parton, so the LO term for this observable is 
of ${\cal O}(\as^3)$.
The LO calculation was
reported
in Ref.~[\ref{Ellis:1987xu}];
it shows that the large-$M_t$
approximation works well as long
as 
$M_H \ltap 2M_t$ and $q_T \ltap M_t$.
Similar results on the validity of the large-$M_t$ approximation
were obtained in the case of the associated production of a Higgs boson plus 
2 jets (2 recoiling partons at large transverse momenta) [\ref{DelDuca:2001fn}].
In the framework of the large-$M_t$ approximation, the NLO QCD corrections
to the transverse-momentum distribution of the SM Higgs boson
were computed in 
Refs.~[\ref{deFlorian:1999zd}--\ref{Anastasiou:2005qj}].
Corrections to the large-$M_t$ approximation are considered in 
Ref.~[\ref{Smith:2005yq}]. 
The numerical programs of 
Refs.~[\ref{deFlorian:1999zd}, \ref{Anastasiou:2005qj}]
can also be used to evaluate arbitrary infrared- and collinear-safe 
observables up to NLO in the large-$q_T$ region and, in the case of
Ref.~[\ref{Anastasiou:2005qj}], up to NNLO when $q_T=0$.

In the small-$q_T$ region ($q_T\ll M_H$), where the bulk of events is produced,
the convergence of the fixed-order expansion is spoiled, since
the coefficients of the perturbative series in $\as(M_H^2)$ are enhanced
by powers of large logarithmic terms, $\ln^m (M_H^2/q_T^2)$. To obtain
reliable perturbative predictions, these terms have to be 
resummed to all orders in $\as$.
The method to systematically perform all-order resummation of
classes of logarithmically-enhanced terms at small $q_T$ is known
[\ref{Dokshitzer:hw}--\ref{Catani:2000vq}].
In the case of the SM Higgs boson, 
resummation has been explicitly worked out at
leading logarithmic (LL), next-to-leading logarithmic (NLL) 
[\ref{Catani:vd}, \ref{Kauffman:cx}]
and next-to-next-to-leading logarithmic (NNLL) [\ref{deFlorian:2000pr}] level.

The fixed-order and resummed approaches at small and large values of 
$q_T$ can then be matched at intermediate values of $q_T$,
to obtain QCD predictions for the entire range of transverse momenta. 
Phenomenological studies of the SM Higgs boson
$q_T$ distribution have been performed in 
Refs.~[\ref{Hinchliffe:1988ap}, \ref{Kauffman:cx},
\ref{Kauffman:1991jt}--\ref{Berger:2002ut}, \ref{Bozzi:2003jy},
\ref{Kulesza:2003wi}--\ref{Lipatov:2005at}],
by combining resummed and fixed-order perturbation theory at different levels
of theoretical accuracy. A comparison of theoretical calculations 
[\ref{Bozzi:2003jy}, \ref{Balazs:2000wv}, \ref{Berger:2002ut},
\ref{Kulesza:2003wn}]
and of results from parton shower Monte Carlo generators
[\ref{PYTHIA}--\ref{Corcella:2004fr}]
is presented in Ref.~[\ref{Balazs:2004rd}].

In the present paper we compute  
the Higgs boson $q_T$ distribution at the LHC
by combining 
the most advanced perturbative information that is available at present:
NNLL resummation at small $q_T$ and NLO perturbation theory at large $q_T$.
The first results of our calculation
were presented in Refs.~[\ref{Bozzi:2003jy}, \ref{bozzi}].
Here we perform a more complete phenomenological study and
present a
discussion of theoretical uncertainties.

The formalism used to obtain these results was briefly described in 
Refs.~[\ref{Catani:2000vq}, \ref{Bozzi:2003jy}] and is 
illustrated in detail in the present paper.
Three distinctive features are anticipated here.
The resummation is performed at the level of the partonic cross section;
this implies that the parton distributions are evaluated at the factorization 
scale $\mu_F$, which has to be chosen of the order of the hard scale $M$.
The resummed terms are embodied in a form factor that is universal: 
it depends only on the flavour of the partons that initiate the hard-scattering
subprocess 
at the Born level (e.g. $q{\bar q}$ annihilation
in the case of Drell--Yan lepton pair production, and $gg$ fusion in the case 
of Higgs boson production).
A constraint of perturbative unitarity is imposed on the resummed terms,
to the purpose of reducing the effect of unjustified higher-order 
contributions at large values
of $q_T$ and, especially, at intermediate values of $q_T$.
The constraint implies that the total cross section
at the nominal fixed-order accuracy (NLO or NNLO) is recovered upon integration
over $q_T$ of the transverse-momentum spectrum (at NLL+LO or NNLL+NLO).

The paper is organized as follows. In Sect.~\ref{sec:res}
the resummation formalism is
discussed in detail. After illustrating
the general aspects of our approach in Sect.~\ref{sec:genfor},
we discuss the structure of the resummed cross section
in Sect.~\ref{sec:rescross}.
The relation to the standard $b$-space resummation
is given in Sect.~\ref{sec:css}.
Sect.~\ref{sec:matc} is devoted to the finite component of the cross section.
In Sect.~\ref{sec:phen} we apply the resummation formalism to
the production of the SM Higgs boson at the LHC. In Sect.~\ref{sec:summ}
we draw our conclusions.
In Appendix \ref{appa} we discuss the details of the exponentiation
in the general multiflavour case.
In Appendix \ref{appb} we illustrate
the calculation of the Bessel integrals required
in the computation of the perturbative expansion of the resummed
cross section.

\section{Transverse-momentum resummation}
\label{sec:res}

The formalism [\ref{Bozzi:2003jy}, \ref{Catani:2000vq}]
that we use to compute the $q_T$ distribution of the Higgs boson
applies to more general hard-scattering processes. Therefore, we describe
it in general terms.

\subsection{The resummation formalism: from small to large values of
$\qt$}
\label{sec:genfor}

We consider the inclusive hard-scattering process
\begin{equation}
\label{process}
h_1(p_1) + h_2(p_2) \to F(M,\qt) + X \;\;,
\end{equation}
where the collision of the two hadrons $h_1$ and $h_2$ with
momenta $p_1$ and $p_2$ produces the triggered final-state system $F$,
accompanied by an arbitrary and undetected final state $X$.
We denote by $\sqrt s$ the centre-of-mass energy of the colliding hadrons
$(s= (p_1+p_2)^2 \simeq 2p_1p_2)$.
The observed final state $F$ is a generic system of non-QCD partons
such as {\em one} or {\em more} vector bosons $(\gamma^*, W, Z, \dots)$, 
Higgs particles, Drell--Yan (DY) lepton pairs and so forth. 
We do not consider the production of strongly interacting particles
(hadrons, jets, heavy quarks, ...), since in this case the resummation
formalism of small-$\qt$ logarithms has not yet been fully developed.

Throughout the paper we limit ourselves to considering  
the case in which only the
total invariant mass $M$ and transverse momentum $\qt$ 
of the system $F$ are measured.
According to the QCD factorization theorem (see Ref.~[\ref{Collins:gx}] and 
references therein),
the corresponding transverse-momentum differential cross 
section\footnote{To be precise, when the system $F$ is not a single
on-shell particle of mass $M$, what we denote by $d{\hat \sigma}_{F}/d q_T^2$
is actually  the differential cross section 
$M^2 d{\hat \sigma}_{F}/dM^2 dq_T^2$.}
$d{\hat \sigma}_{F}/d q_T^2$ can be written as
\begin{equation}
\label{dcross}
\f{d\sigma_F}{d q_T^2}(q_T,M,s)=\sum_{a,b}
\int_0^1 dx_1 \,\int_0^1 dx_2 \,f_{a/h_1}(x_1,\mu_F^2)
\,f_{b/h_2}(x_2,\mu_F^2) \;
\f{d{\hat \sigma}_{F \,ab}}{d q_T^2}(q_T, M,{\hat s};
\as(\mu_R^2),\mu_R^2,\mu_F^2) 
\;,
\end{equation}
where $f_{a/h}(x,\mu_F^2)$ ($a=q_f,{\bar q_f},g$) are the parton densities of 
the colliding hadrons at the factorization scale $\mu_F$,
$d{\hat \sigma}_{F \,ab}/d q_T^2$ are the
partonic cross sections, ${\hat s}=x_1x_2s$ is the partonic centre-of-mass
energy, and $\mu_R$ is the renormalization scale.
Throughout the paper we use parton
densities as defined in the \ms\
factorization scheme, and $\as(q^2)$ is the QCD running coupling in the \ms\
renormalization scheme.

The partonic cross section is computable in QCD perturbation theory as a power
series expansion in $\as$.
We assume that at the parton level the system $F$ is produced with
vanishing $\qt$ (i.e. with no accompanying final-state radiation)
in the lowest-order approximation, so that the
corresponding cross section is 
$d{\hat \sigma}^{(0)}_{F \,c{\bar c}}/d\qt^2 \propto \delta(\qt^2)$.
Since $F$ is colourless, the lowest-order partonic subprocess, 
$c+{\bar c} \to F$, is either
$q{\bar q}$ annihilation ($c=q$), as in the case of $\gamma^*, W$ and $Z$ 
production, or $gg$ fusion ($c=g$), as in the case of the production of the SM
Higgs boson $H$.

As recalled in Sect.~\ref{sec:intro}, higher-order perturbative contributions
to the partonic cross section $d{\hat \sigma}_{F \,ab}/d q_T^2$
contain logarithmic terms of the type $\ln^m (M^2/q_T^2)$ that become
large in the small-$\qt$ region ($\qt \ll M$). 
Therefore, we introduce the following
decomposition of the partonic cross section in Eq.~(\ref{dcross}):
\begin{equation}
\label{resplusfin}
\f{d{\hat \sigma}_{F \,ab}}{d q_T^2}=
\f{d{\hat \sigma}_{F \,ab}^{(\rm res.)}}{d q_T^2}
+\f{d{\hat \sigma}_{F \,ab}^{(\rm fin.)}}{d q_T^2}\, .
\end{equation}
The distinction between the two terms on the right-hand side
is purely theoretical.
The first term, $d{\hat \sigma}_{F \,ab}^{(\rm res.)}$, on the right-hand side
contains all 
the logarithmically-enhanced contributions, $(\as^n/q_T^2)\, \ln^m (M^2/q_T^2)$,
at small
$q_T$, and has to be evaluated by resumming them 
to all orders in $\as$.
The second term, $d{\hat \sigma}_{F \,ab}^{(\rm fin.)}$,
is free of such contributions, and 
can be computed by fixed-order truncation of the perturbative series.
More precisely, we define the `finite' component
$d{\hat \sigma}_{F \,ab}^{(\rm fin.)}$
in such a way that we have\footnote{The notation $\bigl[ X \bigr]_{\rm f.o.}$
means that the quantity $X$ is
computed by truncating its perturbative expansion at a given fixed order in
$\as$.} 
\begin{equation}
\label{deffin}
\lim_{Q_T \to 0} \; \int_0^{Q_T^2} dq_T^2 \;
\Bigl[ \f{d{\hat \sigma}_{F \,ab}^{(\rm fin.)}}{d q_T^2}\Bigr]_{\rm f.o.}
= 0 \;, 
\end{equation}
where the right-hand side vanishes {\em order-by-order} in perturbation
theory. In particular, this implies that any perturbative
contributions proportional to
$\delta(\qt^2)$ have been removed from $d{\hat \sigma}_{F \,ab}^{(\rm fin.)}$
and included in $d{\hat \sigma}_{F \,ab}^{(\rm res.)}$.

The `resummed' component $d{\hat \sigma}_{F \,ab}^{(\rm res.)}$
of the partonic cross section cannot, of course, be evaluated by computing
all the logarithmic contributions in the perturbative series.
However, as discussed in Sect.~\ref{sec:rescross}, these contributions can 
systematically be organized in classes of LL, NLL, $\dots$ terms
and, then, this logarithmic expansion can be truncated at a given 
logarithmic accuracy.

In summary, the $\qt$ distribution in Eq.~(\ref{dcross}) is evaluated,
in practice,
by replacing the partonic cross section on the right-hand side as follows
\begin{equation}
\label{defimp}
\f{d{\hat \sigma}_{F \,ab}}{d q_T^2} \;
\longrightarrow \;
\Bigl[ \f{d{\hat \sigma}_{F \,ab}^{(\rm res.)}}{d q_T^2} \Bigr]_{\rm l.a.}
+ \; \Bigl[
\f{d{\hat \sigma}_{F \,ab}^{(\rm fin.)}}{d q_T^2} \Bigr]_{\rm f.o.} \, .
\end{equation}
The first and second terms on the right-hand side denote the truncation of 
the resummed and finite components at a given logarithmic accuracy and at a
given fixed order, respectively.
The resummed component gives the dominant contribution in the small-$\qt$
region, while the finite component dominates at large values of $\qt$.
The two components 
have to be consistently
matched at intermediate values of $q_T$, so as to obtain
a theoretical prediction with uniform formal accuracy over the
entire range of $\qt$, from $\qt \ll M$ up to $\qt \sim M$.
To this aim, we compute 
$\bigl[ d{\hat \sigma}_{ab}^{(\rm fin.)} \bigr]_{\rm f.o.}$
starting from $\bigl[ d{\hat \sigma}_{ab} \bigr]_{\rm f.o.}$,
the usual perturbative series
for the partonic cross section truncated at a given fixed order in $\as$,
and subtracting from it the perturbative truncation
of the resummed component at the {\em same} fixed order in $\as$:
\begin{equation}
\label{resfin}
\Bigl[ \f{d{\hat \sigma}_{F \,ab}^{(\rm fin.)}}{d q_T^2} \Bigr]_{\rm f.o.} =
\Bigl[\f{d{\hat \sigma}_{F \,ab}^{}}{d q_T^2}\Bigr]_{\rm f.o.}
- \Bigl[ \f{d{\hat \sigma}_{F \,ab}^{(\rm res.)}}{d q_T^2}\Bigr]_{\rm f.o.} \;.
\end{equation}
Moreover, we impose the condition:
\begin{equation}
\label{matchcon}
\left[
\Bigl[ \f{d{\hat \sigma}_{F \,ab}^{(\rm res.)}}{d q_T^2} \Bigr]_{\rm l.a.}
\right]_{\rm f.o.}
= \; \Bigl[
\f{d{\hat \sigma}_{F \,ab}^{(\rm res.)}}{d q_T^2} \Bigr]_{\rm f.o.} \, .
\end{equation}
This matching procedure guarantees that 
the replacement in 
Eq.~(\ref{defimp})
retains
the full information of the perturbative calculation up to
the specified fixed order plus resummation of
logarithmically-enhanced contributions from higher orders.
Equations (\ref{resfin}) and (\ref{matchcon}) indeed
imply that the matching 
is perturbatively exact, in the sense that 
the fixed-order truncation of the right-hand side of Eq.~(\ref{defimp})
exactly reproduces the customary fixed-order truncation of the
partonic cross section in Eq.~(\ref{dcross}).
The (small-$q_T$) resummed and (large-$q_T$)
fixed-order approaches are thus consistently combined 
without double-counting (or neglecting)
of perturbative contributions
and by avoiding the introduction of ad-hoc 
boundaries (such as, for instance, the choice of some intermediate value
of $\qt$ as `switching' point between the resummed 
and fixed-order calculations)
between the large-$q_T$ and small-$q_T$ regions.

The resummed contributions that are present in the term  
$\bigl[ d{\hat \sigma}_{F \,ab}^{(\rm res.)} \bigr]_{\rm l.a.}$
of Eq.~(\ref{defimp}) are necessary and fully justified
at small $\qt$. Nonetheless they can lead to sizeable higher-order
perturbative effects also at large $\qt$, where the small-$\qt$ logarithmic 
approximation is not valid. To reduce the impact of these unjustified
higher-order terms, 
we require that they give no contributions to the most basic quantity,
namely the total cross section,
that is not affected by small-$q_T$ logarithmic terms.
We thus impose that the integral over $q_T$ of Eq.~(\ref{defimp})
exactly reproduces the fixed-order calculation
of the total cross section. Since 
$d{\hat \sigma}_{F \,ab}^{(\rm fin.)}$ is evaluated in fixed-order 
perturbation theory, the perturbative constraint on the total cross
section is achieved by imposing 
the following condition:
\begin{equation}
\label{sigtotconst}
\int_0^{\infty} dq_T^2 \;
\Bigl[ \f{d{\hat \sigma}_{F \,ab}^{(\rm res.)}}{d q_T^2}\Bigr]_{\rm l.a.}
= \int_0^{\infty} dq_T^2 \;
\Bigl[ \f{d{\hat \sigma}_{F \,ab}^{(\rm res.)}}{d q_T^2}\Bigr]_{\rm f.o.}
\;. 
\end{equation}
Equation (\ref{sigtotconst}) can be regarded, in some sense, as a unitarity
constraint. As a matter of fact, the logarithmic contributions that are
resummed in $d{\hat \sigma}_{F \,ab}^{(\rm res.)}$ are, precisely speaking,
plus distributions of the type
$\left[ (\as^n/q_T^2)\, \ln^m (M^2/q_T^2) \right]_+$. Therefore, it is 
quite natural
to require that these resummed terms give a vanishing
contribution to the total cross section.
Note that the bulk of the $q_T$ distribution is in
the region $q_T \ltap M_H$. Since resummed and fixed-order perturbation theory
controls the small-$q_T$ and large-$q_T$ regions respectively,
the total cross section constraint mainly acts on the size of the higher-order
contributions introduced in the intermediate-$q_T$ region by the matching
procedure.

Another distinctive feature of the formalism illustrated so far
is that
we implement perturbative QCD resummation at the level of the partonic cross
section. In the factorization formula (\ref{dcross}), the parton densities are 
thus evaluated at the factorization scale $\mu_F$, as in the customary
perturbative calculations at large $q_T$. Although we are dealing with a 
process characterized by two distinct hard scales, $\qt$ and $M$,
the dominant effects
from the scale region $\qt \ll M$ are explicitly taken into account
through all-order resummation.
Therefore, the central value of $\mu_F$ 
and $\mu_R$ has to be set equal to $M_H$, the `remaining'
typical hard scale of the 
process. Then the theoretical accuracy of the resummed calculation can be 
investigated as in customary fixed-order calculations, 
by varying $\mu_F$ and $\mu_R$
around this central value.

At small values of $q_T$, the perturbative QCD approach has to be supplemented
with non-perturbative contributions, since they become relevant as 
$q_T$ decreases. A discussion on non-perturbative effects on the $q_T$
distribution of the SM Higgs boson is presented in Sect.~\ref{sec:num}.

The resummation and matching formalism, which we have so far illustrated
in quite general terms, is set up to deal with the transverse-momentum region
where $\qt \ltap M$. Resummation of small-$\qt$ logarithms
cannot lead to any theoretical improvements in the large-$\qt$ region, where
those logarithms are not the dominant contributions. When $\qt \gtap M$,
the use of the resummation formalism is no longer justified (recommended),
and we have to use the customary fixed-order perturbative expansion.

\subsection{The resummed component}
\label{sec:rescross}

The method to systematically resum the logarithmically-enhanced 
contributions at small $\qt$ was set up
[\ref{Parisi:1979se}--\ref{Kodaira:1981nh}]
shortly after the first resummed calculation of the DY $\qt$ spectrum
to double logarithmic accuracy [\ref{Dokshitzer:hw}].
The resummation procedure has to be carried out in the 
impact-parameter space, to correctly take into account the 
kinematics constraint of transverse-momentum conservation.
The resummed component
of the transverse-momentum cross section in Eq.~(\ref{resplusfin})
is then obtained by performing the inverse Fourier (Bessel) transformation 
with respect to the impact parameter $b$.
We write\footnote{The subscript $b$, which labels the parton flavour, 
should not be confused with the impact parameter $b$.}
\beeq
\label{resum}
\!\!\! \f{d{\hat \sigma}_{F \,ab}^{(\rm res.)}}{d q_T^2}(q_T,M,{\hat s};
\as(\mu_R^2),\mu_R^2,\mu_F^2) 
&=& \f{M^2}{\hat s} \;\int \f{d^2{\bf b}}{4\pi} \;e^{i {\bf b} \cdot \qt}
\;{\cal W}_{ab}^{F}(b,M,{\hat s};\as(\mu_R^2),\mu_R^2,\mu_F^2)  \\
\label{resum1}
&=&\f{M^2}{\hat s} \;
\int_0^\infty db \; \f{b}{2} \;J_0(b q_T) 
\;{\cal W}_{ab}^{F}(b,M,{\hat s};\as(\mu_R^2),\mu_R^2,\mu_F^2) \;,
\eeeq
where $J_0(x)$ is the 0th-order Bessel function.

The perturbative and process-dependent factor ${\cal W}_{ab}^{F}$
embodies the all-order dependence on 
the large logarithms $\ln M^2b^2$ at large $b$, which correspond to the
$q_T$-space terms $\ln M^2/q_T^2$ that are 
logarithmically enhanced at small $q_T$ 
(the limit $q_T \ll M$ corresponds to $Mb \gg 1$, since $b$
is the variable conjugate to $q_T$). 
Resummation of these large logarithms
is better expressed
by defining the $N$-moments\footnote{Throughout the paper, the $N$-moments
$h_N$ of any function $h(z)$ of the variable $z$ are defined as 
$h_N=\int_0^1 dz \,z^{N-1} \,h(z)$ . } 
${\cal W}_N$ of ${\cal W}$
with respect to $z=M^2/{\hat s}$ at fixed $M$:
\begin{equation}
\label{wndef}
{\cal W}_{ab, \, N}^{F}(b,M;\as(\mu_R^2),\mu_R^2,\mu_F^2) \equiv
\int_0^1 dz \;z^{N-1} \;{\cal W}_{ab}^{F}(b,M,{\hat s}=M^2/z;
\as(\mu_R^2),\mu_R^2,\mu_F^2) \;.
\end{equation}
The resummation structure of ${\cal W}_{ab, \, N}^{F}$ can indeed be 
organized in exponential form, as discussed below.

In the following of this subsection,
the subscripts denoting the flavour indices are 
understood.
More precisely, we present 
the resummation formulae in a simplified form, which is valid when
there is a single species of partons. This simplified form illustrates
more clearly the key structure and the main features of the resummed partonic
cross section. The generalization to considering more species of partons 
does not require any further conceptual steps: it just involves algebraic
complications, which are discussed in Sect.~\ref{sec:css} 
and in Appendix~\ref{appa}.

The logarithmic terms embodied in ${\cal W}_{ab, \, N}^{F}$ are due to
final-state radiation of partons that are soft and/or collinear to the
incoming partons. Their all-order resummation can be organized
[\ref{Catani:2000vq}] in close analogy to the cases of soft-gluon resummed
calculations for hadronic event shapes in hard-scattering processes
[\ref{Catani:1991kz}--\ref{Banfi:2004nk}]
and for threshold contributions to hadronic cross sections
[\ref{threshold}, \ref{Catani:1996yz}].
We write
\begin{align}
\label{wtilde}
{\cal W}_N^F(b,M;\as(\mu_R^2),\mu_R^2,\mu_F^2)
&={\cal H}_N^F\left(M, \as(\mu_R^2);M^2/\mu^2_R,M^2/\mu^2_F,M^2/Q^2
\right) \nonumber \\
&\times \exp\{{\cal G}_N(\as(\mu^2_R),L;M^2/\mu^2_R,M^2/Q^2
)\}
\;\;.
\end{align}
The function ${\cal H}_N^F$ 
does not depend on the impact parameter $b$ and, therefore, it contains
all the perturbative terms that behave as constants in the limit $b \to
\infty$. The function ${\cal G}$ includes the complete dependence on $b$ and,
in particular, it contains 
all the terms that order-by-order in $\as$ are logarithmically
divergent when $b \to \infty$.
This factorization between constant and logarithmic terms involves some degree
of arbitrariness [\ref{shapedis}],
since the argument of the large logarithms can always be rescaled as 
$\ln M^2b^2= \ln Q^2b^2 + \ln M^2/Q^2$, provided that $Q$ is independent 
of $b$ and that $\ln M^2/Q^2={\cal O}(1)$ when $bM \gg 1$.
To parametrize this arbitrariness, on the right-hand side of 
Eq.~(\ref{wtilde}) we have introduced the scale $Q$, such
that $Q \sim M$, and we have defined the large logarithmic 
expansion parameter, $L$, as
\begin{equation}
\label{ldef}
L \equiv \ln \f{Q^2b^2}{b_0^2} \;\;,
\end{equation}
where the coefficient $b_0=2e^{-\gamma_E}$ 
($\gamma_E=0.5772\dots$ is the Euler number) has a kinematical origin
(the use of $b_0$ in Eq.~(\ref{ldef}) in purely conventional: it 
simplifies the algebraic expression of ${\cal G}$). 

The role played by the auxiliary scale $Q$ (which we name the `resummation
scale') in the context of the resummation program is analogous to the role
played by the renormalization (factorization) scale in the context of
renormalization (factorization). Although the resummed cross section 
${\cal W}_{N}^{F}$ does not depend on $Q$ when evaluated to all perturbative
orders, its explicit dependence on $Q$ appears when ${\cal W}_{N}^{F}$ is
computed by truncation of the resummed expression 
at some level of logarithmic accuracy (see below).
As in the case of $\mu_R$ and $\mu_F$, we should set $Q$ at the 
central value $Q=M$; variations of the resummation scale $Q$ around this 
central value can then be used to estimate the uncertainty from
yet uncalculated logarithmic corrections at higher orders.
Note that the resummation scale dependence of ${\cal W}_{N}^{F}$ should not
be confused with the `resummation scheme' dependence considered in 
Ref.~[\ref{Catani:2000vq}]. In fact, as shown in Sect.~\ref{sec:css}, 
${\cal W}_{N}^{F}$ is exactly independent of the resummation scheme.

All the large logarithmic terms $\as^n L^m$
with $1\leq m \leq 2n$ 
are included  
in the form factor $\exp\{{\cal G}\}$. More importantly,
all the logarithmic contributions to ${\cal G}$ with $n+2 \leq m \leq 2n$
are vanishing. This property, which is called exponentiation,
follows
[\ref{Parisi:1979se}--\ref{Kodaira:1981nh}]
from the perturbative dynamics of (abelian and non-abelian)
gauge theories and from kinematics factorization in impact parameter space.
Thus, the exponent ${\cal G}$ can systematically
be expanded as
\begin{align}
\label{gexpan}
{\cal G}_N(\as,L;M^2/\mu^2_R,M^2/Q^2) &=
L \,g^{(1)}(\as L)+
g_N^{(2)}(\as L;M^2/\mu^2_R,M^2/Q^2 ) \nn \\
&+
\f{\as}{\pi} \;g_N^{(3)}(\as L;M^2/\mu^2_R,M^2/Q^2 ) \\
&+ \sum_{n=4}^{+\infty} \left(\f{\as}{\pi}\right)^{n-2} 
\;g_N^{(n)}(\as L;M^2/\mu^2_R,M^2/Q^2 ) 
\;\;, \nn
\end{align}
where $\as=\as(\mu^2_R)$ and the functions $g^{(n)}(\as L)$ are defined such
that $g^{(n)}=0$ when $\as L=0$. Thus
the term $L g^{(1)}$ collects the LL contributions $\as^n L^{n+1}$;
the function $g^{(2)}$ resums
the NLL contributions $\as^n L^{n}$; $g^{(3)}$ controls the NNLL terms
$\as^n L^{n-1}$, and so forth. 
Note that in the context of the resummation approach, 
the parameter $\as L$
is formally considered as being of order unity. Thus, the ratio of two
successive terms in the expansion (\ref{gexpan}) is formally of ${\cal O}(\as)$
(with no $L$ enhancement). In this respect, 
the resummed logarithmic expansion in Eq.~(\ref{gexpan}) is as systematic 
as any customary fixed-order expansion in powers of $\as$.

The function ${\cal H}_N^F$ in Eq.~(\ref{wtilde}) does not contain
large logarithmic terms to be resummed.
It can be expanded in powers of $\as=\as(\mu^2_R)$ as
\begin{align}
\label{hexpan}
{\cal H}_N^F(M,\as;M^2/\mu^2_R,M^2/\mu^2_F,M^2/Q^2)&=
\sigma_F^{(0)}(\as,M)
\Bigl[ 1+ \f{\as}{\pi} \,{\cal H}_N^{F \,(1)}(M^2/\mu^2_R,M^2/\mu^2_F,M^2/Q^2) 
\Bigr. \nn \\
&+ \Bigl.
\left(\f{\as}{\pi}\right)^2 
\,{\cal H}_N^{F \,(2)}(M^2/\mu^2_R,M^2/\mu^2_F,M^2/Q^2) 
\Bigr. \\
&+ \Bigl. \sum_{n=3}^{+\infty} \left(\f{\as}{\pi}\right)^{\!n}
\,{\cal H}_N^{F \,(n)}(M^2/\mu^2_R,M^2/\mu^2_F,M^2/Q^2)
 \; \Bigr] \;\;, \nn
\end{align}
where 
$\sigma_F^{(0)}= \as^p \;\sigma_F^{({\rm LO})}$
is the lowest-order partonic cross section for the
hard-scattering process in Eq.~(\ref{process}).

Two other general aspects of the resummed partonic cross section
${\cal W}_{N}^{F}$ are the factorization scale (and scheme) dependence
and the process dependence. As discussed below, the form factor
$\exp\{{\cal G}\}$ does {\em not} depend on both the factorization scale 
(and scheme) and the specific hard-scattering process.

The hadronic cross section on the left-hand side of Eq.~(\ref{dcross})
is a physical observable and cannot depend on the 
factorization scale $\mu_F$. In practice, the evaluation of 
the right-hand side at a certain perturbative accuracy
introduces the $\mu_F$ dependence of the partonic cross section
$d{\hat \sigma}_{F \,ab}$. This dependence is perturbatively 
balanced by 
the $\mu_F$ dependence of the parton densities $f_{a/h}(x,\mu_F^2)$.
Note that the parton densities in Eq.~(\ref{dcross})
do not depend on the transverse momentum $\qt$
(or on the impact parameter $b$). Recall also that
we implement 
transverse-momentum resummation at the level of the partonic cross section
$d{\hat \sigma}_{F \,ab}$,
by using Eqs.~(\ref{resum}) and (\ref{wtilde}).
Therefore, any $\mu_F$ dependence of the parton
densities cannot introduce 
any {\em logarithmic} dependence on 
$b$ in the form factor $\exp\{{\cal G}\}$.
In other words, the 
perturbative expansion (\ref{hexpan}) of the function ${\cal H}_N^F$
depends on $\mu_F$, while the exponent ${\cal G}$ of the form
factor and its corresponding logarithmic functions $g_N^{(n)}$ in
Eq.~(\ref{gexpan}) do not depend on $\mu_F$ and on the factorization
scheme used to define the parton densities.

As explicitly shown in Sect.~\ref{sec:css}, the form factor 
$\exp\{{\cal G}\}$ in Eq.~(\ref{wtilde}) does not depend on the final-state
system $F$ produced in the hard-scattering process of Eq.~(\ref{process}).
The form factor is process independent: it
is produced by universal soft and collinear radiation from the QCD
partons entering the hard-scattering process
(when the simplification of considering a single parton species is removed,
there are various process-independent form factors for the various partonic
channels). The dependence on the process is fully taken into account by the
hard-scattering function ${\cal H}_N^F$, which embodies contributions produced
by virtual corrections at transverse-momentum scales $\qt \sim M$.

The truncation $\bigl[ {\cal W}_{N}^{F} \bigr]_{\rm l.a.}$
of the resummed cross section at a given logarithmic accuracy is defined 
as follows. At LL accuracy, we include the function $g^{(1)}$ in the 
exponent ${\cal G}$ and we approximate ${\cal H}_N^F$ by
the Born cross section $\sigma_F^{(0)}$.
At NLL accuracy, we include the functions $g^{(1)}$ and $g_N^{(2)}$
and the coefficient ${\cal H}_N^{F \,(1)}$.
At NNLL accuracy, we also include $g_N^{(3)}$ and ${\cal H}_N^{F \,(2)}$.
The reason for including both ${\cal H}_N^{F \,(1)}$ and $g_N^{(2)}$ 
at NLL accuracy is that the combined effect of $\as {\cal H}_N^{F \,(1)}$
and $L g^{(1)}(\as L)$ leads to logarithmic contributions, 
$\as^n L^n$,
that are of the same order as those in $g_N^{(2)}(\as L)$.
An analogous observation 
applies to the inclusion of both 
$g_N^{(3)}$ and ${\cal H}_N^{F \,(2)}$ at NNLL accuracy.

The logarithmic truncation of the resummed component of the cross section
can then be combined, as in Eq.~(\ref{defimp}), 
with the fixed-order expansion of the finite component in Eq.~(\ref{resfin}).
The NLL+LO result is obtained by supplementing NLL resummation with
the LO expansion\footnote{We recall that there is a mismatch of notation
between the $\qt$ distribution at $\qt \sim M$ and the total cross section.
The LO (NLO) term of the finite component of the $\qt$ distribution 
contributes to the total cross section at NLO (NNLO).} at large $\qt$.
The NNLL+NLO result combines NNLL resummation with the NLO expansion
at large $\qt$.
This procedure for combining the resummed and fixed-order approaches 
exactly satisfies the 
matching conditions 
in Eqs.~(\ref{deffin}) and (\ref{matchcon}).
Note that the fulfilment of the matching conditions is not completely
trivial.
For instance, if ${\cal H}_N^{F \,(1)}$ was not included in
$d{\hat \sigma}_{F}^{(\rm res.)}$ at NLL accuracy, the matching condition
in Eq.~(\ref{matchcon}) would be violated at LO
(in other words, Eq.~(\ref{deffin}) would be violated since the 
$\qt$ integral of  
$\bigl[ d{\hat \sigma}_{F \,ab}^{(\rm fin.)}\bigr]_{\rm LO}$
would lead to a non-vanishing finite value when $Q_T \to 0$).

To reduce the impact of unjustified resummed logarithms in the large-$\qt$
region, we use a procedure inspired by that introduced 
in Ref.~[\ref{Catani:1992ua}]
to deal with kinematical constraints when performing
soft-gluon resummation in $e^+e^-$ event shapes. We consider the 
exponent ${\cal G}(\as, L)$ of the form factor in Eqs.~(\ref{wtilde})
and (\ref{gexpan}) and we perform the replacement
\begin{equation}
\label{grepl}
{\cal G}(\as, L) \; \longrightarrow \; {\cal G}(\as, \tL) \;\;.
\end{equation}
In other words, in the argument of ${\cal G}(\as, L)$ we
replace the logarithmic variable $L$ with the variable $\tL$
defined as
\begin{equation}
\label{logdef}
\tL \equiv \ln \left(\f{Q^2b^2}{b_0^2}+ 1 \right) \;\;.
\end{equation}
Comparing the definitions in Eqs.~(\ref{ldef}) and (\ref{logdef}), we see
that in the resummation region $Qb \gg 1$ we have 
$\tL = L +{\cal O}(1/(Qb)^2)$, and thus the replacement in Eq.~(\ref{grepl})
is fully legitimate\footnote{Note that the replacement
in Eq.~(\ref{grepl}) introduces an explicit dependence of
$d{\hat \sigma}_{F}^{(\rm res.)}$ on the resummation scale~$Q$.
Owing to the matching procedure in Eq.~(\ref{resfin}), 
this dependence is balanced by
the $Q$ dependence of the $d{\hat \sigma}_{F}^{(\rm fin.)}$.}
to arbitrary logarithmic accuracy.
Although the variables $L$ and $\tL$ are equivalent to organize the 
resummation formalism in the region $Qb \gg 1$, they lead to a different
behaviour of the form factor at small values of $b$ 
(i.e. large values of $\qt$):
when $Qb \ll 1$, we have $\tL \to 0$ and $\exp \{{\cal G}(\as, \tL)\} \to 1$.
Therefore, performing the replacement in Eq.~(\ref{grepl}), we reduce the
effect produced by the resummed contributions
in the small-$b$ region, where the use of 
the large-$b$ resummation approach is not justified.

In particular, since $\exp \{{\cal G}(\as, \tL)\} = 1$ at $b=0$, 
using Eqs.~(\ref{resum}) and (\ref{wtilde}) we obtain  
the relation
\begin{equation}
\label{restot}
\int_0^{\infty} dq_T^2 \;
\f{d{\hat \sigma}_{F}^{(\rm res.)}}{d q_T^2}(q_T,M,{\hat s};
\as(\mu_R^2),\mu_R^2,\mu_F^2,Q^2) 
= \f{M^2}{\hat s} \;
{\cal H}^F\!\left(M,{\hat s},\as(\mu_R^2);M^2/\mu^2_R,M^2/\mu^2_F,M^2/Q^2 
\right) \;,
\end{equation}
which simply follows from the fact that 
the value at $b=0$ of the ($b$-space) Fourier transformation  
of the $\qt$ distribution is equal to the integral over $\qt$ of the 
$\qt$ distribution itself.
Since the hard cross section ${\cal H}^F$ is evaluated in fixed-order
perturbation theory,
the relation (\ref{restot}) implies that the replacement in Eq.~(\ref{grepl})
also allows us to implement the perturbative constraint (\ref{sigtotconst}) 
on the total cross section.
More precisely, the integral over $\qt$ of the $\qt$
distribution $d{\hat \sigma}_F/dq_T$ at NLL+LO (NNLL+NLO) accuracy
exactly reproduces the calculation of the total cross section at NLO (NNLO).

The purpose of the transverse-momentum resummation program 
[\ref{Parisi:1979se}--\ref{Kodaira:1981nh}]
is to explicitly evaluate
the logarithmic functions $g_N^{(n)}$ of Eq.~(\ref{gexpan}) in terms of few
coefficients that are perturbatively computable. As illustrated 
in Sect.~{\ref{sec:css},
this goal is achieved by showing that 
the all-order resummation formula (\ref{gexpan}) has the  
following integral representation: 
\begin{equation}
\label{gformfact}
{\cal G}_N(\as(\mu^2_R),L;M^2/\mu^2_R,M^2/Q^2)
 =   - \int_{b_0^2/b^2}^{Q^2} \frac{dq^2}{q^2} 
\left[ A(\as(q^2)) \;\ln \frac{M^2}{q^2} + {\widetilde B}_N(\as(q^2)) \right]  
\;\;, 
\end{equation}
where $A(\as)$ and ${\widetilde B}_N(\as)$ are perturbative functions
\beeq
\label{Afun}
A(\as) &=& {\asp} A^{(1)}+\left(\asp \right)^2 A^{(2)} 
+\left(\asp \right)^3 A^{(3)} +
\sum_{n=4}^\infty \left( \frac{\as}{\pi} \right)^n A^{(n)} 
\;\;, \\
\label{Btfun}
{\widetilde B}_N(\as) &= & {\asp} {\widetilde B}_N^{(1)}+
\left(\asp \right)^2 {\widetilde B}_N^{(2)} +
\sum_{n=3}^\infty \left( \frac{\as}{\pi} \right)^n {\widetilde B}_N^{(n)}
\;\;. 
\eeeq
The coefficients $A^{(n)}$ and ${\widetilde B}_N^{(n)}$ are related to the
customary coefficients of the Sudakov form factors and of the parton anomalous
dimensions. This relation is discussed in Sect.~\ref{sec:css}.

Using Eq.~(\ref{resum}), the resummed component 
$d{\hat \sigma}_{F}^{(\rm res.)}/{dq_T^2}$
of the $\qt$ distribution is fully determined by the functions 
${\cal H}_N^F$ and ${\cal G}_N$ in Eq.~(\ref{wtilde}). These
functions are in turn specified by the perturbative coefficients
${\cal H}_N^{F \,(n)}$ (see Eq.~(\ref{hexpan})), $A^{(n)}$ and 
${\widetilde B}_N^{(n)}$ (see Eqs.~(\ref{gformfact})--(\ref{Btfun})),
which can be extracted from the logarithmic terms in
the perturbative expansion of the 
$\qt$ distribution at the $n$-th relative order in $\as$.
Therefore, the customary fixed-order computation of the
$\qt$ distribution is sufficient to obtain the full information
that is necessary 
to explicitly perform resummation at the required logarithmic
accuracy.

By inspection of the $q^2$ integration in Eq.~(\ref{gformfact}),
it is evident that the exponent ${\cal G}_N$ of the process-independent form
factor in Eq.~(\ref{wtilde}) has
the logarithmic structure of Eq.~(\ref{gexpan}).
The functions $g_N^{(n)}$ depend on the coefficients in Eqs.~(\ref{Afun}) and
(\ref{Btfun}), and the functional dependence is completely specified by
Eq.~(\ref{gformfact}).
More precisely (see Eqs.~(\ref{g1fun})--(\ref{g3fun})),
the LL function $g_N^{(1)}$ depends on $A^{(1)}$, the NLL function 
$g_N^{(2)}$ depends also on $A^{(2)}$ and ${\widetilde B}_N^{(1)}$, 
the NNLL function $g_N^{(3)}$ depends 
also on $A^{(3)}$ and ${\widetilde B}_N^{(2)}$, and so forth.
Starting from the integral representation in Eq.~(\ref{gformfact}),
the explicit functional form of the functions $g_N^{(n)}$ 
(for arbitrary values of $n$) can easily be computed by using the 
method that is described in Appendix~C of Ref.~[\ref{Catani:2003zt}].

The LL, NLL and NNLL functions $g_N^{(1)}$, $g_N^{(2)}$ and $g_N^{(3)}$
have the following explicit expressions\footnote{Note that the functional form
of the functions $g_N^{(n)}$ is exactly the same as that of the functions
that appear in the calculation of the energy--energy correlation in $e^+e^-$
annihilation [\ref{deFlorian:2004mp}].}:
\begin{align}
\label{g1fun}
g^{(1)}(\as L) &= \f{A^{(1)}}{\beta_0} \f{\la+\ln(1-\la)}{\la} \;\;,  \\
\label{g2fun}
g_N^{(2)}\!\left(\as L;\f{M^2}{\mu^2_R},\f{M^2}{Q^2} \right)
&= \f{{\overline B}_N^{(1)}}{\beta_0}
\ln(1-\la) -\f{A^{(2)}}{\beta_0^2} 
\left( \f{\la}{1-\la} +\ln(1-\la)\right) \nn \\ 
&+ \f{A^{(1)}}{\beta_0} 
\left( \f{\la}{1-\la} +\ln(1-\la)\right) \ln\f{Q^2}{\mu_R^2}  \nn \\
& +\f{A^{(1)} \beta_1}{\beta_0^3} \left( \f{1}{2} \ln^2(1-\la)+ 
\f{\ln(1-\la)}{1-\la} + \f{\la}{1-\la}  \right) \;,  \\
\label{g3fun}
g_N^{(3)}\!\left(\as L;\f{M^2}{\mu^2_R},\f{M^2}{Q^2} \right)
&= -\f{A^{(3)}}{2 \beta_0^2} \f{\la^2}{(1-\la)^2}
-\f{{\overline B}_N^{(2)}}{\beta_0} \f{\la}{1-\la}  
+\f{A^{(2)} \beta_1}{\beta_0^3} \left( \f{\la (3\la-2)}{2(1-\la)^2}
 - \f{(1-2\la) \ln(1-\la)}{(1-\la)^2} \right) \nn \\
& + \f{{\overline B}_N^{(1)} \beta_1}{\beta_0^2} \left( \f{\la}{1-\la} + 
\f{\ln(1-\la)}{1-\la} \right) - \f{A^{(1)}}{2} \f{\la^2}{(1-\la)^2}  
\ln^2\f{Q^2}{\mu_R^2} \nn \\
&+ \ln\f{Q^2}{\mu_R^2} \left( {\overline B}_N^{(1)} \f{\la}{1-\la} + 
\f{A^{(2)}}{\beta_0}
 \f{\la^2}{(1-\la)^2} + A^{(1)} \f{\beta_1}{\beta_0^2}
 \left( \f{\la}{1-\la} + \f{1-2\la}{(1-\la)^2} \ln(1-\la) \right) \right) 
 \nn \\
&   +A^{(1)} \left( \f{\beta_1^2}{2 \beta_0^4} \f{1-2\la}{(1-\la)^2} 
\ln^2(1-\la) 
+ \ln(1-\la) \left[  \f{\beta_0 \beta_2 -\beta_1^2}{\beta_0^4} 
+\f{\beta_1^2}{\beta_0^4 (1-\la)}  \right] \right.  \nn \\
& \left. + \f{\la}{2 \beta_0^4 (1-\la)^2} ( \beta_0 \beta_2 (2-3\la)+
\beta_1^2 \la) \right) \;,       
\end{align}
where
\begin{equation}
\lambda = \frac{1}{\pi} \,\beta_0 \,\as(\mu_R^2) \,L \;\;,
\end{equation}
\begin{equation}
{\overline B}_N^{(n)} = {\widetilde B}_N^{(n)} + 
A^{(n)} \ln\f{M^2}{Q^2}\;\;,
\end{equation}
and $\beta_n$ are the coefficients of the QCD $\beta$ function: 
\begin{equation}
\label{rge}
\f{d \ln \as(\mu^2)}{d \ln \mu^2} = \beta(\as(\mu^2)) = 
- \sum_{n=0}^{+\infty} \beta_n \left( \f{\as}{\pi} \right)^{n+1}\;\;.
\end{equation}
The explicit expression of the first three coefficients, $\beta_0$,
$\beta_1$ and $\beta_2$, is [\ref{beta2}]
\begin{align}
\beta_0 &= \frac{1}{12} \left( 11 C_A - 2 N_f \right) \;\;,
\quad\quad \beta_1=  \frac{1}{24} 
\left( 17 C_A^2 - 5 C_A N_f - 3 C_F N_f \right) \;\;,
\nonumber \\
\label{bcoef}
\beta_2 &= \frac{1}{64} \left( \f{2857}{54} C_A^3
- \f{1415}{54} C_A^2 N_f - \f{205}{18} C_A C_F N_f + C_F^2 N_f
+ \f{79}{54} C_A N_f^2 + \f{11}{9} C_F N_f^2 \right) \;\;,
\end{align}
where $N_f$ is the number of QCD massless flavours and the $SU(N_c)$ colour
factors are $C_A=N_c$  and $C_F=(N_c^2-1)/(2N_c)$.

\setcounter{footnote}{0}

Note that the functions $g_N^{(n)}(\as L)$ in Eqs.~(\ref{g1fun})--(\ref{g3fun}) 
are singular at the point $\lambda=1$,
which in terms of the impact parameter  
corresponds to the value 
$b^2=b_L^2=(b_0^2/Q^2) \exp\{\pi/(\beta_0\as(\mu_R^2))\}$ 
(i.e. $b_L \sim 1/\Lambda_{QCD}$, where $\Lambda_{QCD}$ is the momentum
scale of the Landau pole in QCD).
These singularities, which are related (see Eq.~(\ref{gformfact}) when $b \sim
1/\Lambda_{QCD}$)
to the divergent behaviour of the perturbative running coupling  
$\as(q^2)/\pi \sim [\beta_0 \ln (q^2/\Lambda_{QCD}^2)]^{-1}$
near the Landau pole, 
signal the onset of
non-perturbative phenomena at very large values of $b$ or, equivalently,
in the region of very small transverse momenta.

This type of singularities\footnote{Note that these singularities are not
related to the presence of factorially-growing coefficients, such as those
due to renormalon
singularities, at very high perturbative orders. A concise discussion
on this point can be found in Sect.~3.1 of Ref.~[\ref{Catani:1996yz}], in the
related context of threshold resummation.} 
is a common feature of all-order resummation 
formulae of soft-gluon contributions. Within a perturbative framework, these
singularities have to be 
regularized. A possible regularization 
procedure consists in introducing a `minimal prescription', such as those
introduced in Refs.~[\ref{Catani:1996yz}] 
(in the case of threshold resummation) and 
[\ref{Laenen:2000de}, \ref{Kulesza:2003wn}] (in the case of $b$-space or joint
resummation).
In the case of $b$-space resummation, other procedures are to use the 
`$b_*$ prescription' of Ref.~[\ref{Collins:va}], by freezing the integration 
over $b$ below a fixed upper limit, or more simply, to introduce a 
cut-off at a very large (but smaller than $b_L$) value of $b$ 
[\ref{Qiu:2000ga}].
Admittedly, when the non-perturbative contributions are sizeable, they have to
be properly included, according to the prescription used 
to regularize the singularities.

\subsection{Sudakov form factor, universal form factor 
and perturbative coefficients}
\label{sec:css}

The $b$-space resummation approach was fully formalized by Collins, Soper and
Sterman [\ref{Collins:1981uk}, \ref{Collins:1984kg}]
in terms of perturbative coefficients. Considering the generic
hard-scattering process in Eq.~(\ref{process}), the 
transverse-momentum differential cross section in Eq.~(\ref{dcross})
is written as 
\begin{equation}
\label{sigcss}
\f{d\sigma_F}{d q_T^2}(q_T,M,s) = \f{M^2}{s} \;
\int_0^\infty db \; \f{b}{2} \;J_0(b q_T) 
\;W^{F}(b,M,s) + \dots \;,
\end{equation}
where the dots on the right-hand side stand for terms that are not
logarithmically enhanced at small $\qt$ (large $b$).
Note that Eq.~(\ref{sigcss}) regards the hadronic cross section (and not the
partonic cross section in Eq.~(\ref{resum1})). Therefore, the $b$-space 
function $W^{F}(b,M,s)$, which embodies the logarithmically-enhanced terms,
depends on the parton densities of the colliding hadrons. The all-order
resummation of the large logarithms $\ln (M^2b^2)$ in the region $Mb \gg 1$
is accomplished by showing that the $N$-moments $W_N(b,M)$ of $W(b,M,s)$
with respect to $z=M^2/s$ at fixed $M$
can be recast in the following form [\ref{Collins:1984kg}, 
\ref{Catani:2000vq}]:
\beeq
\label{wcss}
W_{N}^{F}(b,M) &=& \sum_c \sigma_{c{\bar c}, \,F}^{(0)}(\as(M^2),M) 
\;H_c^{F}(\as(M^2)) \;S_c(M,b) \nn \\
&\times&\sum_{a,b}
C_{ca, \,N}(\as(b_0^2/b^2)) \;C_{{\bar c}b, \,N}(\as(b_0^2/b^2)) 
\; f_{a/h_1, \,N}(b_0^2/b^2) \;f_{b/h_2, \,N}(b_0^2/b^2) \;\;,
\eeeq
where $f_{a/h, \,N}(\mu^2)$ are the $N$-moments of the parton density
$f_{a/h}(z,\mu^2)$, and $\sigma_{c{\bar c}, \,F}^{(0)}$ is the 
lowest-order cross section for the partonic subprocess $c+{\bar c} \to F$. 
The function $S_c(M,b)$ is the Sudakov form factor of the quark 
($c=q, {\bar q}$) or of the gluon ($c=g$), and it has the following 
expression\footnote{In Ref.~[\ref{Collins:1984kg}] the upper limit of the 
integral in Eq.~(\ref{formfact}) is set to $C_2M^2$, where $C_2$ is an arbitrary
factor. The scale $C_2M^2$ is thus related to the resummation scale $Q^2$ in
Eq.~(\ref{gformfact}).}:
\begin{equation}
\label{formfact}
S_c(M,b) = \exp \left\{ - \int_{b_0^2/b^2}^{M^2} \frac{dq^2}{q^2} 
\left[ A_c(\as(q^2)) \;\ln \frac{M^2}{q^2} + B_c(\as(q^2)) \right] \right\} 
\;\;. 
\end{equation}
The functions $A, B, C$ and $H^{F}$
in Eqs.~(\ref{wcss}) and (\ref{formfact}) are perturbative series in $\as$:
\beeq
\label{aexp}
A_c(\as) &=& \sum_{n=1}^\infty \left( \frac{\as}{\pi} \right)^n A_c^{(n)} 
\;\;, \\
\label{bexp}
B_c(\as) &= &\sum_{n=1}^\infty \left( \frac{\as}{\pi} \right)^n B_c^{(n)}
\;\;, \\
\label{cexp}
C_{ab}(\as,z) &=& \delta_{ab} \,\delta(1-z) + 
\sum_{n=1}^\infty \left( \frac{\as}{\pi} \right)^n C_{ab}^{(n)}(z) 
\;\;, \\
\label{hexp}
H_c^{F}(\as) &=& 1 +  
\sum_{n=1}^\infty \left( \frac{\as}{\pi} \right)^n H_c^{F \,(n)} \;\;.
\eeeq
The functions $A_c, B_c$ and $C_{ab}$ are process independent,
while $H_c^{F}$ depends on the specific hard-scattering process.

The resummation formulae (\ref{wcss}) and (\ref{formfact}) are
invariant under the following `resummation scheme'  
transformations [\ref{Catani:2000vq}]:
\beeq
H_c^{F}(\as) & \to & H_c^{F}(\as) \; \left[ \, h(\as) \, \right]^{-1}
\;, \nn \\
\label{restranf}
B_c(\as) & \to & B_c(\as) - \beta(\as) \;\frac{d\ln h(\as)}{d\ln \as} 
\;,\\
C_{ab}(\as,z) & \to & C_{ab}(\as,z) \;
\left[ \, h(\as) \, \right]^{1/2} \;. \nn
\eeeq
The invariance can easily be proven by using the following
renormalization-group identity (see Eq.~(\ref{rge})):
\begin{equation}
\label{rgiden}
h(\as(b_0^2/b^2)) = h(\as(M^2)) \; \exp \left\{ 
-\int_{b_0^2/b^2}^{M^2} \frac{dq^2}{q^2} 
\;\beta(\as(q^2)) \;\frac{d \ln h(\as(q^2))}{d \ln \as(q^2)} 
\right\} \;\;,
\end{equation}
which is valid for any perturbative function $h(\as)$.

The physical origin of the 
resummation scheme invariance of Eq.~(\ref{wcss}) is discussed in 
Ref.~[\ref{Catani:2000vq}]. The invariance implies that the
factors $H_c^{F}, S_c$ (more precisely, the function $B_c$) 
and $C_{ab}$ are not unambiguously
computable order by order in perturbation theory. 
In other words, these factors can be unambiguously defined only
by choosing a `resummation scheme'.
The choice of a resummation scheme amounts to
defining $H_c^{F}$ (or $C_{ab}$) for a {\em single} process.
More precisely,
$H_c^{F}$ has to be defined for two processes: one process that 
is controlled, at the lowest perturbative order, 
by $q{\bar q}$ annihilation $(c=q,{\bar q})$ and another process that  
is controlled
by $gg$ fusion $(c=g)$.
Having done that, 
the process-dependent
factor $H_c^{F}$ and the universal (process-independent) factors 
$S_c$ and $C_{ab}$ 
are unambiguously determined for any other process of the type
in Eq.~(\ref{process}).

Note that Eq.~(\ref{wcss}) is usually presented in a form where
$H_c^{F}(\as)=1$. Such a form is certainly consistent since, by choosing
$h(\as)=H_c^{F}(\as)$ and using
the invariance under the transformation in Eq.~(\ref{restranf}), it is always
possible to set $H_c^{F}(\as)=1$ on a process-dependent basis.
Note that this procedure does not correspond to the definition of a   
resummation scheme. Indeed, the corresponding Sudakov form factor
$S_c^F$ and the functions $C_{ab}^F$
turn out to be process-dependent quantities,
as pointed out by the explicit and general calculation 
of $B_c^{(2)}$ and $C_{ab}^{(1)}(z)$ in Ref.~[\ref{deFlorian:2000pr}].
For example, in the case of $gg$ fusion processes, the Sudakov form factors
for the production of a scalar and a pseudoscalar Higgs boson turn out to be
different and to have even a different dependence on the mass of the 
top quark. 

Comparing the partonic and the hadronic cross sections in Eqs.~(\ref{resum1})
and (\ref{wcss}), we see that the resummed factors ${\cal W}_{ab}^{F}$ and
$W^{F}(b,M)$ are related by
\begin{equation}
\label{wrelation}
W_{N}^{F}(b,M) = \sum_{a,b}
{\cal W}_{ab, \,N}^{F}(b,M;\as(\mu_R^2),\mu_R^2,\mu_F^2)
\; f_{a/h_1, \,N}(\mu_F^2) \;f_{b/h_2, \,N}(\mu_F^2) \;\;.
\end{equation}
To express the resummed partonic cross section ${\cal W}_{ab}^{F}$
in terms of the perturbative coefficients in Eqs.~(\ref{aexp})--(\ref{hexp}),
we have to use Eq.~(\ref{wcss}) and substitute the parton densities
$f_{a/h, \,N}(b_0^2/b^2)$ for the same parton densities evaluated
at the factorization scale $\mu_F$. The substitution can be done by using 
\begin{equation}
\label{apev}
f_{a/h, \,N}(\mu^2) = 
\sum_{b} \; U_{ab, \,N}(\mu^2, \mu_0^2) \;f_{b/h, \,N}(\mu_0^2) 
\;\;,
\end{equation}
where the QCD evolution operator $U_{ab, \,N}(\mu^2, \mu_0^2)$ 
fulfils the evolution equations
\begin{equation}
\label{apeq}
\frac{ d \,U_{ab, \,N}(\mu^2, \mu_0^2)}{d \ln \mu^2} =
\sum_{c} \gamma_{ac, \,N}(\as(\mu^2)) \;U_{cb, \,N}(\mu^2, \mu_0^2) \;\;,
\end{equation}
and $\gamma_{ab, \,N}(\as)$ are the parton anomalous dimensions or,
more precisely, the $N$-moments of the customary Altarelli--Parisi
splitting functions $P_{ab}(\as,z)$ [\ref{book}]:
\begin{equation}
\label{apexp}
\gamma_{ab, \,N}(\as) = \int_0^1 dz \; z^{N-1} \;P_{ab}(\as,z) =
\sum_{n=1}^\infty \left( \frac{\as}{\pi} \right)^n \gamma_{ab, \,N}^{(n)} 
\;\;.
\end{equation}
We finally obtain [\ref{Catani:2000vq}]
\beeq
\label{calw}
{\cal W}_{ab, \,N}^{F}(b,M;\as(\mu_R^2),\mu_R^2,\mu_F^2) \!\!
&=&\!\! \sum_c \sigma_{c{\bar c}, \,F}^{(0)}(\as(M^2),M) 
\;H_c^{F}(\as(M^2)) \;S_c(M,b) 
\nn \\
\!\!&\times&\!\!\sum_{a_1, \,b_1}
C_{ca_1, \,N}(\as(b_0^2/b^2)) \;C_{{\bar c}b_1, \,N}(\as(b_0^2/b^2)) \\
\!\!&\times&\!\!U_{a_1a, \,N}(b_0^2/b^2,\mu_F^2) 
\;U_{b_1b, \,N}(b_0^2/b^2,\mu_F^2) \nn
\;\;,
\eeeq
which relates the resummed partonic cross section in Eq.~(\ref{resum1})
to the perturbative coefficients in Eqs.~(\ref{aexp})--(\ref{hexp})
and the anomalous dimensions coefficients in Eq.~(\ref{apexp}).

In the following we explicitly show how Eq.~(\ref{calw}) is related to the
exponential structure of Eq.~(\ref{wtilde}) in the case with a single
species of partons.
The general case with partons of different flavours
is discussed in Appendix~\ref{appa}. Here we only anticipate that the
generalization of Eq.~(\ref{wtilde}) 
to the multiflavour case\footnote{In the multiflavour case, 
Eq.~(\ref{wtilde}) directly applies to
the flavour non-singlet components of the resummed partonic cross section.}
simply involves a
sum of exponential terms, namely
\begin{align}
\label{wtildeflav}
{\cal W}_{ab, \,N}^F(b,M;\as(\mu_R^2),\mu_R^2,\mu_F^2)
&=\sum_{\{I\}} {\cal H}_{ab, \,N}^{\{I\}, \,F}\left(M, 
\as(\mu_R^2);M^2/\mu^2_R,M^2/\mu^2_F,M^2/Q^2
\right) \nonumber \\
&\times \exp\{{\cal G}_{\{I\}, \,N}(\as(\mu^2_R),L;M^2/\mu^2_R,M^2/Q^2
)\}
\;\;,
\end{align}
where the index $\{I\}$ labels a set of flavour indices (which is precisely
specified in Appendix~\ref{appa}).

Within the simplified treatment in which there is a 
single species of partons, the resummed partonic cross section
in Eq.~(\ref{calw}) can easily be recast in the factorized exponential form 
of Eqs.~(\ref{wtilde}) and (\ref{gformfact}). To this aim, we first
use the identity (\ref{rgiden}) with $h(\as) = C_N(\as)$ to replace
$C_{N}(\as(b_0^2/b^2))$ in Eq.~(\ref{calw}) in terms of 
$C_{N}(\as(M^2))$. Then, we insert in Eq.~(\ref{calw}) the solution
of the evolution equation (\ref{apeq}):
\begin{equation}
\label{apsol}
U_{N}(b_0^2/b^2,\mu_F^2) = \exp \left\{ 
-\int_{b_0^2/b^2}^{\mu_F^2} \frac{dq^2}{q^2} 
\; \gamma_{N}(\as(q^2)) \right\} 
\;\;. 
\end{equation}
We finally obtain the exponential form in Eq.~(\ref{gformfact}), where
the perturbative function $A(\as)$ is exactly the perturbative function in
Eq.~(\ref{aexp}), and the function ${\widetilde B}_{N}(\as)$ is given 
as follows in terms of the perturbative functions in Eqs.~(\ref{rge}),
(\ref{bexp}), (\ref{cexp}) and (\ref{apexp}):
\begin{equation}
\label{btilde}
{\widetilde B}_{N}(\as) = B(\as) + 2 \beta(\as) 
\;\frac{d \ln C_N(\as)}{d \ln \as} + 2 \gamma_{N}(\as) \;\;.
\end{equation}
The expression of the hard-process function ${\cal H}_{N}^F$ in 
Eq.~(\ref{wtilde}) is
\begin{align}
\label{hexpr}
&{\cal H}_{N}^F(M,\as(\mu_R^2);M^2/\mu^2_R,M^2/\mu^2_F,M^2/Q^2)=
 \sigma_F^{(0)}(\as(M^2),M) \;H^{F}(\as(M^2))
\;C_{N}^2(\as(M^2)) \nn \\
& \quad \quad \times \exp \left\{ 
\int_{M^2}^{Q^2} \frac{dq^2}{q^2} 
\left[ A(\as(q^2)) \;\ln \frac{M^2}{q^2} + {\widetilde B}_{N}(\as(q^2))
\right] 
+ \int_{\mu_F^2}^{M^2} \frac{dq^2}{q^2} \;2 \,\gamma_{N}(\as(q^2)) \right\}
\;\;.
\end{align}

Note that, as discussed in Sect.~\ref{sec:rescross}, the form factor
$\exp\{{\cal G}\}$ and, hence, the perturbative functions $A(\as)$ and
${\widetilde B}_{N}(\as)$ in Eq.~(\ref{gformfact}) do not depend on the
factorization scale $\mu_F$. As a consequence, the functions $A(\as)$
and ${\widetilde B}_{N}(\as)$ are also independent of the factorization scheme
used to define the parton densities. Since, as is well known, the anomalous
dimensions $\gamma_{ab, \,N}(\as)$ do depend on the factorization scheme,
the relation (\ref{btilde}) implies that both the perturbative functions
$B_c(\as)$ and $C_{ab}(\as)$ depend on the factorization scheme in such a way
that ${\widetilde B}_{N}(\as)$ turns out to be factorization-scheme
independent.

As anticipated in Sect.~\ref{sec:rescross}, the form factor
$\exp\{{\cal G}\}$ does not depend on the final-state system $F$
produced in the hard-scattering process. From Eqs.~(\ref{gformfact}) and
(\ref{btilde}), this independence is a simple
consequence of the process independence of each of the perturbative functions
$A_c(\as)$, $B_c(\as)$, $C_{ab}(\as)$ and $\gamma_{ab, \,N}(\as)$.

The relation (\ref{btilde}) also implies that the form factor 
$\exp\{{\cal G}\}$ does not depend on the resummation scheme used
to express the various factors in the resummation formulae 
(\ref{wcss}) and (\ref{formfact})
(we recall that the customary Sudakov form factor $S_c(M,b)$ in 
Eq.~(\ref{formfact}) does instead depend on the resummation scheme).
It is indeed straightforward to show that
the function ${\widetilde B}_{N}(\as)$ in Eq.~(\ref{btilde}) is invariant
under the resummation-scheme transformations in Eq.~(\ref{restranf}).

Unlike the form factor $\exp\{{\cal G}\}$, the non-logarithmic function
${\cal H}_{N}^F$ in Eq.~(\ref{hexpr}) explicitly depends on 
the factorization scale $\mu_F$, on the factorization scheme
(through $C_{ab, \,N}(\as)$ and $\gamma_{ab, \,N}(\as)$) and on the 
final-state system $F$ (through $\sigma_F^{(0)}$ and $H^{F}$). Nonetheless,
${\cal H}_{N}^F$ does not depend on the resummation scheme, since
the factor $H^F(\as) C_{N}^2(\as)$ is invariant under the
transformations in Eq.~(\ref{restranf}).

The universal (i.e. independent of the process and of the factorization and
resummation schemes) perturbative function $A_c(\as)$ in Eqs.~(\ref{Afun}) and
(\ref{aexp}) is known up to ${\cal O}(\as^2)$. The LL and NLL coefficients
$A_c^{(1)}$ and $A_c^{(2)}$ are [\ref{Kodaira:1981nh}, \ref{Catani:vd}]
\begin{equation}
\label{acoeff}
A_c^{(1)}= C_c \;\;, \quad
\quad A_c^{(2)}= \frac{1}{2}\; C_c \left[ \left( \f{67}{18} - \f{\pi^2}{6}
\right) C_A - \f{5}{9} N_f \right] \;\;,
\end{equation}
where $C_c= C_F$ if $c=q,{\bar q}$ and $C_c= C_A$ if $c=g$.
The NNLL coefficient $A_c^{(3)}$ is not yet known.
In our quantitative study of transverse-momentum resummation at NNLL
accuracy (see Sect.~\ref{sec:phen}),
we assume that the value of $A_c^{(3)}$
is the same as the one [\ref{Vogt:2000ci}, \ref{mvv}]
that appears in resummed calculations of soft-gluon contributions near
partonic threshold. This assumption is based on the fact that 
the two coefficients $A_c^{(1)}$ and $A_c^{(2)}$ in Eq.~(\ref{acoeff})
are exactly equal to those of the related perturbative
function that controls threshold resummation [\ref{threshold}] in the
\ms\ factorization scheme. Note, however, that the two soft-gluon
functions $A_c(\as)$ do not necessarily coincide at high perturbative orders
since, for instance, the soft-gluon function for transverse-momentum 
resummation is universal while the soft-gluon function for
threshold resummation depends on the factorization scheme.

The first-order coefficient ${\widetilde B}_{c, \,N}^{(1)}$ of the universal
perturbative function ${\widetilde B}_{N}(\as)$ in Eqs.~(\ref{Btfun})
and (\ref{btilde}) is
\begin{equation}
\label{b1tilde}
{\widetilde B}_{c, \,N}^{(1)} = B_c^{(1)} + 2 \gamma_{cc, \,N}^{(1)} \;\;,
\end{equation}
where [\ref{Kodaira:1981nh}, \ref{Catani:vd}]
\begin{equation}
\label{b1coeff}
B_q^{(1)}= B_{\bar q}^{(1)} = - \f{3}{2} \; C_F\;\;, \quad
\quad B_g^{(1)}= - \f{1}{6} \left( 11 C_A - 2 N_f \right) \;\;.
\end{equation}
Note that, since the LO anomalous dimensions $\gamma_{cc, \,N}^{(1)}$ are
universal, the NLL coefficients $B_c^{(1)}$ in Eq.~(\ref{b1coeff})
are themselves independent of the factorization and resummation schemes.

The universal second-order coefficient ${\widetilde B}_{c, \,N}^{(2)}$ 
in Eq.~(\ref{btilde}) is
\begin{equation}
\label{b1tilde2}
{\widetilde B}_{c, \,N}^{(2)} = B_c^{(2)} - 2 \beta_0 \;C_{cc, \,N}^{(1)}
+ 2 \gamma_{cc, \,N}^{(2)} \;\;,
\end{equation}
or, equivalently, by performing the inverse Mellin transformation to $z$-space:
\begin{equation}
\label{b1tilde2z}
{\widetilde B}_{c}^{(2)}(z) = \delta(1-z) \;B_c^{(2)} 
- 2 \beta_0 \;C_{cc}^{(1)}(z)
+ 2 P_{cc}^{(2)}(z) \;\;.
\end{equation}
The value of the quark coefficient ${\widetilde B}_{q}^{(2)}$ can be 
obtained by using the results of Ref.~[\ref{Davies:1984hs}] for
the coefficients $B_{q}^{(2)}$ and $C_{qq}^{(1)}(z)$ of the DY process.
These results are confirmed 
by the general (process-independent) calculation of 
Ref.~[\ref{deFlorian:2000pr}], which considers both the 
$q{\bar q}$-annihilation and the gluon fusion channels. From the results of
Ref.~[\ref{deFlorian:2000pr}] we obtain the value of the gluon
coefficient ${\widetilde B}_{g}^{(2)}$, and we can also explicitly check
the universality of both ${\widetilde B}_{q}^{(2)}$ and 
${\widetilde B}_{g}^{(2)}$. To write down the expression of 
${\widetilde B}_{c}^{(2)}$, we recall that the second-order term 
$P_{cc}^{(2)}(z)$ of the Altarelli--Parisi
splitting functions $P_{cc}(\as,z)$ has the following general dependence
on $z$:
\begin{equation}
\label{psecond}
P_{cc}^{(2)}(z) =  \f{1}{(1-z)_+}  \;A_{c}^{(2)} \;
+ \delta(1-z) \;\f{1}{2} \,\gamma_{c}^{(2)} + P_{cc}^{(2) {\rm reg}}(z) \;\;,
\end{equation}
where $A_{c}^{(2)}$ is the coefficient in Eq.~(\ref{acoeff}),
$1/(1-z)_+$ is the customary `plus'-distribution and 
$P_{cc}^{(2) {\rm reg}}(z)$ denotes all the remaining and less singular
(when $z \to 1$) contributions to $P_{cc}^{(2)}(z)$. The explicit expressions
of  $P_{cc}^{(2) {\rm reg}}(z)$ and of the constants $\gamma_{c}^{(2)}$
can be found in the literature (see e.g. Ref.~[\ref{book}]).
Using the notation of Eq.~(\ref{psecond}), the universal NNLL coefficient 
${\widetilde B}_{c}^{(2)}$ is [\ref{deFlorian:2000pr}]
\begin{equation}
\label{btilde2z}
{\widetilde B}_{c}^{(2)}(z) = \f{2}{(1-z)_+}  \;A_{c}^{(2)} \;
+\delta(1-z) \;\beta_0 C_c \f{\pi^2}{6} + 2  P_{cc}^{(2) {\rm reg}}(z)
+ 2 \beta_0 {\hat P}_{cc}^{\epsilon}(z) \;\;,
\end{equation}
where
\begin{equation}
{\hat P}_{qq}^{\epsilon}(z) = - \frac{1}{2} \,C_F \,(1-z) \;\;, 
\quad \quad {\hat P}_{gg}^{\epsilon}(z)=0 \;\;.
\end{equation}

The first-order coefficients $C_{qg}^{(1)}$ and $C_{gq}^{(1)}$ 
in Eq.~(\ref{cexp}) do not depend on
the process and on the resummation scheme, and were first computed in
Refs.~[\ref{Davies:1984hs}] and [\ref{Kauffman:cx}], respectively.
Their expressions in the $\msbar$ factorization scheme are
\begin{equation}
\label{c1coeff}
C_{qg}^{(1)}(z) = C_{{\bar q}g}^{(1)}(z) = \frac{1}{2} \,z (1-z) \;\;,
\quad \quad
C_{gq}^{(1)}(z) = C_{g{\bar q}}^{(1)}(z) = \frac{1}{2} \,C_F \; z \;\;.
\end{equation}

The flavour-diagonal first-order coefficients $C_{qq}^{(1)}$ and 
$C_{gg}^{(1)}$ and the coefficients $H_q^{F \,(1)}$ and $H_g^{F \,(1)}$
depend on the resummation scheme.
The dependence on the resummation scheme is cancelled in the perturbative
coefficients of the hard-process function ${\cal H}_{N}^F$.
For example, by expanding Eq.~(\ref{hexpr}) in powers of $\as(\mu_R^2)$,
we obtain the following expression for the first-order coefficient 
${\cal H}_N^{F \,(1)}$ of Eq.~(\ref{hexpan}):
\begin{equation}
\label{ch1coef}
{\cal H}_N^{F \,(1)}(M^2/\mu^2_R,M^2/\mu^2_F,M^2/Q^2) =
H^{F \,(1)} + 2 C_N^{(1)} - p \beta_0 \ell_R + 2 \gamma_N^{(1)} \ell_F
- \left( \frac{1}{2} A^{(1)} \ell_Q + {\widetilde B}_N^{(1)} \right) \ell_Q
\;\;,
\end{equation}
where we have defined
\begin{equation}
\label{elldef}
\ell_R = \ln \f{M^2}{\mu_R^2} \;\;, \quad 
\ell_F = \ln \f{M^2}{\mu_F^2} \;\;, \quad 
\ell_Q = \ln \f{M^2}{Q^2} \;\;.
\end{equation}
The coefficient ${\cal H}_N^{F \,(1)}$ depends on the process and is
explicitly known for several processes (see Ref.~[\ref{deFlorian:2000pr}] 
and references therein).

To complete the resummation program at NNLL, the coefficient 
${\cal H}_N^{F \,(2)}$ is also needed. This coefficient is not known 
in analytic form for any hard-scattering process. Nonetheless, within our
resummation formalism, it can be determined for any hard-scattering process
whose corresponding total cross section is known at NNLO.
This point is discussed in detail at the end of Sect.~\ref{sec:matc}.

\subsection{The finite component}
\label{sec:matc}

The finite component 
$d{\hat \sigma}_{F \,ab}^{(\rm fin.)}/d q_T^2$
of the transverse-momentum cross section is computed at a given 
fixed order in $\as$ according to Eq.~(\ref{resfin}).
To implement Eq.~(\ref{resfin}), we have to subtract
$\bigl[ d{\hat \sigma}_{F \,ab}^{(\rm res.)}\bigr]_{\rm f.o.}$
from $\bigl[ d{\hat \sigma}_{F \,ab} \bigr]_{\rm f.o.}$.

As discussed in Sects.~\ref{sec:genfor} and \ref{sec:rescross},
the finite component $d{\hat \sigma}_{F \,ab}^{(\rm fin.)}/d q_T^2$
does not contain any perturbative contributions 
proportional to $\delta(q_T^2)$ (these contributions and all the
logarithmically-enhanced terms at small $q_T$ are included in 
$d{\hat \sigma}_{F \,ab}^{(\rm res.)}/d q_T^2$). Therefore,
when computing $\bigl[ d{\hat \sigma}_{F \,ab}^{(\rm fin.)}\bigr]_{\rm f.o.}$
according to the subtraction procedure in Eq.~(\ref{resfin}),
we can consistently neglect any terms proportional to $\delta(q_T^2)$
both in $\bigl[ d{\hat \sigma}_{F \,ab} \bigr]_{\rm f.o.}$
and in $\bigl[ d{\hat \sigma}_{F \,ab}^{(\rm res.)}\bigr]_{\rm f.o.}$.
This is formally equivalent to the evaluation of both 
$\bigl[ d{\hat \sigma}_{F \,ab} \bigr]_{\rm f.o.}$ and 
$\bigl[ d{\hat \sigma}_{F \,ab}^{(\rm res.)}\bigr]_{\rm f.o.}$
in the large-$q_T$ region (or, more precisely, in the region where 
$q_T \neq 0$). 
The expansions
of $\bigl[ d{\hat \sigma}_{F \,ab}^{(\rm fin.)}\bigr]_{\rm f.o.}$
at the first and at the second perturbative order thus give
\beeq
\label{finlo}
\Bigl[ \f{d{\hat \sigma}_{F \,ab}^{(\rm fin.)}}{d q_T^2} \Bigr]_{\rm LO} 
&=&
\Bigl[\f{d{\hat \sigma}_{F \,ab}^{}}{d q_T^2}\Bigr]_{\rm LO}
- \Bigl[ \f{d{\hat \sigma}_{F \,ab}^{(\rm res.)}}{d q_T^2}\Bigr]_{\rm LO} \;,
\\
\label{finnlo}
\Bigl[ \f{d{\hat \sigma}_{F \,ab}^{(\rm fin.)}}{d q_T^2} \Bigr]_{\rm NLO} 
&=&
\Bigl[\f{d{\hat \sigma}_{F \,ab}^{}}{d q_T^2}\Bigr]_{\rm NLO}
- \Bigl[ \f{d{\hat \sigma}_{F \,ab}^{(\rm res.)}}{d q_T^2}\Bigr]_{\rm NLO} \;,
\eeeq
where the subscript LO (NLO) denotes the perturbative truncation of the various
cross sections
at the leading order (next-to-leading order) in the region where 
$q_T \neq 0$. The extension of Eqs.~(\ref{finlo}) and (\ref{finnlo}) 
at still higher perturbative order is straightforward.

The contributions $\bigl[ d{\hat \sigma}_{F \,ab} \bigr]_{\rm f.o.}$
on the right-hand side of Eqs.~(\ref{finlo}) and (\ref{finnlo})
are obtained by computing the  customary
perturbative series for the partonic cross section  
at a given fixed order (f.o.=LO, NLO, ...) in $\as$.
The fixed-order truncation
$\bigl[ d{\hat \sigma}_{F \,ab}^{(\rm res.)}\bigr]_{\rm f.o.}$ of the resummed
component is obtained by perturbatively expanding the resummed component 
$d{\hat \sigma}_{ab}^{(\rm res.)}$ in Eqs.~(\ref{resum1}).
To this purpose, 
we define the perturbative coefficients
${\widetilde \Sigma}^{(n)}$ as follows:
\beeq
\label{calwexp}
{\cal W}_{ab}^{F}(b,M,{\hat s};\as,\mu_R^2,\mu_F^2,Q^2) \!\!\!&=&\!\!\!\!
\sum_c \sigma_{c{\bar c}, \,F}^{(0)}(\as,M) 
\Bigl\{ \delta_{ca} \;\delta_{{\bar c}b} \;\delta(1-z) \Bigr.\nn \\
&+& \left.
\sum_{n=1}^{\infty} \left( \frac{\as}{\pi} \right)^n
\left[ {\widetilde \Sigma}_{c{\bar c} \ito 
ab}^{F \,(n)}\left(z,\tL; 
\f{M^2}{\mu_R^2},\f{M^2}{\mu_F^2},\f{M^2}{Q^2}\right) \!\!\!
\right. \right.  \\
&+&\!\!\!\!\left. \Bigl.
 {\cal H}_{c{\bar c} \ito 
ab}^{F \,(n)}\!\left(z; 
\f{M^2}{\mu_R^2},\f{M^2}{\mu_F^2},\f{M^2}{Q^2}\right)
\right] \Bigr\} \;, \!\! \nn
\eeeq
where $z=M^2/{\hat s}$, $\as=\as(\mu_R^2)$, 
$\sigma_{c{\bar c}, \,F}^{(0)}(\as,M)= \as^{p_{c F}} \,
\sigma_{c{\bar c}, \,F}^{(\rm LO)}(M)$ and, in general, the power $p_{c F}$
depends on the lowest-order partonic subprocess $c+{\bar c} \to F$.
In Eq.~(\ref{calwexp}), ${\cal W}_{ab}^{F}$ is
the resummed cross section 
on the right-hand side of Eq.~(\ref{resum1}). 
Note, however, that Eq.~(\ref{calwexp}) depends
on the resummation scale $Q^2$. The dependence on the resummation scale has
been introduced in Eqs.~(\ref{resum1}) and (\ref{wtilde}) through 
the replacement in Eqs.~(\ref{grepl}). The perturbative coefficient
${\widetilde \Sigma}^{(n)}$ on the right-hand side of Eq.~(\ref{calwexp})
is a polynomial of degree $2n$ in the logarithmic variable $\tL$ defined in
Eq.~(\ref{logdef}). The coefficients ${\widetilde \Sigma}^{(n)}$ vanish
by definition when $\tL=0$ (i.e. when $b=0$), and the $b$-independent part
of ${\cal W}_{ab, \,N}^{F}(b,M)$ is embodied in the coefficients 
${\cal H}^{(n)}$.

The perturbative
expansion of Eq.~(\ref{wtilde}) or, more precisely, of Eq.~(\ref{calw})
gives
\begin{equation}
\label{sigt1}
{\widetilde \Sigma}_{c{\bar c} \ito 
ab}^{F \,(1)}(z,\tL)
=
\Sigma_{c{\bar c} \ito 
ab}^{F \,(1;2)}(z)
\; \tL^2
+ \Sigma_{c{\bar c} \ito 
ab}^{F \,(1;1)}(z) \;\tL \;,
\end{equation}
\begin{equation}
\label{sigt2}
{\widetilde \Sigma}_{c{\bar c} \ito 
ab}^{F \,(2)}(z,\tL)
=
\Sigma_{c{\bar c} \ito 
ab}^{F \,(2;4)}(z)
\; \tL^4
+ \Sigma_{c{\bar c} \ito 
ab}^{F \,(2;3)}(z) \;\tL^3 
+ \Sigma_{c{\bar c} \ito 
ab}^{F \,(2;2)}(z) \;\tL^2 
+ \Sigma_{c{\bar c} \ito 
ab}^{F \,(2;1)}(z) \;\tL \;,
\end{equation}
where the dependence on the scale ratios 
$M^2/\mu_R^2, M^2/\mu_F^2$ and $M^2/Q^2$ is understood.  The extension of
Eqs.~(\ref{sigt1}) and (\ref{sigt2}) to the higher order terms
${\widetilde \Sigma}_{c{\bar c} \ito 
ab}^{F \,(n)}(z,\tL)$ with $n \geq 3$, is straightforward.
The $b$-independent coefficients 
$\Sigma^{F \,(1;k)}(z), {\cal H}^{F \,(1)}(z),
\Sigma^{F \,(2;k)}(z)$
and ${\cal H}^{F \,(2)}(z)$
are more easily presented by considering their 
$N$-moments with respect to the variable $z$. We have
\begin{equation}
\label{sigt12}
\Sigma_{c{\bar c} \ito 
ab, \,N}^{F \,(1;2)} =
- \,\frac{1}{2} A_c^{(1)} \;\delta_{ca} \delta_{{\bar c}b} \;,
\end{equation}
\begin{equation}
\label{sigt11}
\Sigma_{c{\bar c} \ito 
ab, \,N}^{F \,(1;1)}(M^2/Q^2) = - \left[
\delta_{ca} \delta_{{\bar c}b} \, \left( B_c^{(1)} + A_c^{(1)} \, 
\ell_Q \right) + \delta_{ca} \gamma_{{\bar c}b, \,N}^{(1)}
+ \delta_{{\bar c}b} \gamma_{ca, \,N}^{(1)}
\right] \;,
\end{equation}
\beeq
\label{ht1}
{\cal H}_{c{\bar c} \ito 
ab, \,N}^{F \,(1)}\!\left( 
\f{M^2}{\mu_R^2},\f{M^2}{\mu_F^2},\f{M^2}{Q^2}\right) &=&
\delta_{ca} \delta_{{\bar c}b} \left[ H_c^{F \,(1)}
- \left( B_c^{(1)}+\frac{1}{2} A_c^{(1)} \ell_Q \right) \ell_Q
- p_{c F} \beta_0 \ell_R \right] \nn \\
&+& \delta_{ca} C_{{\bar c}b, \,N}^{(1)} +  
\delta_{{\bar c}b} C_{ca, \,N}^{(1)}
+ \left( \delta_{ca} \gamma_{{\bar c}b, \,N}^{(1)}
+ \delta_{{\bar c}b} \gamma_{ca, \,N}^{(1)}
\right) \left( \ell_F - \ell_Q \right) \;,
\eeeq
\begin{equation}
\label{sigt24}
\Sigma_{c{\bar c} \ito 
ab, \,N}^{F \,(2;4)} =
\frac{1}{8}  \left( A_c^{(1)} \right)^2\;\delta_{ca} \delta_{{\bar c}b} \;,
\end{equation}
\begin{equation}
\label{sigt23}
\Sigma_{c{\bar c} \ito 
ab, \,N}^{F \,(2;3)}(M^2/Q^2) = - \,A_c^{(1)} \left[ 
\frac{1}{3} \,\beta_0 \;\delta_{ca} \delta_{{\bar c}b}
+ \frac{1}{2} \,\Sigma_{c{\bar c} \ito 
ab, \,N}^{F \,(1;1)}(M^2/Q^2) \right] \;,
\end{equation}
\beeq
\label{sigt22}
\Sigma_{c{\bar c} \ito 
ab, \,N}^{F \,(2;2)}\left( 
\f{M^2}{\mu_R^2},\f{M^2}{\mu_F^2},\f{M^2}{Q^2}\right) &=&
- \,\frac{1}{2} \,A_c^{(1)} \left[ {\cal H}_{c{\bar c} \ito 
ab, \,N}^{F \,(1)}\!\left( 
\f{M^2}{\mu_R^2},\f{M^2}{\mu_F^2},\f{M^2}{Q^2}\right) -
\beta_0 \;\delta_{ca} \delta_{{\bar c}b}
\left( \ell_R - \ell_Q \right) \right] \nn \\
&-&\,\frac{1}{2} \sum_{a_1,b_1} 
\Sigma_{c{\bar c} \ito a_1b_1, \,N}^{F \,(1;1)}(M^2/Q^2)
\left[ \delta_{a_1a} \gamma_{b_1b, \,N}^{(1)}
+ \delta_{b_1b} \gamma_{a_1a, \,N}^{(1)}
\right] \\
&-&\,\frac{1}{2} \left[ A_c^{(2)} \;\delta_{ca} \delta_{{\bar c}b} + 
\left( B_c^{(1)} + A_c^{(1)} \, \ell_Q - \beta_0 \right)
\,\Sigma_{c{\bar c} \ito ab, \,N}^{F \,(1;1)}(M^2/Q^2) \right] 
\;, \nn 
\eeeq
\beeq
\label{sigt21}
\Sigma_{c{\bar c} \ito 
ab, \,N}^{F \,(2;1)}\!\!\!\!&&\!\!\!\!\!\!\!\!\!\!\!\left( 
\f{M^2}{\mu_R^2},\f{M^2}{\mu_F^2},\f{M^2}{Q^2}\right) =
\Sigma_{c{\bar c} \ito 
ab, \,N}^{F \,(1;1)}(M^2/Q^2) \,\beta_0 \left( \ell_Q - \ell_R \right)
\nn \\
&-& \sum_{a_1,b_1}
{\cal H}_{c{\bar c} \ito 
a_1b_1, \,N}^{F \,(1)}\!\left( 
\f{M^2}{\mu_R^2},\f{M^2}{\mu_F^2},\f{M^2}{Q^2}\right) 
\left[ \delta_{a_1a} \delta_{b_1b} \left(
B_c^{(1)} + A_c^{(1)} \, \ell_Q \right) +
\delta_{a_1a} \gamma_{b_1b, \,N}^{(1)}
+ \delta_{b_1b} \gamma_{a_1a, \,N}^{(1)}
\right] \nn \\
&-&\left[
\delta_{ca} \delta_{{\bar c}b} \, \left( B_c^{(2)} + A_c^{(2)} \, 
\ell_Q \right) - \beta_0 \left( \delta_{ca} C_{{\bar c}b, \,N}^{(1)} +  
\delta_{{\bar c}b} C_{ca, \,N}^{(1)}
\right)
+ \delta_{ca} \gamma_{{\bar c}b, \,N}^{(2)}
+ \delta_{{\bar c}b} \gamma_{ca, \,N}^{(2)}
\right] \;,
\eeeq
\beeq
\label{ht2}
&& \!\!\!\!\!\!\!\!\!\!\!\!
{\cal H}_{c{\bar c} \ito 
ab, \,N}^{F \,(2)}\!\left( 
\f{M^2}{\mu_R^2},\f{M^2}{\mu_F^2},\f{M^2}{Q^2}\right) =
\delta_{ca} \delta_{{\bar c}b}  \,H_c^{F \,(2)}
+\delta_{ca} \, C_{{\bar c}b, \,N}^{(2)} +  
\delta_{{\bar c}b}\, C_{ca, \,N}^{(2)}
+ C_{ca, \,N}^{(1)} \,C_{{\bar c}b, \,N}^{(1)} \nn \\
&+& \!\!H_c^{F \,(1)} \left( \delta_{ca} \,C_{{\bar c}b, \,N}^{(1)} 
+ \delta_{{\bar c}b} \,C_{ca, \,N}^{(1)} \right) + \frac{1}{6}\,A_c^{(1)}  
\,\beta_0 \,\ell_Q^3\;\delta_{ca} \delta_{{\bar c}b} 
+ \frac{1}{2} \left[ A_c^{(2)}\;\delta_{ca} \delta_{{\bar c}b} +
\beta_0 \,\Sigma_{c{\bar c} \ito 
ab, \,N}^{F \,(1;1)}(M^2/Q^2) \right] \ell_Q^2 \nn \\
&-&\left[
\delta_{ca} \delta_{{\bar c}b} \, \left( B_c^{(2)} + A_c^{(2)} \, 
\ell_Q \right) - \beta_0 \left( \delta_{ca} C_{{\bar c}b, \,N}^{(1)} +  
\delta_{{\bar c}b} C_{ca, \,N}^{(1)}
\right)
+ \delta_{ca} \gamma_{{\bar c}b, \,N}^{(2)}
+ \delta_{{\bar c}b} \gamma_{ca, \,N}^{(2)}
\right] \ell_Q \nn \\
&+& \frac{1}{2} \,\beta_0 \,\left( \delta_{ca} \gamma_{{\bar c}b, \,N}^{(1)}
+ \delta_{{\bar c}b} \gamma_{ca, \,N}^{(1)} \right) \ell_F^2
+ \left( \delta_{ca} \gamma_{{\bar c}b, \,N}^{(2)}
+ \delta_{{\bar c}b} \gamma_{ca, \,N}^{(2)} \right) \ell_F 
- {\cal H}_{c{\bar c} \ito 
ab, \,N}^{F \,(1)}\!\left( 
\f{M^2}{\mu_R^2},\f{M^2}{\mu_F^2},\f{M^2}{Q^2}\right) \beta_0 \ell_R\nn \\
&+& \f{1}{2} \;\sum_{a_1,b_1} \left[
{\cal H}_{c{\bar c} \ito 
a_1b_1, \,N}^{F \,(1)}\!\left( 
\f{M^2}{\mu_R^2},\f{M^2}{\mu_F^2},\f{M^2}{Q^2}\right) 
+ \delta_{ca_1} \delta_{{\bar c}b_1}  \,H_c^{F \,(1)}
+\delta_{ca_1} \, C_{{\bar c}b_1, \,N}^{(1)} +  
\delta_{{\bar c}b_1}\, C_{ca_1, \,N}^{(1)}\right] \nn \\
&\times& \left[ \left( \delta_{a_1a} \gamma_{b_1b, \,N}^{(1)}
+ \delta_{b_1b} \gamma_{a_1a, \,N}^{(1)} \right) \left( \ell_F - \ell_Q \right)
- \delta_{a_1a} \delta_{b_1b} \left( \left(
B_c^{(1)} + \f{1}{2} \,A_c^{(1)} \, \ell_Q \right) \ell_Q 
+ p_{c F} \,\beta_0 \,\ell_R \right)
\right] \nn \\
&-&\delta_{ca} \delta_{{\bar c}b} \,p_{c F} \left( 
\f{1}{2} \,\beta_0^2 \,\ell_R^2 + \beta_1 \,\ell_R \right)
\;\;.
\eeeq
The right-hand side of Eqs.~(\ref{sigt12})--(\ref{ht2}) is expressed in terms
of the resummation-scheme independent coefficients given in 
Sect.~\ref{sec:css} and of the logarithms $\ell_Q, \ell_F$ and $\ell_R$ 
defined in Eq.~(\ref{elldef}). To explicitly exhibit the independence
of the resummation scheme we can, for example, rewrite
the contribution in the third line of Eq.~(\ref{sigt21}) in terms
of the resummation-scheme independent coefficients 
${\widetilde B}_{c \,N}^{(2)}$ (see Eq.~(\ref{b1tilde2}))
and $C_{ab, \,N}^{(1)}$ with $a \neq b$ (see Eq.~(\ref{c1coeff})):
\beeq
&&\!\!\!\!\!\!\!\!\!\!\!\!\left[
\delta_{ca} \delta_{{\bar c}b} \, \left( B_c^{(2)} + A_c^{(2)} \, 
\ell_Q \right) - \beta_0 \left( \delta_{ca} C_{{\bar c}b, \,N}^{(1)} +  
\delta_{{\bar c}b} C_{ca, \,N}^{(1)}
\right)
+ \delta_{ca} \gamma_{{\bar c}b, \,N}^{(2)}
+ \delta_{{\bar c}b} \gamma_{ca, \,N}^{(2)}
\right] = 
\delta_{ca} \delta_{{\bar c}b} \, \left( {\widetilde B}_{c \,N}^{(2)} 
+ A_c^{(2)} \, \ell_Q \right)\nn \\
&&\!\!\! + \;
\delta_{ca} \left( 1-\delta_{{\bar c}b} \right)
\left( \gamma_{{\bar c}b, \,N}^{(2)} - \beta_0 C_{{\bar c}b, \,N}^{(1)} \right)
+ \delta_{{\bar c}b} \left( 1-\delta_{ca} \right)
\left( \gamma_{ca, \,N}^{(2)} - \beta_0 C_{ca, \,N}^{(1)} \right) \;\;.
\eeeq

Inserting Eqs.~(\ref{calwexp})--(\ref{sigt2}) in Eq.~(\ref{resum1}),
performing the integral over the impact parameter $b$, and removing the
contributions proportional to $\delta(q_T^2)$ (for example, all the
contributions coming from ${\cal H}_{c{\bar c} \ito ab}^{F \,(n)}$
in Eq.~(\ref{calwexp})), we obtain the following expressions
for the fixed-order contributions
$\bigl[ d{\hat \sigma}_{F \,ab}^{(\rm res.)}\bigr]_{\rm f.o.}$
on the right-hand side of Eqs.~(\ref{finlo}) and (\ref{finnlo}):
\beeq
\label{reslo}
&&\Bigl[ \;\f{d{\hat \sigma}_{F \,ab}^{(\rm res.)}}{d q_T^2}(q_T,M,
{\hat s}=\f{M^2}{z};
\as(\mu_R^2),\mu_R^2,\mu_F^2,Q^2) \;\Bigr]_{\rm LO} 
= \frac{\as(\mu_R^2)}{\pi} \;\frac{z}{Q^2} 
\;\sum_c \sigma_{c{\bar c}, \,F}^{(0)}(\as(\mu_R^2),M) \nn \\
&& \quad \quad \quad \quad \quad \quad \quad \quad \times \left[
\Sigma_{c{\bar c} \ito 
ab}^{F \,(1;2)}(z) \;{\widetilde I}_2(q_T/Q)
+ \Sigma_{c{\bar c} \ito 
ab}^{F \,(1;1)}\left(z; \f{M^2}{Q^2}\right) \;{\widetilde I}_1(q_T/Q)
\right] \;.
\eeeq
\beeq
\label{resnlo}
&&\!\!\!\!\!\! \!\!\!\!\!\!\!\!\! 
\Bigl[ \;\f{d{\hat \sigma}_{F \,ab}^{(\rm res.)}}{d q_T^2}(q_T,M,
{\hat s}=\f{M^2}{z};
\as(\mu_R^2),\mu_R^2,\mu_F^2,Q^2) \;\Bigr]_{\rm NLO} 
= \Bigl[ \;\f{d{\hat \sigma}_{F \,ab}^{(\rm res.)}}{d q_T^2}(q_T,M,{\hat s};
\as(\mu_R^2),\mu_R^2,\mu_F^2,Q^2) \;\Bigr]_{\rm LO} \nn \\
&&\quad  + \left( \frac{\as(\mu_R^2)}{\pi} \right)^2 \;\frac{z}{Q^2}
 \;\sum_c \sigma_{c{\bar c}, \,F}^{(0)}(\as(\mu_R^2),M) 
\sum_{k=1}^4 \Sigma_{c{\bar c} \ito ab}^{F \,(2;k)}\left(z; 
\f{M^2}{\mu_R^2},\f{M^2}{\mu_F^2},\f{M^2}{Q^2}\right)
\;{\widetilde I}_k(q_T/Q) \;,
\eeeq

On the right-hand side of Eqs.~(\ref{reslo}) and (\ref{resnlo}),
the dependence on $q_T$ is fully embodied in the functions
${\widetilde I}_n(q_T/Q)$, which are obtained by the following Bessel
transformation:
\begin{equation}
\label{integrals}
{\widetilde I}_n(q_T/Q) = Q^2 \int_0^\infty db \;\f{b}{2} \,J_0(b q_T) 
\; \ln^n\left( \frac{Q^2b^2}{b_0^2}+1 \right) \;.
\end{equation}
The term $\ln^n(1+ Q^2b^2/b_0^2)={\widetilde L}^n$ 
in the integrand comes from the replacement
$L \to {\widetilde L}$ (see Eq.~(\ref{grepl})). In customary implementations
of $b$-space resummation, one has to consider the Bessel transformation
of powers of $\ln^n(Q^2b^2/b_0^2)=L^n$, which can be expressed in terms of
powers of  $\ln^n(Q^2/q_T^2)$. The functions ${\widetilde I}_n(q_T/Q)$
have instead a more involved functional dependence on $q_T$. As shown in
Appendix~\ref{appb}, this functional dependence can be expressed in terms
of $K_\nu(q_T/Q)$, the modified Bessel function of imaginary argument that is
defined by the following integral representation:
\begin{equation}
\label{interep}
K_\nu(x)=\int_0^\infty dt \; e^{-x \cosh t}\cosh \nu t \;\;.
\end{equation}

We conclude this section with some observations on the hard-scattering 
function ${\cal H}_{c{\bar c} \ito 
ab}^{F}$. This function
is resummation-scheme independent, but it depends on the specific
hard-scattering subprocess $c + {\bar c} \to F$. The coefficients
${\cal H}_{c{\bar c} \ito ab}^{F \,(n)}$ of its perturbative expansion
can be determined by performing a customary perturbative calculation
of the $q_T$ distribution in the limit $q_T \to 0$.
Moreover, as discussed in Sect.~\ref{sec:rescross}, 
within our resummation formalism
${\cal H}^{F}$ controls the strict perturbative normalization of the 
corresponding total cross section (i.e. the integral
of the $q_T$ distribution). This property can be exploited to determine  
the coefficients ${\cal H}_{c{\bar c} \ito ab}^{F \,(n)}$ in a different
manner, that is, from the perturbative calculation of the total cross 
section.  

To illustrate this point we consider the total cross section,
${\hat \sigma}_{F \,ab}^{{\rm {tot}}}$, at the partonic level,
\begin{equation}
\label{restotp}
{\hat \sigma}_{F \,ab}^{{\rm {tot}}}(M,{\hat s};
\as(\mu_R^2),\mu_R^2,\mu_F^2)
 = \int_0^{\infty} dq_T^2 \;
\f{d{\hat \sigma}_{F \,ab}}{d q_T^2}(q_T,M,{\hat s};
\as(\mu_R^2),\mu_R^2,\mu_F^2) 
 \;,
\end{equation}
and we evaluate the $q_T$ spectrum on right-hand side according
to the decomposition in terms of `resummed' and `finite' components
(see Eq.~(\ref{resplusfin})). Then we use Eq.~(\ref{restot}) to integrate the
resummed component over $q_T$, and we obtain
\beeq
\label{sigtotrel}
{\hat \sigma}_{F \,ab}^{{\rm {tot}}}
 = \f{M^2}{\hat s} \;{\cal H}_{ab}^F 
+ \int_0^{\infty} dq_T^2 \;
\f{d{\hat \sigma}_{F \,ab}^{(\rm fin.)}}{d q_T^2}  \;\;.
\eeeq
This expression is valid order by order in QCD perturbation theory.
Once the perturbative coefficients of the fixed-order expansions
of ${\hat \sigma}_{F \,ab}^{{\rm {tot}}}$,
${\cal H}_{ab}^F$ and  $d{\hat \sigma}_{F \,ab}^{(\rm fin.)}/d q_T^2$
are all known, the relation (\ref{sigtotrel}) has to be regarded as an identity,
which can explicitly be checked. 
Note, however, that since the fixed-order truncation
$\bigl[ d{\hat \sigma}_{F \,ab}^{(\rm fin.)}/{d q_T^2} \bigr]_{\rm f.o.}$ 
does not contain any contributions proportional to $\delta(q_T^2)$,
$\bigl[ d{\hat \sigma}_{F \,ab}^{(\rm fin.)}/d q_T^2 \Bigr]_{\rm LO}$
does not explicitly depend on the coefficient ${\cal H}_{ab}^{F \,(1)}$
(see Eqs.~(\ref{finlo}) and (\ref{reslo})). Analogously, 
$\bigl[ d{\hat \sigma}_{F \,ab}^{(\rm fin.)}/d q_T^2 \Bigr]_{\rm NLO}$
does not explicitly depend on the coefficient ${\cal H}_{ab}^{F \,(2)}$
(see Eqs.~(\ref{finnlo}) and (\ref{resnlo})), and so forth.
Therefore, Eq.~(\ref{sigtotrel}) can be used to determine the $\rm N^nLO$
coefficient ${\cal H}_{ab}^{F \,(n)}$ from the knowledge of 
${\hat \sigma}_{F \,ab}^{{\rm {tot}}}$ at $\rm N^nLO$ and of
$d{\hat \sigma}_{F \,ab}^{(\rm fin.)}/d q_T^2$ at $\rm N^{n-1}LO$, without
the need of explicitly computing the small-$q_T$ behaviour of the spectrum
$d{\hat \sigma}_{F \,ab}/d q_T^2$ at $\rm N^nLO$.
For example, at NLO Eq.~(\ref{sigtotrel}) gives
\beeq
\label{h1nlo}
\f{\as}{\pi} \;\f{M^2}{\hat s} \sum_c \sigma_{c{\bar c}, \,F}^{(0)}(\as,M)
&&\!\!\!\!\!\! \!\!\!\!\!\!{\cal H}_{c{\bar c} \ito 
ab}^{F \,(1)}\!\left(\f{M^2}{\hat s}; 
\f{M^2}{\mu_R^2},\f{M^2}{\mu_F^2},\f{M^2}{Q^2}\right)
= 
\Bigl[ {\hat \sigma}_{F \,ab}^{{\rm {tot}}}(M,{\hat s};
\as,\mu_R^2,\mu_F^2) \Bigr]_{\rm NLO} \!\! \\
&\!-& \!\!\Bigl[ {\hat \sigma}_{F \,ab}^{{\rm {tot}}}(M,{\hat s};
\as) \Bigr]_{\rm LO}
- \int_0^{\infty} \!dq_T^2 
\Bigl[ \f{d{\hat \sigma}_{F \,ab}^{(\rm fin.)}}{d q_T^2}(q_T,M,
{\hat s}; \as,\mu_R^2,\mu_F^2,Q^2) 
\Bigr]_{\rm LO} \,, \nn
\eeeq
where $\as=\as(\mu_R^2)$ and we have used
\begin{equation}
\Bigl[ {\hat \sigma}_{F \,ab}^{{\rm {tot}}}(M,{\hat s};
\as) \Bigr]_{\rm LO} = \delta(1 - M^2/\hat s)
\;\sum_c \sigma_{c{\bar c}, \,F}^{(0)}(\as,M) \;\delta_{ca} 
\;\delta_{{\bar c}b} \;\;.
\end{equation}
At NNLO Eq.~(\ref{sigtotrel}) gives
\beeq
\label{h2nnlo}
&&\left( \f{\as}{\pi} \right)^2\;\f{M^2}{\hat s} 
\sum_c \sigma_{c{\bar c}, \,F}^{(0)}(\as,M)
\; {\cal H}_{c{\bar c} \ito 
ab}^{F \,(2)}\!\left(\f{M^2}{\hat s}; 
\f{M^2}{\mu_R^2},\f{M^2}{\mu_F^2},\f{M^2}{Q^2}\right) \nn \\
&& \quad = \left\{
\Bigl[ {\hat \sigma}_{F \,ab}^{{\rm {tot}}} \Bigr]_{\rm NNLO} 
- \Bigl[ {\hat \sigma}_{F \,ab}^{{\rm {tot}}} \Bigr]_{\rm NLO} 
\right\}
- \int_0^{\infty} dq_T^2 
\left\{
\Bigl[ \f{d{\hat \sigma}_{F \,ab}^{(\rm fin.)}}{d q_T^2} 
\Bigr]_{\rm NLO} - 
\Bigl[ \f{d{\hat \sigma}_{F \,ab}^{(\rm fin.)}}{d q_T^2} 
\Bigr]_{\rm LO}
\right\}\;\;,
\eeeq
and the generalization at still higher orders $n$ $(n > 2)$ is
\beeq
\label{hn}
\left( \f{\as}{\pi} \right)^n\;\f{M^2}{\hat s} 
\sum_c \sigma_{c{\bar c}, \,F}^{(0)}(\as,M)
\; {\cal H}_{c{\bar c} \ito 
ab}^{F \,(n)} &=& \left\{
\Bigl[ {\hat \sigma}_{F \,ab}^{{\rm {tot}}} \Bigr]_{\rm N^nLO} 
- \Bigl[ {\hat \sigma}_{F \,ab}^{{\rm {tot}}} \Bigr]_{\rm N^{n-1}LO} 
\right\} \\
&-& \int_0^{\infty} dq_T^2 
\left\{
\Bigl[ \f{d{\hat \sigma}_{F \,ab}^{(\rm fin.)}}{d q_T^2} 
\Bigr]_{\rm N^{n-1}LO} - 
\Bigl[ \f{d{\hat \sigma}_{F \,ab}^{(\rm fin.)}}{d q_T^2} 
\Bigr]_{\rm N^{n-2}LO}
\right\} . \nn
\eeeq
In our study of the transverse-momentum spectrum of the Higgs boson
at NNLL accuracy (see Sect.~\ref{sec:phen}),
we use Eq.~(\ref{h2nnlo}) to obtain a numerical value
for the corresponding perturbative 
coefficient~${\cal H}^{(2)}$.

\section{The $\qt$ spectrum of the Higgs boson at the LHC}
\label{sec:phen}

In this section we apply the resummation formalism 
described in Sect.~{\ref{sec:res}
to the production of the SM Higgs boson at the LHC.

We consider the gluon fusion production mechanism $gg \to H$, whose Born 
level cross section in Eqs.~(\ref{hexpan}) and (\ref{calwexp}) is
\begin{equation}
\sigma^{(0)}_{c {\bar c}, H}(\as, M_H) = \delta_{cg} \,\delta_{{\bar c}g}
\;\as^2 \;\sigma^{(0)}(M_H;M_t,M_b) \;\;,
\end{equation}
where $M_t$ and $M_b$ denote the masses of the top and bottom quark,
which circulate in the heavy-quark loop that couples to the Higgs boson.
In our numerical study we use $M_t=175$~GeV and $M_b=4.75$~GeV. The expression
of $\sigma^{(0)}(M_H;M_t,M_b)$ can be found, for instance, in Eq.~(3)
of Ref.~[\ref{Catani:2003zt}].
Though the Born cross section is evaluated exactly, 
i.e. including its dependence
on the top-- and bottom--quark masses,
the computation of the higher-order QCD corrections is performed in 
the framework of the large-$M_t$ approximation. More precisely, we proceed as
in Ref.~[\ref{Catani:2003zt}]: we first compute $d\sigma_H/dq_T$ 
in the large-$M_t$ limit and then we rescale the result by the factor
$\sigma^{(0)}(M_H;M_t,M_b)/\sigma^{(0)}_\infty$, where $\sigma^{(0)}_\infty$
is obtained from $\sigma^{(0)}(M_H;M_t,M_b)$ by setting $M_b=0$ and 
$M_t/M_H\to \infty$. As recalled in Sect.~\ref{sec:intro}, this implementation
of the large-$M_t$ approximation is expected to produce an uncertainty that is
smaller than the uncertainties from yet uncalculated perturbative terms
from higher orders.

We compute the Higgs boson differential cross section $d\sigma/dq_T$ at the
LHC ($pp$ collisions at ${\sqrt s}=14$~TeV)
and present quantitative results at NLL+LO and NNLL+NLO 
accuracy.

As discussed in Sect.~\ref{sec:rescross},
at NLL+LO accuracy the resummed component in Eq.~(\ref{wtilde}) is evaluated 
by including the functions $g^{(1)}$ and $g_N^{(2)}$ in Eq.~(\ref{gexpan}) 
and the coefficient ${\cal H}_N^{F(1)}$ in Eq.~(\ref{hexpan}),
and then it is matched with 
the fixed-order contribution evaluated at the LO 
(i.e. at ${\cal O}(\as^3)$) in the large-$q_T$ region.
The functions $g^{(1)}$ and $g_N^{(2)}$ are process independent
and given in terms of the universal coefficients $A^{(1)}, A^{(2)}$ and 
${\widetilde B}_N^{(1)}$ (see Sect.~\ref{sec:css}).
The flavour off-diagonal part of ${\cal H}_{gg\ito ab,N}^{H(1)}$
is also process independent and given by Eq.~(\ref{ht1});
setting $\mu_F=Q=M_H$, we simply have
\begin{equation}
{\cal H}_{gg\ito gq,N}^{H(1)}={\cal H}_{gg\ito qg,N}^{H(1)}=C^{(1)}_{gq,N}
=\f{1}{2(N+1)} \,C_F\, ,
\end{equation}
where the 
coefficient $C^{(1)}_{gq,N}$
is the Mellin transformation of Eq.~(\ref{c1coeff}).
Of course, these process-independent coefficients are exact, i.e. they are
not affected by the large-$M_t$ approximation.
The flavour diagonal coefficient ${\cal H}_{gg\ito gg,N}^{H(1)}$ is instead
process dependent; therefore it depends on $M_t$ and, 
in the large-$M_t$ approximation, it is given by
[\ref{Yuan:1991we}, \ref{Kauffman:cx}]
\begin{equation}
\label{ht1h}
{\cal H}_{gg\to gg,N}^{H(1)}=H_g^{H(1)}+2C_{gg, \,N}^{(1)}=
\f{1}{2}\left[ (5+\pi^2)C_A-3C_F \right] = \f{1}{2}\, (11+3\pi^2)\, ,
\end{equation}
where, for simplicity, the scale-dependent terms have been dropped 
(i.e. we have set $\mu_R=\mu_F=Q=M_H$ in Eq.~(\ref{ht1})).

At NNLL+NLO accuracy the function $g_N^{(3)}$ and the 
coefficient ${\cal H}_N^{H(2)}$ have also to be included in the resummed
component of the $q_T$ cross section, and the finite component has to include
the fixed-order contribution to the cross section evaluated at the NLO 
(i.e. at ${\cal O}(\as^4)$) in the large-$q_T$ region.
The process-independent function $g_N^{(3)}$ depends on the universal  
coefficients $A^{(3)}$ and ${\widetilde B}_N^{(2)}$ (see Sect.~\ref{sec:css}).
The scale-independent part of the coefficient ${\cal H}_N^{H(2)}$
(its scale-dependent part can be obtained from Eq.~(\ref{ht2}))
is not known in analytic form. We thus exploit Eq.~(\ref{h2nnlo}),
which follows from the constraint of perturbative unitarity, to extract
the numerical value of ${\cal H}_N^{H(2)}$ from the 
knowledge of the total cross section at the NNLO [\ref{NNLOtotal}].
The scale-independent
part of ${\cal H}_{gg \ito gg, \,N}^{F \,(2)}$ can be written~as
\begin{align}
\label{ht2hc}
{\cal H}^{H(2)}_{gg \ito gg, \,N} {\Big |}_{\mu_R=\mu_F=Q=M_H}
&= H_g^{H \,(2)} + 2 \, C_{gg, \,N}^{(2)}   
+ \left( C_{gg, \,N}^{(1)} \right)^2 + 2 \,H_g^{H \,(1)} \,C_{gg, \,N}^{(1)} 
\\
\label{h2app}
&= \left(\f{19}{16}+\f{1}{3}N_f\right)\ln\f{M_H^2}{M_t^2}+c_N \;\;,
\end{align}
where the $M_t$-dependent contribution on the right-hand side
is obtained from the results in
Refs.~[\ref{Kramer:1996iq}, \ref{Chetyrkin:1997iv}], and $c_N$ does not depend
on $M_t$ in the large-$M_t$ approximation. Since from Eq.~(\ref{ht1h})
we know that $C_{gg, \,N}^{(1)}$ is actually independent of $N$, the $N$
dependence of $c_N$ can only follow from that of $C_{gg, \,N}^{(2)}$
in Eq.~(\ref{ht2hc}). Using Eq.~(\ref{h2nnlo}) and the NNLO total cross
section, we find that the flavour off-diagonal terms 
in ${\cal H}_{c{\bar c} \ito 
ab, \,N}^{F \,(2)}$ can numerically be neglected, and that the coefficient
$c_N$ in Eq.~(\ref{h2app}) can numerically be approximated by an
$N$-independent value, $c_N \simeq 178.75$.
This numerical approximation is pretty good,
since
the integral of the NNLL+NLO spectrum reproduces
the NNLO total cross section to better than $1\%$ accuracy
in a wide Higgs mass range, 100~GeV$\ltap M_H \ltap
300$~GeV, at the LHC.

We recall that the functions
$g_N^{(k)}(\lambda)$ are singular when $\lambda\to 1$ 
(see Eqs.~(\ref{g1fun})--(\ref{g3fun})).
The singular behaviour is related to the presence of the Landau pole
in the perturbative running of the QCD coupling $\as(q^2)$.
As mentioned at the end of Sect.~\ref{sec:rescross},
a practical implementation of the resummation procedure
requires a prescription to deal with these
singularities. In our numerical study
we follow Ref.~[\ref{Laenen:2000de}] and deform the integration contour
in the complex $b$ space.
In particular we choose the two integration branches as
\begin{equation}
b=(\cos\phi\pm i\sin\phi)t\, ,~~~~~t\in \{0,\infty\}\, .
\end{equation}
We have checked that the result is very mildly dependent
on the choice of $\phi$. We have also used the simpler procedure
of integrating over the real $b$-axis, using a sharp cut-off at 
a large value of $b$, and checking the independence of the actual value of the 
cut-off. We found that the numerical
differences between the results obtained by these two procedures 
are negligible.

Our complete calculation of the $q_T$ spectrum of the Higgs boson at the LHC
is implemented in the numerical code {\bf {\tt HqT}},
which can be downloaded from [\ref{code}] together with some accompanying
notes. This code is a slightly modified and numerically improved version
of the code used in Ref.~[\ref{Bozzi:2003jy}]:
the most important difference regards the computation of the finite component.
In Ref.~[\ref{Bozzi:2003jy}] we used the Monte Carlo program
of Ref.~[\ref{deFlorian:1999zd}] to compute the fixed-order contribution
to the $q_T$ cross section at LO and NLO.
Here we have implemented the analytic calculation of Glosser and Schmidt
[\ref{Glosser:2002gm}].
Although the two methods are in principle equivalent, the use of the 
analytic calculation allows us to achieve a 
faster numerical stability in the small-$q_T$ region.
In the next subsection we present a selection of numerical results
that can be obtained with our code. We also include 
a discussion of theoretical uncertainties.

\subsection{Numerical results at the LHC}
\label{sec:num}

To compute the hadronic cross section,
we use the MRST2004 set [\ref{Martin:2004ir}] of parton distribution functions.
As for the perturbative order of the parton densities and $\as$,
at variance with Ref.~[\ref{Bozzi:2003jy}],
we adopt here the following choice.
At NLL+LO we use NLO parton densities and
2-loop $\as$, whereas at NNLL+NLO we use NNLO parton densities 
and 3-loop $\as$.
This choice is perfectly consistent in the small $q_T$ region, 
since the corresponding partonic cross section is dominated by the resummed
component evaluated at NLL and NNLL accuracy, respectively. 
The choice is fully justified also at intermediate values of $q_T$, 
where the calculation of the partonic cross section is driven by the 
small-$q_T$ resummation and strongly constrained by the total cross section
at NLO and NNLO, respectively.
At large values of $q_T$, $q_T\sim M_H$, our evaluation of the 
partonic cross section is dominated by the fixed-order contributions
at LO and NLO, respectively. Therefore, our choice introduces
a formal mismatch with respect to the customary use of
parton densities and $\as$.
However, as shown and discussed later in this subsection, 
this formal mismatch does not lead
to any inconsistencies at the quantitative level.

\setcounter{footnote}{1}
 
\begin{figure}[htb]
\begin{center}
\begin{tabular}{c}
\epsfxsize=15truecm
\epsffile{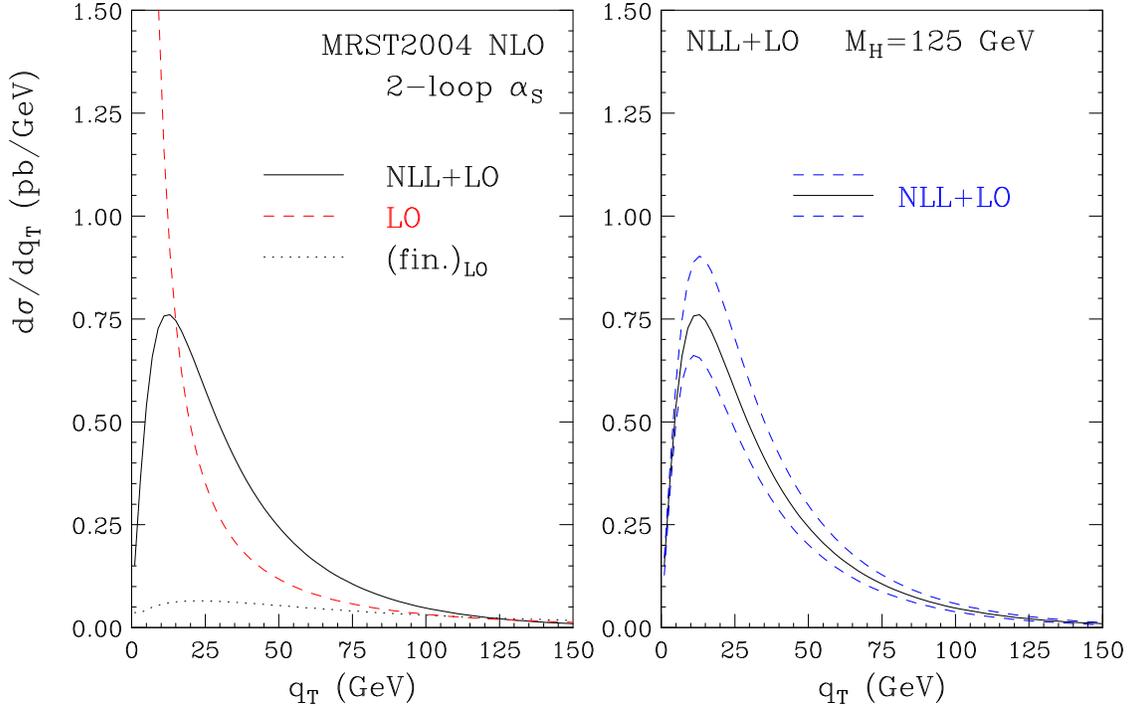}\\
\end{tabular}
\end{center}
\caption{\label{fig:nlllo}
{\em 
The $q_T$ spectrum at the LHC with $M_H=125$~GeV: ({\it left}) setting
$\mu_R=\mu_F=Q=M_H$, the results at 
NLL+LO accuracy are compared with the LO spectrum and the finite component 
of the LO spectrum; ({\it right})~the uncertainty band from variations
of the scales $\mu_R$ and $\mu_F$ at 
NLL+LO accuracy.}}
\end{figure}

The NLL+LO spectrum with $M_H=125$ GeV is shown in Fig.~\ref{fig:nlllo}.
In the left-hand side, the full NLL+LO result (solid line)
is compared with the LO one (dashed line)
at the default scales $\mu_F=\mu_R=Q=M_H$.
We see that the LO calculation diverges to $+\infty$ as $q_T\to 0$. 
The effect of the resummation, which is relevant below $q_T\sim 100$~GeV,
leads to a physically well-behaved distribution: it has a kinematical
peak at $q_T \sim 12$~GeV and vanishes
as $q_T\to 0$.
The LO finite component of the spectrum (dotted line),
which is defined in Eq.~(\ref{finlo}), 
is also shown: as expected it dominates when $q_T \sim M_H$ and 
vanishes as $q_T\to 0$.
Note, however, that the contribution of the finite component is sizeable
in the intermediate-$q_T$ region (about 20\% at $q_T \sim 50$~GeV)
and not yet negligible at small values of $q_T$ 
(about 8\% around the peak region). This underlies the importance
of a careful and consistent matching between the resummed and fixed-order
calculations.
In the right-hand side of Fig.~\ref{fig:nlllo} we show the NLL+LO band as
obtained by varying $\mu_F$ and $\mu_R$
simultaneously and independently
in the range $0.5M_H\leq \mu_F,\mu_R\leq 2M_H$ with the constraint
$0.5 \leq \mu_F/\mu_R \leq 2$ (the resummation scale is kept fixed at $Q=M_H$).
The scale dependence increases from about $\pm 15\%$ at the peak
to about $\pm 20\%$ at $q_T=100$~GeV.
The integral over $q_T$ of the NLL+LO spectrum
is in agreement with the value of the NLO 
total cross section to better than $1\%$, thus proving the numerical accuracy
of the code.

\begin{figure}[htb]
\begin{center}
\begin{tabular}{c}
\epsfxsize=15truecm
\epsffile{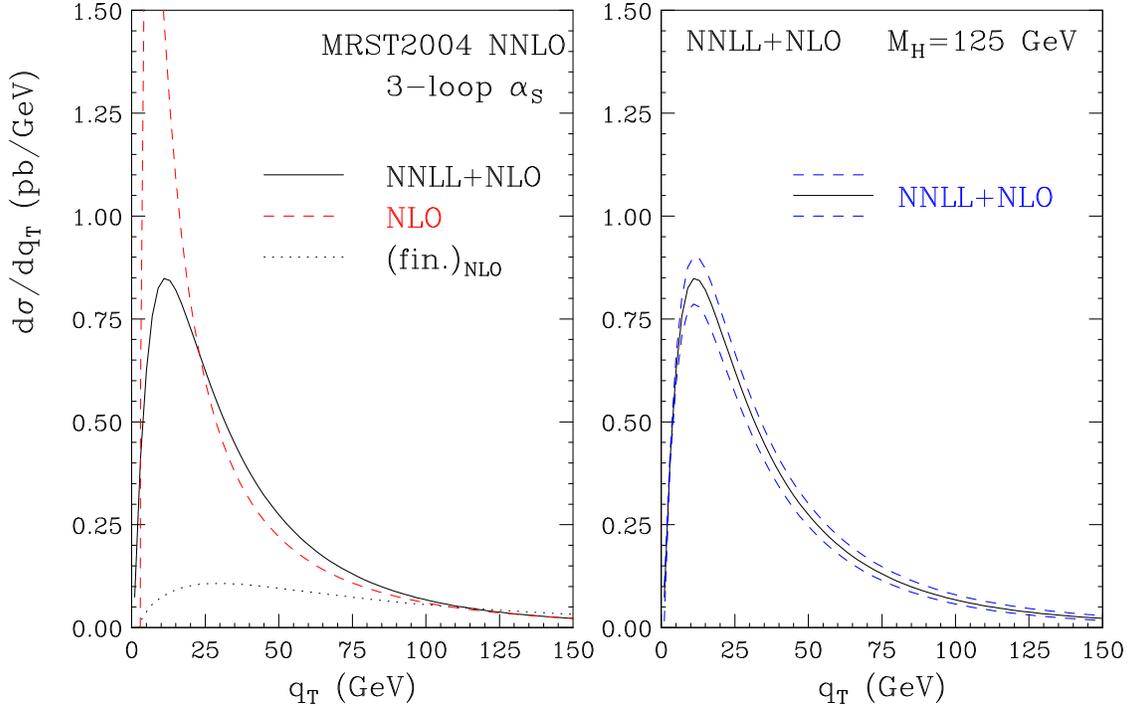}\\
\end{tabular}
\end{center}
\caption{\label{fig:nnllnlo}
{\em 
The $q_T$ spectrum at the LHC with $M_H=125$~GeV: ({\it left}) setting
$\mu_R=\mu_F=Q=M_H$, the results at 
NNLL+NLO accuracy are compared with the NLO spectrum and the finite component 
of the NLO spectrum; ({\it right})~the uncertainty band from variations 
of the scales $\mu_R$ and $\mu_F$ at NNLL+NLO accuracy. }}
\end{figure}

The NNLL+NLO results at the LHC are shown in Fig.~\ref{fig:nnllnlo}.
In the left-hand side, the full result (solid line)
is compared with the NLO one (dashed line) at the
default scales $\mu_F=\mu_R=Q=M_H$.
The NLO result diverges to $-\infty$ as $q_T\to 0$ and, at small values of 
$q_T$, it has an unphysical peak (the top of the peak is above the vertical
scale of the plot) that is produced by the numerical compensation of negative
leading logarithmic and positive subleading logarithmic contributions.
The resummed result is physically well-behaved at small $q_T$.
The NLO finite component of the spectrum (dotted line),
which is defined in Eq.~(\ref{finnlo}), 
vanishes smoothly as $q_T\to 0$; its contribution amounts to about
10\% in the peak region, about 17\%
at $q_T \sim 25$~GeV and about 35\% 
at $q_T \sim 50$~GeV.
This shows both the quality and the relevance of the matching procedure.

We find that the contribution of $A^{(3)}$ (recall from Sect.~\ref{sec:css}
that we are using an educated guess
on the value of the coefficient $A^{(3)}$) to the resummed component
can safely be neglected.
The coefficient ${\cal H}_N^{H(2)}$ contributes significantly, and enhances
the $q_T$ distribution by roughly $20\%$ in the region of intermediate and
small values of $q_T$.
The NNLL resummation effect starts to be visible
below $q_T\sim 100$~GeV, and 
it increases the NLO result by about $25\%$ at $q_T=50$~GeV.

The right-hand side of Fig.~\ref{fig:nnllnlo} shows the scale dependence 
computed as in Fig.~\ref{fig:nlllo}. The scale dependence is now about 
$8\%$ at the peak and increases to about $20\%$ at $q_T=100$~GeV.

To better illustrate the main features of the dependence on the scales
$\mu_R$ and $\mu_F$, we present numerical results at two fixed values
of $q_T$ in Figs.~\ref{fig:scale50} and~\ref{fig:scale15}.
In Fig.~\ref{fig:scale50} we show our results at $q_T=50$~GeV and
$M_H=125$~GeV. 
The scale dependence is analysed by varying the
factorization and renormalization scales around
the default value $M_H$.  The plot on the left corresponds to the simultaneous
variation of both scales, $\mu_F=\mu_R=\chi \, M_H$, whereas the plot in the
centre (on the right) corresponds to the
variation of the factorization (renormalization) scale 
$\mu_F=\chi_F \, M_H$ ($\mu_R=\chi_R \, M_H$) by fixing the other scale
at the default value $M_H$.
\begin{figure}[htb]
\begin{center}
\begin{tabular}{c}
\epsfxsize=15truecm
\epsffile{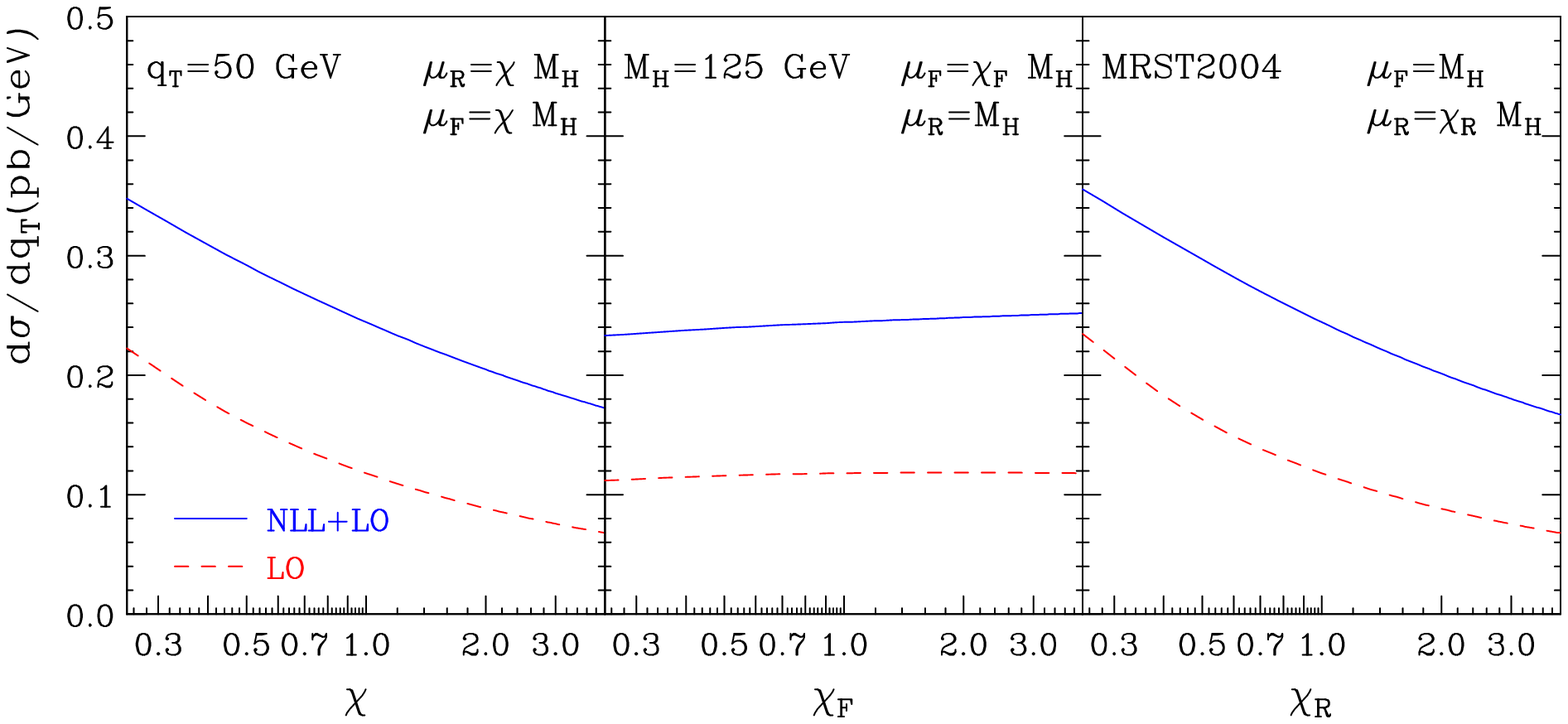}\\
\epsfxsize=15truecm
\epsffile{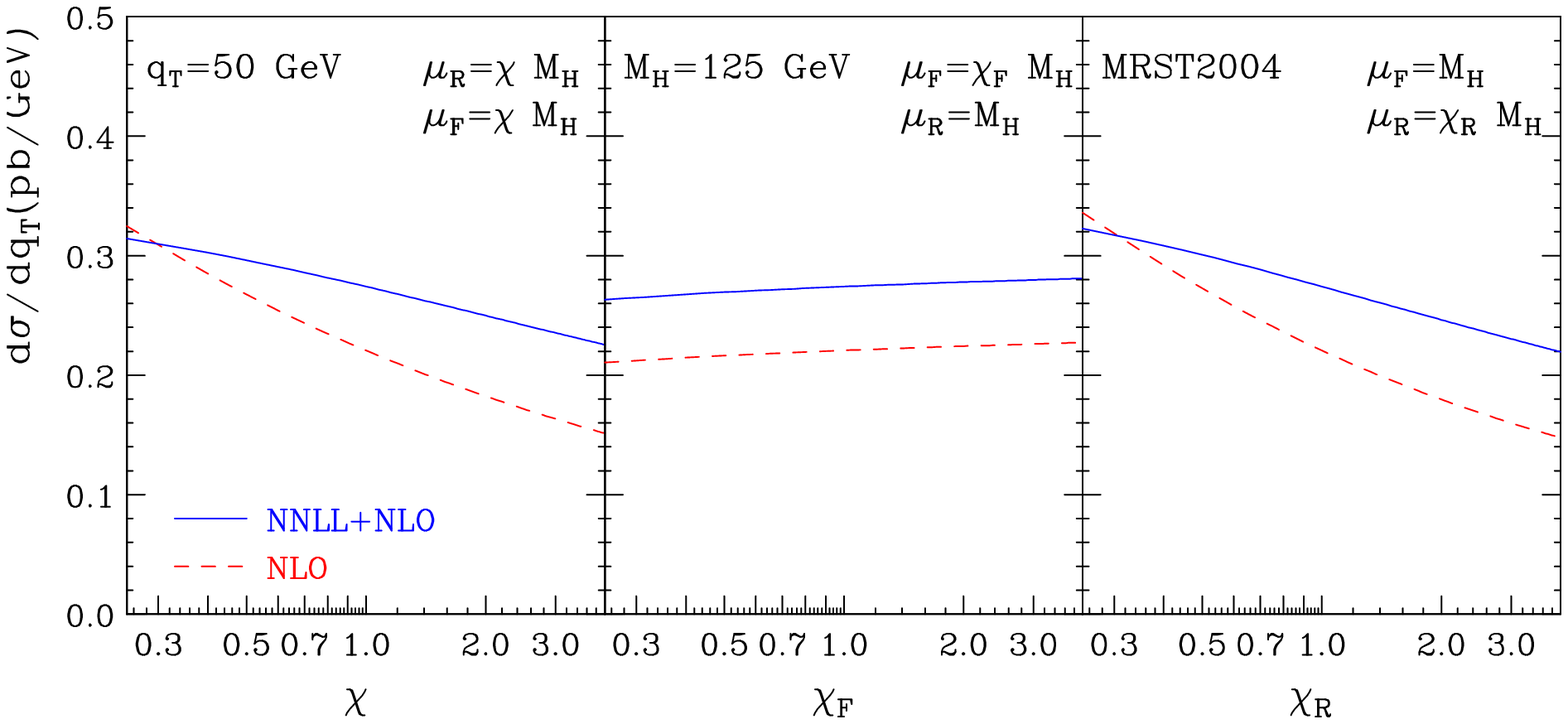}
\end{tabular}
\end{center}
\caption{\label{fig:scale50}{\em Scale dependence of the LHC cross section
for Higgs boson production ($M_H=125$~GeV) at $q_T=50$~GeV. Results 
at a) (upper) LO, NLL+LO 
and b) (lower) NLO, NNLL+NLO accuracy.}}
\end{figure}

As expected from the QCD running of $\as$, the cross sections typically
decrease when $\mu_R$ increases around the characteristic hard scale $M_H$,
at fixed $\mu_F=M_H$.
In the case of variations of $\mu_F$ at fixed $\mu_R=M_H$, we observe 
the opposite behaviour. This is not unexpected, since when $M_H=125$~GeV
the cross section is mainly sensitive to partons
with momentum fraction $x \sim 10^{-2}$, and in this $x$-range scaling
violations of the parton densities are (moderately) positive. 
Varying the two scales simultaneously ($\mu_F=\mu_R$) leads to a partial
compensation of the two different behaviours.
As a result, the
scale dependence is mostly driven by the renormalization scale, because the
lowest-order contribution to the process is proportional to $\as^3$, a
(relatively) high power of $\as$.

Comparing the LO with the NLL+LO results and the NLO with the NNLL+NLO results,
we see that the
scale dependence of the resummed results (solid lines) is smaller than 
that of the corresponding fixed-order results (dashed lines):
the LO and NLL+LO curves have a comparable slope, but the NLL+LO results are
higher; the NLO and NNLL+NLO results have smaller differences, but the slope
of the NNLL+NLO curve is flatter.
In summary, resummation reduces the scale dependence of the fixed-order
calculations also in the region of intermediate values of $q_T$.

\begin{figure}[htb]
\begin{center}
\begin{tabular}{c}
\epsfxsize=15truecm
\epsffile{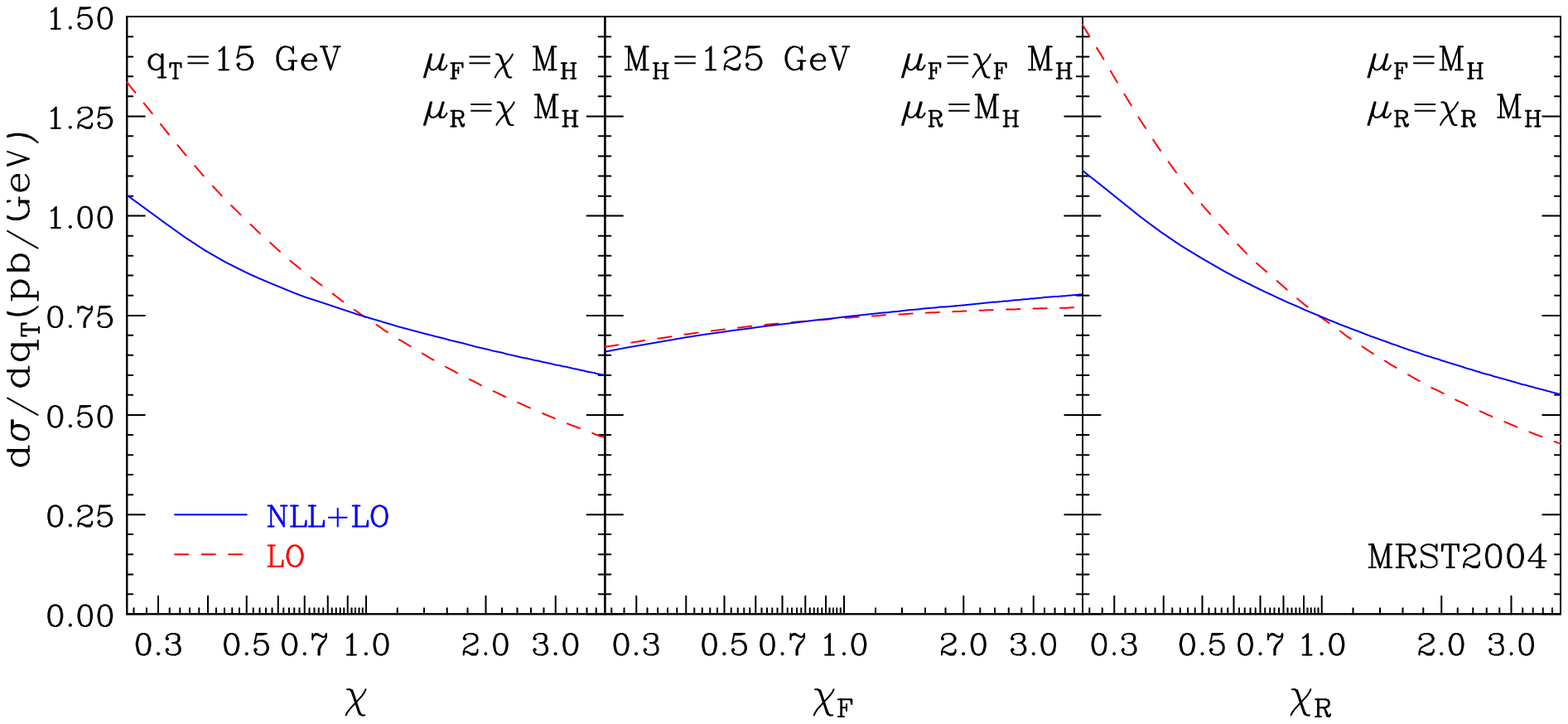}\\
\epsfxsize=15truecm
\epsffile{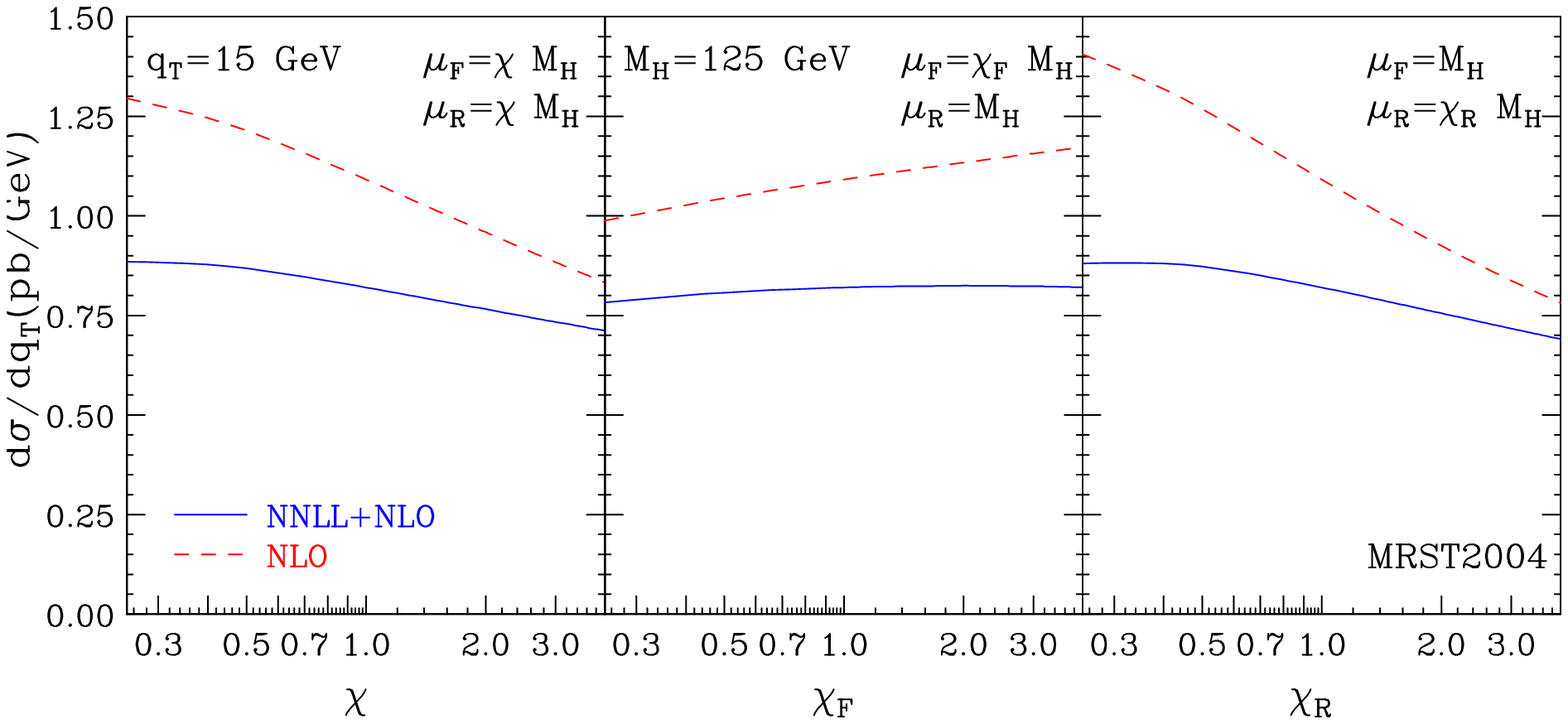}
\end{tabular}
\end{center}
\caption{\label{fig:scale15}{\em Scale dependence of the LHC cross section
for Higgs boson production ($M_H=125$~GeV) at $q_T=15$~GeV. Results 
at a) (upper) LO, NLL+LO 
and b) (lower) NLO, NNLL+NLO accuracy.}}
\end{figure}

In Fig.~\ref{fig:scale15} we report analogous results at a smaller value of 
$q_T$,  namely $q_T=15$~GeV. 
The qualitative behaviour is similar to the one in 
Fig.~\ref{fig:scale50}.
In this region of small transverse momenta the fixed-order result is no longer 
reliable (see Figs.~\ref{fig:nlllo} and \ref{fig:nnllnlo}), but its 
relative scale dependence does not increase and is even smaller than 
at $q_T=50$~GeV. This is due to the fact that the fixed-order cross section
is much larger than at higher values of $q_T$. The slope of the resummed
results (solid lines) is sizeably flatter than that of the corresponding
fixed-order results (dashed lines).
We also notice a slight reduction in the scale 
dependence of the resummed results compared to Fig.~\ref{fig:scale50}, 
especially at NNLL+NLO accuracy.

In Fig.~\ref{fig:bands} the NLL+LO and NNLL+NLO bands shown
in Figs.~\ref{fig:nlllo} and \ref{fig:nnllnlo} are compared.
We see that the NNLL+NLO band (solid lines) is smaller 
than the NLL+LO one (dashed lines) and overlaps with the latter at 
$q_T \ltap 100$~GeV.
This suggests a good convergence of the resummed perturbative expansion. 
This result is confirmed by the inset plot, that shows the NNLL+NLO band 
normalized to the NLL+LO result at central value of the scales.
This $q_T$-dependent K factor,
\begin{equation}
K(q_T)=\f{d\sigma_{NNLL+NLO}(\mu_F,\mu_R)}{d\sigma_{NLL+LO}(\mu_F=\mu_R=M_H)}
\, ,
\end{equation}
is stable, around the values 1.1--1.2, 
in the central region of the inset plot, and it
increases (decreases) drastically when $q_T\gtap 50$~GeV ($q_T\ltap 2$~GeV). 
In the large-$q_T$ region, the effect of perturbative 
higher-order corrections is known to be important
[\ref{deFlorian:1999zd}--\ref{Glosser:2002gm}].
At very small values of $q_T$, non-perturbative effects are definitely 
expected to be relevant.
We observe that a naive rescaling of the NLL+LO result by a constant
(i.e. independent of $q_T$)
$K$ factor would not reproduce the NNLL+NLO result over the entire 
$q_T$-range.

\begin{figure}[htb]
\begin{center}
\begin{tabular}{c}
\epsfxsize=12truecm
\epsffile{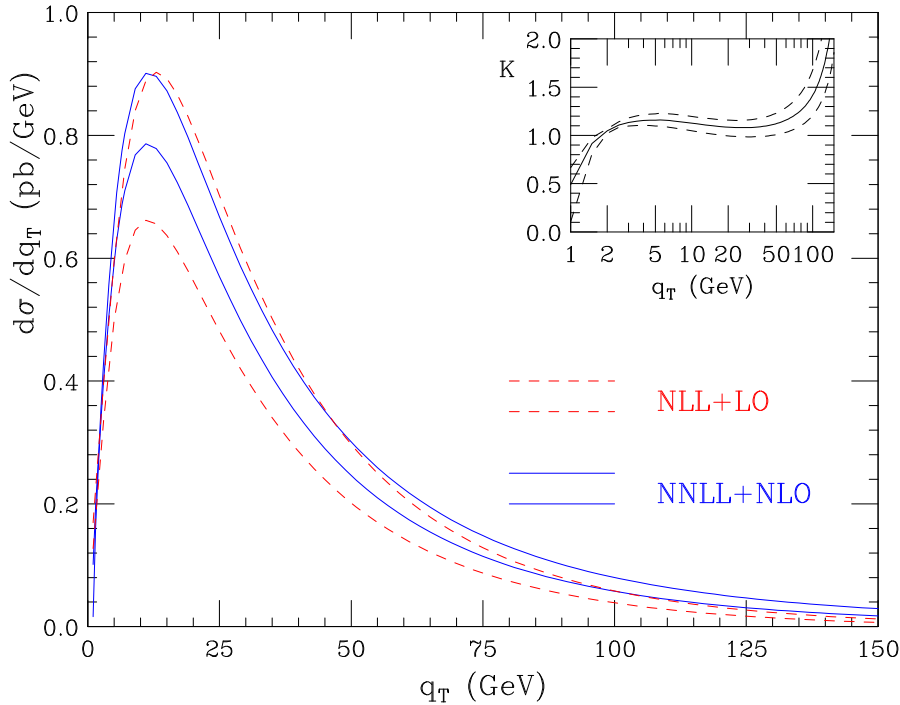}\\
\end{tabular}
\end{center}
\caption{\label{fig:bands}
{\em 
Comparison of the NLL+LO and NNLL+NLO bands ($M_H=125$~GeV). 
The inset plot shows the NNLL+NLO 
band normalized to the central value of the NLL+LO result.}}
\end{figure}

The nice convergence of the resummed perturbative expansion suggested by
Fig.~\ref{fig:bands}
should be contrasted with the results in Fig.~\ref{fig:lonlo}, 
where the corresponding fixed-order
bands, computed as in Fig.~\ref{fig:bands}, are shown.
The results in Fig.~\ref{fig:lonlo} have no physical significance in the 
small-$q_T$ region, owing to the non-convergence of the fixed-order expansion
herein. When $q_T\gtap 25$~GeV, we see that the scale dependence of the
NLO (LO) result is larger than the one of the corresponding NNLL+NLO 
(NLL+LO) result in Fig.~\ref{fig:bands}.
More importantly, we see that the LO and NLO bands do not overlap. This 
implies that the scale dependence enclosed by these bands certainly
underestimates the true theoretical uncertainty from missing higher-order
terms. Equivalently, we can say that the uncertainty of these fixed-order
calculations is more reliably estimated by performing scale variations
over a range of scales that is wider than that used in Fig.~\ref{fig:lonlo}. 
All this indicates a
poor convergence of the fixed-order perturbative expansion at intermediate
values of $q_T$

\begin{figure}[htb]
\begin{center}
\begin{tabular}{c}
\epsfxsize=12truecm
\epsffile{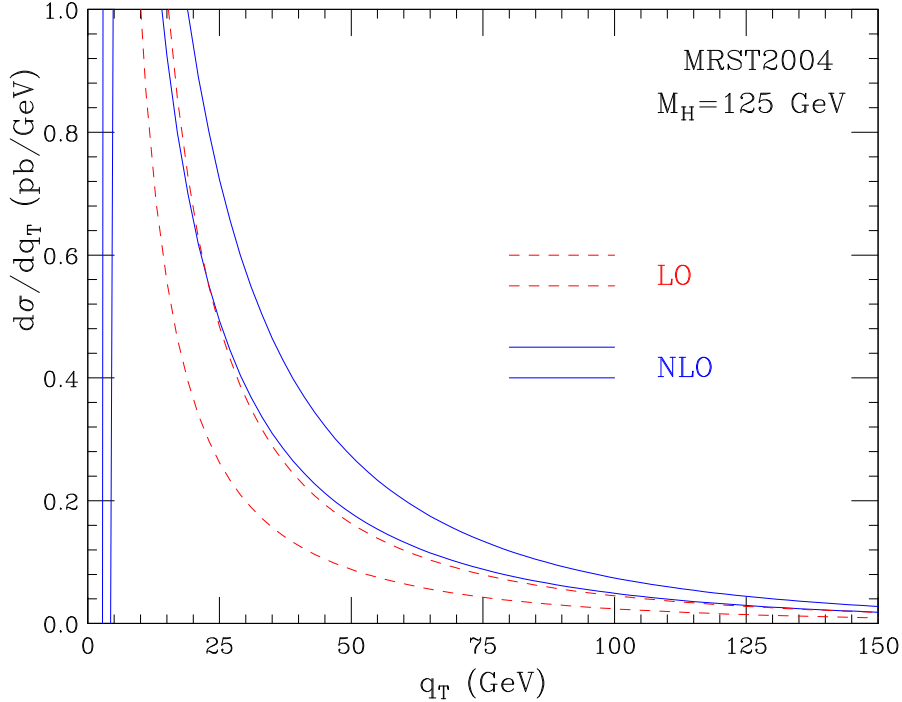}\\
\end{tabular}
\end{center}
\caption{\label{fig:lonlo}
{\em 
Comparison of the LO and NLO bands ($M_H=125$~GeV).}}
\end{figure}

As mentioned at the beginning of this subsection, in our resummed calculations
at NLL+LO and NNLL+NLO accuracy we use parton densities and $\as$ at 
perturbative orders that are different from those customarily used in 
fixed-order calculations at LO and NLO, respectively.
Indeed, the consistent procedure at large values of $q_T$ would be
to use LO densities with 1-loop $\as$ at the LO, and NLO densities 
with 2-loop $\as$ at the NLO. We have also explained why our procedure is
justified in the intermediate-$q_T$ region, and we have postponed the
discussion on the large-$q_T$ region.
To come back to this point, 
in Fig.~\ref{fig:hierarchy} we compare our NLL+LO and NNLL+NLO results
with the customary NLO results, which are obtained by using NLO parton 
densities and 2-loop $\as$. We also include the corresponding bands,
computed from scale variations. In the left-hand side 
we see that in the intermediate-$q_T$ region our NLL+LO result catches 
the bulk of the NLO effect. Obviously, at large $q_T$, the inclusion of NLO 
corrections is necessary. In the right-hand side, the calculations
at NNLL+NLO accuracy and at the NLO are compared. In spite of the fact that the
two calculations use different parton densities and $\as$, the corresponding
bands show a very good overlap when $q_T \sim M_H$.
We thus conclude that, within the NLO theoretical uncertainty, the two
calculations are perfectly compatible at the quantitative level
in the large-$q_T$ region, $q_T \sim M_H$.

\begin{figure}[htb]
\begin{center}
\begin{tabular}{c}
\epsfxsize=15truecm
\epsffile{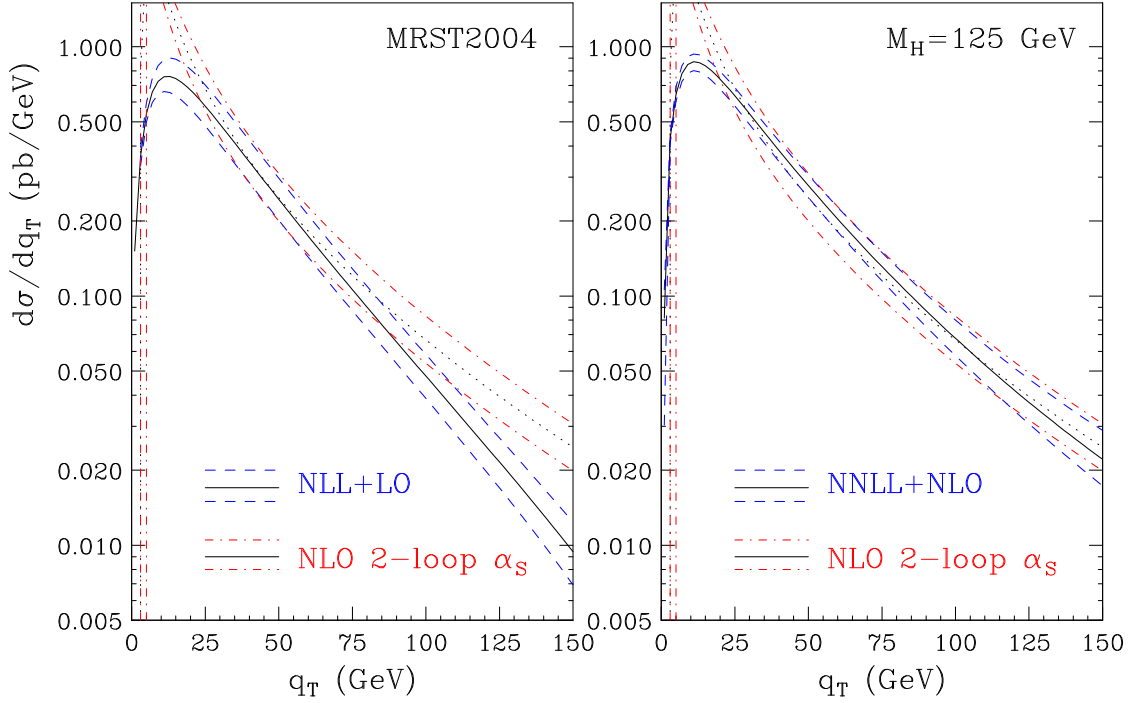}\\
\end{tabular}
\end{center}
\caption{\label{fig:hierarchy}{\em Comparison of the NLL+LO (left) and 
NNLL+NLO (right) bands with the NLO band computed by using NLO parton densities
and 2-loop $\as$.}}
\end{figure}

\begin{figure}[htb]
\begin{center}
\begin{tabular}{c}
\epsfxsize=15truecm
\epsffile{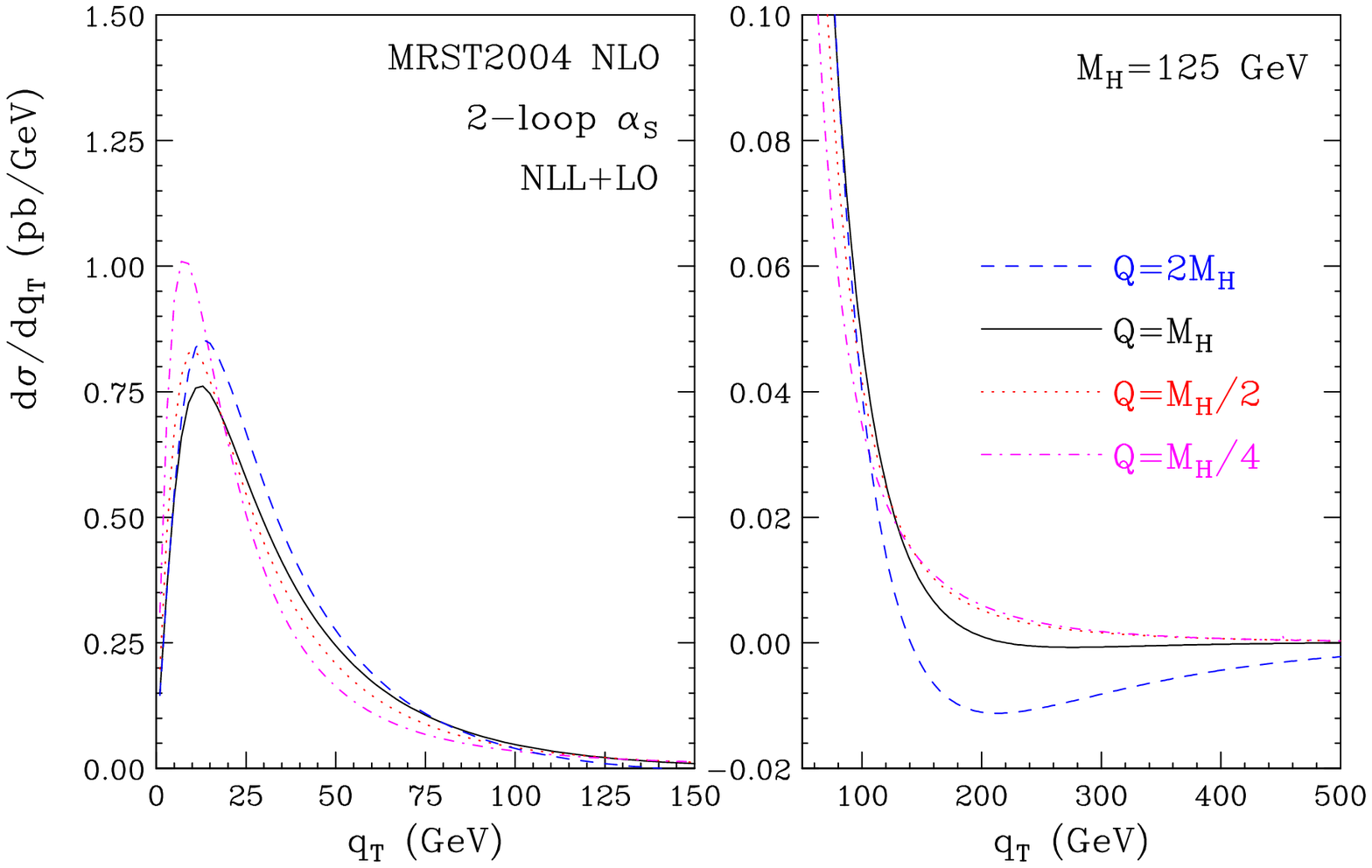}\\
\end{tabular}
\end{center}
\caption{\label{fig:nlladep}{\em NLL+LO spectra for different choices of 
the resummation scale $Q$ at fixed $\mu_R=\mu_F=~M_H$.}}
\end{figure}

\begin{figure}[htb]
\begin{center}
\begin{tabular}{c}
\epsfxsize=15truecm
\epsffile{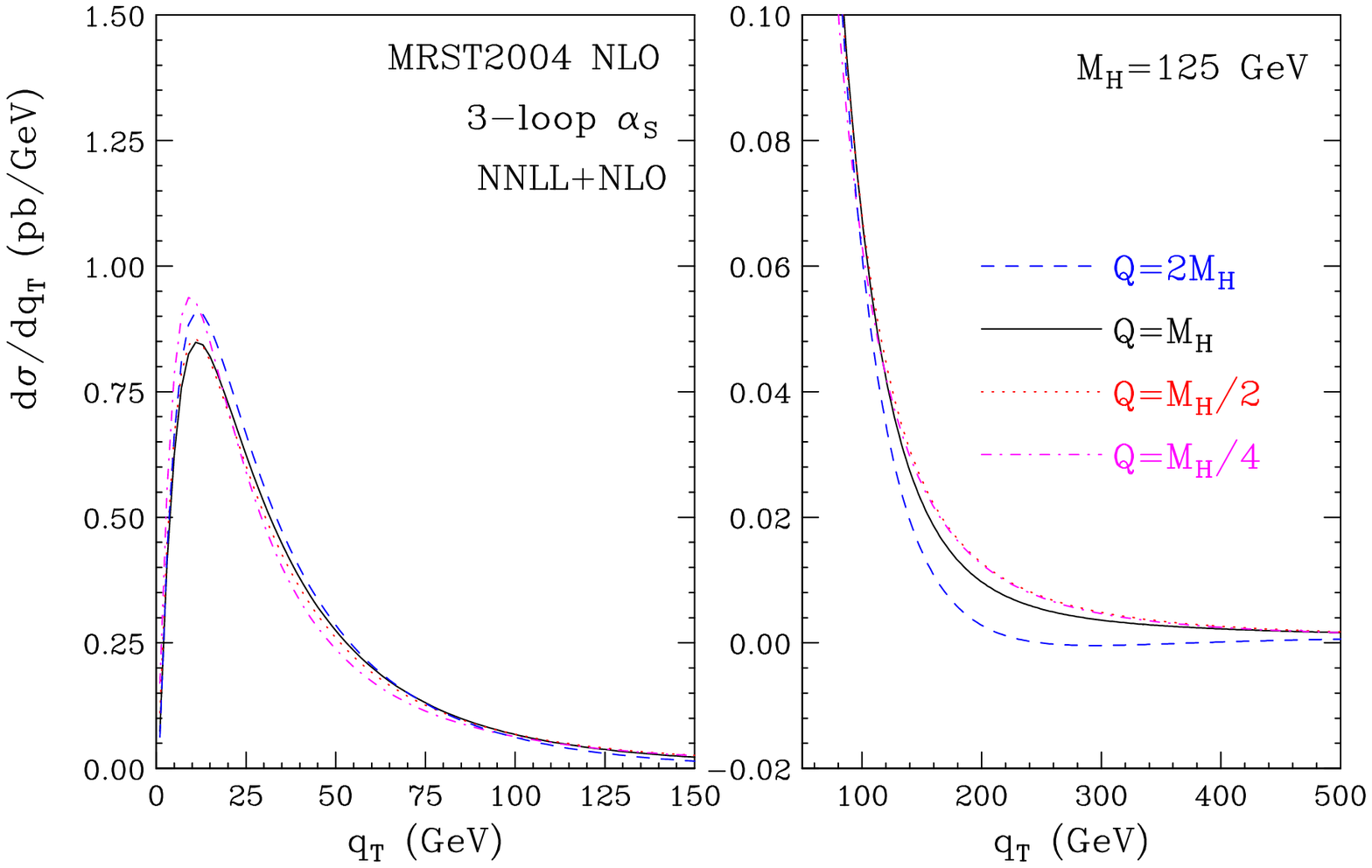}\\
\end{tabular}
\end{center}
\caption{\label{fig:nnlladep}{\em NNLL+NLO spectra for different choices of 
the resummation scale $Q$ at fixed $\mu_R=\mu_F=M_H$.}}
\end{figure}

In Fig.~\ref{fig:nlladep} (Fig.~\ref{fig:nnlladep}) we plot 
the NLL+LO (NNLL+NLO)
spectra for different choices of the resummation scale $Q$.
We remind the reader that the
resummation scale $Q$ has to be chosen of the order of $M_H$.
Variations of the resummation scale around $M_H$ can be studied to estimate 
the uncertainty of the resummed calculation arising from not yet computed
terms at higher logarithmic accuracy. 
In our quantitative study we consider
four different values of $Q$, $Q=2M_H, M_H, M_H/2, M_H/4$.

We first comment on the behaviour at large transverse momenta, which is best
visible looking at the plots on the right of Figs.~\ref{fig:nlladep} and
\ref{fig:nnlladep}. We see that the NLL+LO cross section can become negative 
if $Q=2M_H$. This behaviour should not be regarded as particularly worrisome: 
it takes place when $q_T>M_H$, where the use of the resummation formalism is 
not anymore justified. In general, the cross section has a better
behaviour at large $q_T$ when the resummation scale has the values
$Q=M_H,M_H/2,M_H/4$.
In particular, at large-$q_T$ the results of the fixed-order calculation
at LO (NLO) accuracy are very well approximated by the NLL+LO (NNLL+NLO)
calculation with $Q=M_H/2$; the line corresponding to the LO (NLO) results
is not shown in the plot on the right of Fig.~\ref{fig:nlladep} 
(Fig.~\ref{fig:nnlladep}),
since it is hardly distinguishable from the dotted and dot-dashed lines.
The fact that the fixed-order behaviour at large $q_T$ is approximated better 
when $Q$ is smaller is not unexpected.
By varying $Q$, we smoothly set the transverse-momentum scale below 
which the resummed logarithmic terms are mostly effective; 
when $Q$ is smaller, the resummation effects are confined to a range of 
smaller values of $q_T$.

To quantify the resummation-scale uncertainty on the cross section
at small and intermediate values of $q_T$, we 
proceed as in the case of the renormalization and factorization scales, 
and we vary $Q$ by a factor of 2 up and down from a reference value.
We choose the reference value $Q=M_H/2$, because of the better 
quality of the behaviour of the corresponding cross section at large $q_T$.
From Fig.~\ref{fig:nlladep}, we see
that at NLL+LO accuracy a scale variation between $1/4 M_H$ and $M_H$ produces
a variation of the cross section of about $\pm 15\%$ in the region around
the peak.
At NNLL+NLO accuracy (Fig.~\ref{fig:nnlladep}) the resummation-scale 
dependence is much reduced: when
$Q$ varies between $M_H/4$ and $M_H$ the change in the cross section 
at the peak is about $\pm 5\%$, i.e., smaller than 
the corresponding uncertainty from variations of the renormalization and
factorization scales (see Fig.~\ref{fig:nnllnlo}).

\begin{figure}[htb]
\begin{center}
\begin{tabular}{c}
\epsfxsize=15truecm
\epsffile{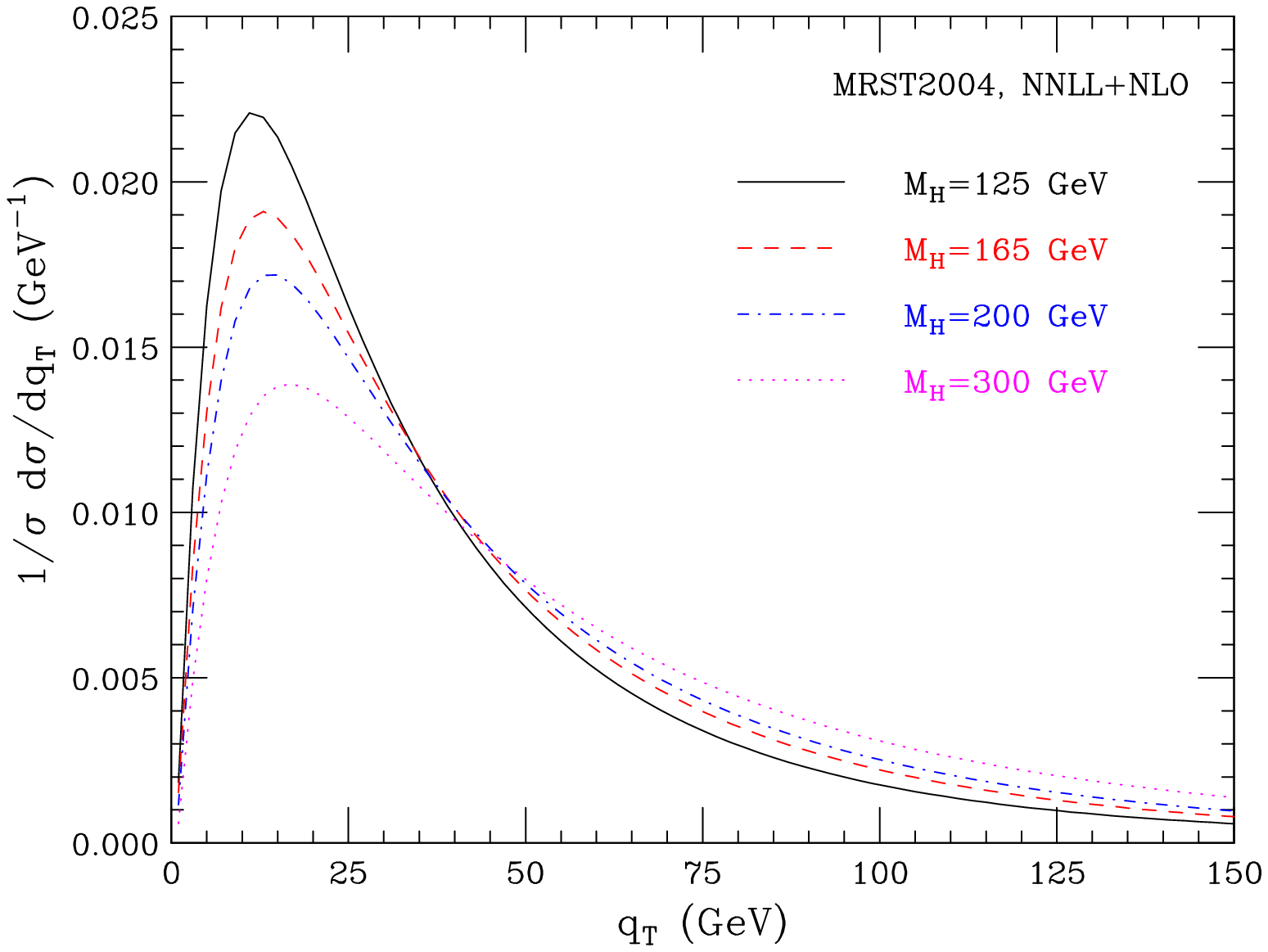}\\
\end{tabular}
\end{center}
\caption{\label{fig:masses1}{\em NNLL+NLO spectra for different values of the
Higgs boson mass. The scales are set at the default value 
$\mu_F=\mu_R=Q=M_H$.}}
\end{figure}

Throughout this section we used the  MRST2004 set [\ref{Martin:2004ir}] 
of parton distribution functions at NLO and NNLO. The NLO and NNLO parton
densities from Alekhin are currently being updated [\ref{Alekhin:2002fv}].
The CTEQ [\ref{Pumplin:2002vw}] and GRV [\ref{Gluck:1998xa}] groups
do not include sets of NNLO parton densities. The parton distribution sets
of MRST, Alekhin and CTEQ include estimates of experimental uncertainties,
which lead to effects below to about 5\% on the total cross section
for Higgs boson production at the LHC. We do not expect significantly 
different results in the case of the $q_T$ cross section at the LHC, 
and we refer
to Ref.~[\ref{Catani:2003zt}] for results and discussions about the effects
of available parton densities on the total cross section.

The numerical results presented so far refer to the value $M_H=125$~GeV of
the Higgs boson mass. By varying $M_H$, the typical features of the results
are unchanged, the main difference being the decrease of the cross section
as $M_H$ increases.
In Fig.~\ref{fig:masses1} we plot the NNLL+NLO spectra, normalized to the total
cross section, for different values of the Higgs boson mass,
$M_H=125,165,200$ and 300~GeV. For reference, the corresponding values of the 
NNLO total cross sections are $\sigma_{NNLO}=38.43, 24.37, 17.78$ and 10.03~pb.
As expected, the $q_T$ distribution becomes harder as $M_H$ increases.
The average value, 
$\langle q_T \rangle$, of the transverse momentum
increases almost linearly with increasing $M_H$, and it is very roughly 
approximated by an effective lowest-order expression, 
$\langle q_T \rangle \sim C_A \as(M_H^2) \,M_H$.

The quantitative predictions
presented up to now are obtained in a purely
perturbative framework. It is known (see e.g. Ref.~[\ref{Collins:va}] 
and references therein)
that the transverse-momentum distribution
is affected by non-perturbative (NP) effects,
which become important as $q_{T}$ becomes small.
In impact parameter space, 
these effects are associated to the large-$b$ region.
In our perturbative study the integral over the impact parameter turns out 
to be dominated by the region where $b\ltap 0.1$--$0.2$ GeV$^{-1}$, 
larger values of $b$
being strongly suppressed by the resummation of the logarithmic terms
in the gluon form factor.
Thus we do not expect particularly-large 
NP effects in the case of Higgs boson production at the LHC.
This expectation is in agreement with the findings in 
Refs.~[\ref{Balazs:2000wv}--\ref{Kulesza:2003wn}].

A customary way of modelling NP effects
in the case of DY lepton-pair production
is to introduce an NP transverse-momentum smearing of the distribution.
This is implemented by multiplying the $b$-space perturbative form factor
by an NP form factor.
Several different parametrizations 
of the NP form factor
are available in the literature
[\ref{Qiu:2000ga}, \ref{DSW}--\ref{Konychev:2005iy}];
the corresponding NP parameters are obtained
from global fits to DY data.

In the case of Higgs boson production, the estimate of NP effects 
is obviously more
uncertain, since we cannot exploit available experimental data.
In Ref.~[\ref{Assamagan:2004mu}] we studied the impact of NP contributions 
on the $\qt$ spectrum of the Higgs boson, by applying the
DY NP corrections of
Refs.~[\ref{DSW}--\ref{BLNY}] to
our resummed results at NLL accuracy. We also considered the effect of
rescaling the DY NP coefficients by the factor $C_A/C_F$,  
to take into account the different colour charges of the initial-state partons
($q{\bar q}$ in the DY process, $gg$ in Higgs boson production) in the
hard-scattering subprocess. Alternatively, we used the NP coefficients
extracted in Ref.~[\ref{Kulesza:2003wi}] from a fit of data on 
$\Upsilon$ production, a production process that is more sensitive to the 
gluon content of the colliding hadrons. 
All these different quantitative implementations of NP corrections, 
although certainly not fully justified, can give an idea of the
size of the NP effects on the Higgs boson spectrum.

The results of Ref.~[\ref{Assamagan:2004mu}] show that 
the impact of the NP effects on the NLL resummed distribution
is definitely below $10\%$ for $q_T\gtap 10$~GeV, and it decreases 
very rapidly as $q_T$
increases. Moreover, when $q_T\ltap 10$~GeV, different parametrizations of the
NP terms can lead to sizeably different relative effects, as a 
consequence of our present ignorance on the absolute value of the NP
contributions.  

In view of these results, in the present paper we limit ourselves
to considering a simple parametrization of the NP contributions.
We multiply the $b$-space resummed component 
${\cal W}_{ab}^{H}(b,M,{\hat s})$
on the right-hand side of Eq.~(\ref{resum1}) by a NP factor, $S_{NP}$,
which includes a gaussian smearing of the form
\begin{equation}
\label{npff}
S_{NP}=\exp\{-g_{NP} \;b^2\}\, .
\end{equation}
The NP coefficient $g_{NP}$ is varied in the range suggested by the study of
Ref.~[\ref{Kulesza:2003wi}]: $g_{NP}=1.67$--$5.64~{\rm GeV}^2$.
Note that this procedure, with these values of $g_{NP}$, well approximates
the quantitative spread of NP effects found in Ref.~[\ref{Assamagan:2004mu}]
at NLL accuracy.
In Fig.~\ref{fig:NP} we plot
the effect of the NP smearing on our best perturbative predictions, as given
by the results at NNLL+NLO accuracy.
The inner plot shows the relative deviation from the
NNLL+NLO perturbative result, as defined by the ratio
\begin{equation}
\Delta=\f{d\sigma_{NNLL+NLO}^{NP}-d\sigma_{NNLL+NLO}}{d\sigma_{NNLL+NLO}} \,,
\end{equation}
where $d\sigma_{NNLL+NLO}^{NP}$ is the NNLL+NLO cross section, 
$d\sigma_{NNLL+NLO}$, supplemented with the NP form factor.
We see that the NP effects give deviations from the purely perturbative result
that are below $10\%$ for $q_T\gtap 5$ GeV. 
\begin{figure}[htb]
\begin{center}
\begin{tabular}{c}
\epsfxsize=12truecm
\epsffile{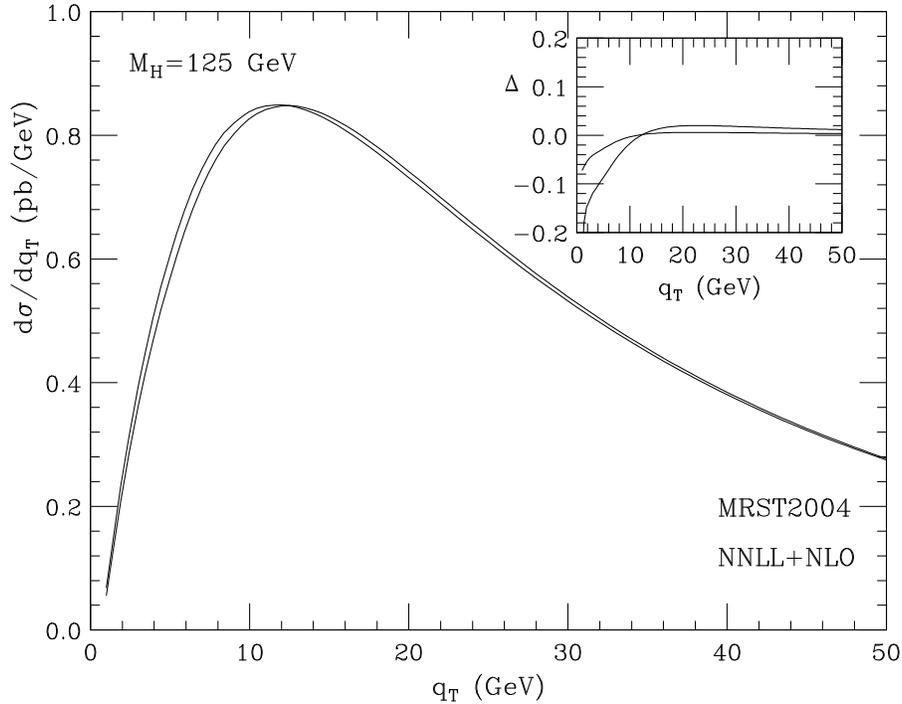}\\
\end{tabular}
\end{center}
\caption{\label{fig:NP}{\em The NNLL+NLO perturbative results supplemented
with the NP form factor in Eq.~(\ref{npff}). The upper (lower) curve at small
$q_T$ is obtained with $g_{NP}=1.67~{\rm GeV}^2 \,(g_{NP}=5.64~{\rm GeV}^2)$.}}
\end{figure}
Comparing the inset plots in Figs.~\ref{fig:bands} and \ref{fig:NP},
we also notice that the inclusion of higher-order contributions
(going from NLL+LO to NNLL+NLO) and of NP contributions have a qualitatively 
similar effect at intermediate and small values of transverse momenta: 
both contributions make the distribution harder.
At the quantitative level, $\Delta$ is much smaller than $K-1$ when
$q_T \gtap 10$~GeV, while $\Delta$ and $K-1$ are comparable when 
$q_T \ltap 10$~GeV.
This points towards a non-trivial interplay between higher-order perturbative 
effects and NP effects at fixed value of the Higgs boson mass.

In summary, the comparison of the NLL+LO and NNLL+NLO results
from small (around the peak region) to intermediate (say, roughly,
$q_T \ltap M_H/3$) values of transverse momenta
shows a nice convergence of the resummed QCD predictions for
the $q_T$ spectrum of the Higgs boson at the LHC.
From this comparison and from the effects of variations of the renormalization,
factorization and resummation scales, we conclude that the perturbative QCD
uncertainty of the NNLL+NLO results is {\em uniformly} 
of about $10\%$ {\em over} this range of transverse momenta.
The perturbative and NP uncertainty increases at smaller values of $q_T$
(see Figs.~\ref{fig:bands} and \ref{fig:NP}); the perturbative uncertainty
increases also at larger values of $q_T$
[\ref{deFlorian:1999zd}--\ref{Glosser:2002gm}].
The perturbative uncertainty on the NNLO cross section [\ref{NNLOtotal}],
as estimated in the same manner (i.e. by comparing the NLO and NNLO results,
and performing scale variations), 
is about 15\% [\ref{Catani:2003zt}].
Our results on the $q_T$ spectrum are thus fully consistent 
with those on the total cross section, since the bulk of the 
events is concentrated at small and intermediate values of the Higgs boson
$q_T$.

\section{Conclusions}
\label{sec:summ}

In this paper we have considered the
transverse-momentum spectrum of generic systems of high-mass $M$
produced in hadron--hadron collisions.
Following our previous work on the subject
[\ref{Catani:2000vq}, \ref{Bozzi:2003jy}],
we have illustrated and discussed in detail a perturbative QCD formalism
that allows us to resum the large logarithmic contributions in the 
small-$q_T$ region ($q_T\ll M$)
and to consistently match the ensuing result to the fixed-order 
contributions 
in the large-$q_T$ region ($q_T\sim M$).
The main features
of our approach,
that make it different from other implementations of $b$-space 
resummation 
presented in the literature, are summarized below.
\begin{itemize}
\item The resummation is performed at the level of
the partonic cross section. 
The parton distributions are thus
evaluated at the
factorization scale $\mu_F$, which has to be chosen of the order of the 
hard scale $M$.
The resummation formula is then organized in a form that is in close 
analogy with the case of event shapes variables
in hard-scattering processes
[\ref{Catani:1991kz}--\ref{Banfi:2004nk}]
and threshold resummation in hadronic collisions [\ref{threshold}, 
\ref{Catani:1996yz}]:
the various classes of logarithmic
contributions are controlled by the QCD coupling $\as(\mu_R^2)$
evaluated at the renormalization scale $\mu_R$.
This procedure
naturally
allows us to perform a systematic study of renormalization- and 
factorization-scale dependence, as is 
customarily
done in fixed-order calculations.
This should be 
compared
with the other implementations of
$b$-space resummation,
where the scale at which the parton distributions are evaluated
is of the order of $1/b$, which also 
necessarily requires
an extrapolation of the parton 
distributions in the NP region.

\item The large logarithmic contributions are exponentiated in the form 
factor $\exp\{{\cal G}_N\}$,
where the function ${\cal G}_N$ (see Eq.~(\ref{gexpan}))
is universal: it does not depend on the produced high-mass system,
and it only depends on the flavour of the partons involved in the
hard-scattering subprocess.
More precisely (see Appendix~\ref{appa}),
various process-independent form factors 
control 
the various partonic channels.
The process dependence, as well as the factorization-scale and 
factorization-scheme 
dependence is fully
included in the hard-scattering coefficient ${\cal H}_N$ 
(see Eq.~(\ref{wtilde})).

\item 
We impose a constraint of perturbative unitarity
through the replacement in 
Eq.~(\ref{grepl}): the $b$-space form factor 
$\exp\{{\cal G}_N({\widetilde L})\}$ is equal to unity at $b=0$.
This constraint has a twofold 
purpose.
On one hand, it avoids the introduction of
unjustified higher-order contributions in the small-$b$ region, which are 
present [\ref{Frixione:1998dw}] in standard implementations
of $b$-space resummation.
On the other hand, it allows us to recover the total cross section at the 
nominal fixed-order accuracy
upon integration over $q_T$.
Note that, as a consequence, perturbative uncertainties at intermediate values
of $q_T$ are reduced.

\end{itemize}

The resummation formalism 
has been applied
to the production of the SM Higgs boson in $pp$ collisions. 
We combined
the most advanced perturbative information that is available at present 
for this process:
NNLL resummation at small $q_T$ and fixed-order perturbation theory 
at NLO at large $q_T$.
We developed a numerical code, named {\tt HqT} [\ref{code}], that performs the 
calculation at NLL+LO and NNLL+NLO accuracy.
In Sect.~\ref{sec:num} we have presented a selection of results that can be 
obtained by our program at LHC energies.
Owing to the unitarity constraint, the integral of our spectra at NLL+LO 
(NNLL+NLO) correctly
reproduces the total NLO (NNLO) cross sections.
The results show a high stability with respect to 
scale variations and an increasing stability when going from NLL+LO to 
NNLL+NLO accuracy. As summarized at the end of Sect.~\ref{sec:num},
this suggests that the
uncertainty from missing higher-order perturbative contributions is under good
control.

\appendix
\section{Appendix:  Exponentiation in the multiflavour case}
\label{appa}

In Sects.~\ref{sec:rescross} and \ref{sec:css}, we have discussed the 
exponentiation structure
of the resummed component of the $q_T$ distribution.
To simplify the notation and the presentation, we have limited ourselves
to illustrating the case in which the partonic scattering involves a single
flavour of partons. This appendix is devoted to generalize the exponentiation
to the case with partons of different flavours.

To obtain Eq.~(\ref{wtildeflav}), the multiflavour analogue 
of Eq.~(\ref{wtilde}), we start from 
the representation in Eq.~(\ref{calw}) of the resummed partonic cross section
${\cal W}_{ab, \,N}^{F}$, and then we proceed as in Sect.~\ref{sec:css}. 
The main difference with respect to the steps in 
Eqs.~(\ref{apsol})-(\ref{hexpr}) is that the solution of the QCD evolution
equations (\ref{apeq}) has the customary form\footnote{In this appendix we use
the boldface notation $\bom X$ to denote the flavour space matrix whose matrix
elements are  $X_{ab} = \left( {\bom X} \right)_{ab}$.}  
\begin{equation}
\label{apsolmat}
{\bom U}_{N}(b_0^2/b^2,Q^2) = P \exp \left\{ 
\int^{b_0^2/b^2}_{Q^2} \frac{dq^2}{q^2} 
\; {\bom \gamma}_{N}(\as(q^2)) \right\} 
\;\;, 
\end{equation}
where the symbol $P$ on the right-hand side denotes the path ordering expansion
of the exponential matrix. Because of its matrix structure,
the exponential in Eq.~(\ref{apsolmat}) has only a formal meaning.
To recast Eq.~(\ref{apsolmat}) in a true exponential form,
we can perform a systematic logarithmic expansion of the solution
of the Altarelli--Parisi equations, by using a well-known
method that dates back, at least,
to Ref.~[\ref{Furmanski:1981cw}].

The evolution operator in Eq.~(\ref{apsolmat}) can be written in the following
form [\ref{Furmanski:1981cw}] (see also Ref.~[\ref{Vogt:2004ns}] for technical
details):
\begin{equation}
\label{vuv1}
{\bom U}_{N}(b_0^2/b^2,Q^2) = {\bom V}_{N}(\as(b_0^2/b^2))
\;{\bom U}_{N}^{(\rm LO)}(\as(b_0^2/b^2),\as(Q^2)) 
\;{\bom V}_{N}^{-1}(\as(Q^2)) \;\;,
\end{equation}
where ${\bom U}_{N}^{(\rm LO)}$ is determined by the lowest-order
anomalous dimensions ${\bom \gamma}_N^{(1)}$,
\begin{equation}
\label{u0ev}
\frac{d {\bom U}_{N}^{(\rm LO)}(\as,\as^\prime) }{d \ln \as} 
= - \frac{1}{\beta_0}
\;{\bom \gamma}_N^{(1)} \;{\bom U}_{N}^{(\rm LO)}(\as,\as^\prime) \;\;,
\end{equation}
and the operator ${\bom V}_{N}$ fulfils
the following differential equation:
\begin{equation}
\label{vevol}
\frac{d {\bom V}_{N}(\as) }{d \ln \as} = \frac{1}{\beta(\as)}
\;{\bom \gamma}_N(\as) \;{\bom V}_{N}(\as) + {\bom V}_{N}(\as) \;
\frac{1}{\beta_0} \;{\bom \gamma}_N^{(1)} \;.
\end{equation}
The evolution equation (\ref{u0ev}) can be solved by diagonalizing
the anomalous dimensions matrix ${\bom \gamma}_N^{(1)}$, which has
three different eigenvalues ${\gamma}_{i, \,N}^{(1)}$:
one eigenvalue in the flavour non-singlet sector $(i={\rm NS})$,
and two eigenvalues in the flavour singlet sector $(i=\pm)$.
The solution of Eq.~(\ref{u0ev}) is
\begin{equation}
\label{u0sol}
{\bom U}_{N}^{(\rm LO)}(\as(b_0^2/b^2),\as(Q^2)) = \sum_{i={\rm NS},\pm}
\left[ 
\f{\as(Q^2)}{\as(b_0^2/b^2)} \right]^{{\gamma}_{i, \,N}^{(1)}/\beta_0} 
\;{\bom E_N^{(i)}}\;\;,
\end{equation}
where ${\bom E_N^{(i)}}$ denotes the projector onto the flavour 
eigenspace corresponding to the eigenvalue ${\gamma}_{i, \,N}^{(1)}$.
By inspection of Eq.~(\ref{vevol}), we see that it can be solved
by performing a perturbative expansion,
\begin{equation}
\label{vexp}
{\bom V}_{N}(\as) = {\bom 1} + \sum_{n=1}^{\infty}
\left( \f{\as}{\pi} \right)^n \;{\bom V}_{N}^{(n)} \;\;,
\end{equation}
and the perturbative coefficients ${\bom V}_{N}^{(n)}$ 
are obtained in terms of the anomalous dimensions coefficients
${\bom \gamma}_N^{(k+1)}$ and the $\beta$ function coefficients $\beta_k$
with $k \leq n$. For example, the first-order coefficient ${\bom V}_{N}^{(1)}$
is given by
\begin{equation}
\label{v1res}
{\bom V}_{N}^{(1)} = \sum_{i,j= {\rm NS},\pm} 
\f{1}{{\gamma}_{j, \,N}^{(1)}- {\gamma}_{i, \,N}^{(1)}- \beta_0}
\;{\bom E_N^{(i)}} \;\left({\bom \gamma}_N^{(2)} 
-\f{\beta_1}{\beta_0} \,{\bom \gamma}_N^{(1)} \right)  {\bom E_N^{(j)}}
 \;\;.
\end{equation}

We now come back to the right-hand side of Eq.~(\ref{calw}). The evolution
operator ${\bom U}_{N}(b_0^2/b^2,\mu_F^2)$
is rewritten as ${\bom U}_{N}(b_0^2/b^2,\mu_F^2)=
{\bom U}_{N}(b_0^2/b^2,Q^2) \;{\bom U}_{N}(Q^2,\mu_F^2)$. Then 
${\bom U}_{N}(b_0^2/b^2,Q^2)$ is replaced by the expression in 
Eq.~(\ref{vuv1}). Equation (\ref{calw}) thus becomes
\beeq
\label{calw1}
&&{\cal W}_{ab, \,N}^{F}(b,M;\as(\mu_R^2),\mu_R^2,\mu_F^2) 
=\sum_c \sigma_{c{\bar c}, \,F}^{(0)}(\as(M^2),M) 
\;H_c^{F}(\as(M^2))  
\nn \\
&&\quad \times \sum_{a_2, \,b_2} \left[ {\bom V}_{N}^{-1}(\as(Q^2))
\; {\bom U}_{N}(Q^2,\mu_F^2)\right]_{a_2 a} \;
\left[ {\bom V}_{N}^{-1}(\as(Q^2)) 
\;{\bom U}_{N}(Q^2,\mu_F^2) \right]_{b_2 b}
\nn \\
&&\quad \times \Bigl\{ \;S_c(M,b) \; \sum_{a_1, \,b_1}
{\widetilde C}_{ca_1, \,N}(\as(b_0^2/b^2)) \;
{\widetilde C}_{{\bar c}b_1, \,N}(\as(b_0^2/b^2)) \Bigr. \\
&&\quad \times \Bigl. \;U_{a_1a_2, \,N}^{(\rm LO)}(\as(b_0^2/b^2),\as(Q^2)) 
\;U_{b_1b_2, \,N}^{(\rm LO)}(\as(b_0^2/b^2),\as(Q^2)) \Bigr\}\nn
\;\;,
\eeeq
where we have defined the perturbative function
\begin{equation}
\label{tildec}
{\bom {\widetilde C}}_{N}(\as) = {\bom C}_{N}(\as) \;{\bom V}_{N}(\as) 
= {\bom 1} + \sum_{n=1}^{\infty}
\left( \f{\as}{\pi} \right)^n \;{\bom {\widetilde C}}_{N}^{(n)} \;\;,
\end{equation}
and inside the curly brackets we have collected all the factors,
$S_c, {\bom {\widetilde C}}_{N}$ and ${\bom U}_{N}^{(\rm LO)}$, that 
depend on the impact parameter~$b$. These factors contain the 
logarithmically-enhanced contributions that have to be resummed and organized
in exponential form. 
The factor $S_c$ can be rewritten as
\begin{equation}
\label{scexp}
S_c(M,b) = S_c(M,b_0/Q) \;\exp \left\{
{\cal G}_{c}(\as(\mu^2_R),L;M^2/\mu^2_R,M^2/Q^2
)\right\}
\end{equation}
where (see Eq.~(\ref{formfact}))
\begin{equation}
\label{calgc}
{\cal G}_{c}(\as(\mu^2_R),L;M^2/\mu^2_R,M^2/Q^2) =
- \int_{b_0^2/b^2}^{Q^2} \frac{dq^2}{q^2} 
\left[ A_c(\as(q^2)) \;\ln \frac{M^2}{q^2} + B_c(\as(q^2)) \right]
\;\;.
\end{equation}
The factor ${\bom U}_{N}^{(\rm LO)}$ is 
\begin{equation}
\label{uloexp}
U_{ab, \,N}^{(\rm LO)}(\as(b_0^2/b^2),\as(Q^2))= \sum_{i={\rm NS},\pm}
E_{ab, \,N}^{(i)} \;\exp \left\{
{\cal G}_{i, \,N}(\as(\mu^2_R),L;M^2/\mu^2_R,M^2/Q^2
)\right\}
\;\;,
\end{equation}
where (see Eq.~(\ref{u0sol}))
\begin{equation}
\label{calgi}
{\cal G}_{i, \,N}(\as(\mu^2_R),L;M^2/\mu^2_R,M^2/Q^2) =
\f{{\gamma}_{i, \,N}^{(1)}}{\beta_0} \; \ln \f{\as(Q^2)}{\as(b_0^2/b^2)} 
= \f{{\gamma}_{i, \,N}^{(1)}}{\beta_0} 
\;\int_{b_0^2/b^2}^{Q^2} \frac{dq^2}{q^2} \;\beta(\as(q^2)) \;\;.
\end{equation}
The factor ${\widetilde C}_{ca, \,N}$ can be written as
\begin{equation}
\label{wtildecexp}
{\widetilde C}_{ca, \,N}(\as(b_0^2/b^2)) =
{\widetilde C}_{ca, \,N}(\as(Q^2)) \;
\exp\{{\cal G}_{ca, \,N}(\as(\mu^2_R),L;M^2/\mu^2_R,M^2/Q^2
)\}
\end{equation}
where (see Eq.~(\ref{rgiden}))
\begin{equation}
\label{calgca}
{\cal G}_{ca, \,N}(\as(\mu^2_R),L;M^2/\mu^2_R,M^2/Q^2) =
-\int_{b_0^2/b^2}^{Q^2} \frac{dq^2}{q^2} 
\;\beta(\as(q^2)) 
\;\frac{d \ln {\widetilde C}_{ca, \,N}(\as(q^2))}{d \ln \as(q^2)}\;\;.
\end{equation}
Note a key point: Eq.~(\ref{wtildecexp}) does not regard the flavour matrix 
${\bom {\widetilde C}}_{N}$, but rather its matrix element  
${\widetilde C}_{ca, \,N}$. Therefore, its right-hand side involves a true
$c$-number exponential instead of a formal matrix exponential.

Inserting Eqs.~(\ref{scexp}), (\ref{uloexp}) and (\ref{wtildecexp}) 
in Eq.~(\ref{calw1}), we eventually obtain the final exponentiated result in
Eq.~(\ref{wtildeflav}), namely
\begin{align}
{\cal W}_{ab, \,N}^F(b,M;\as(\mu_R^2),\mu_R^2,\mu_F^2)
&=\sum_{\{I\}} {\cal H}_{ab, \,N}^{\{I\}, \,F}\left(M, 
\as(\mu_R^2);M^2/\mu^2_R,M^2/\mu^2_F,M^2/Q^2
\right) \nonumber \\
&\times \exp\{{\cal G}_{\{I\}, \,N}(\as(\mu^2_R),L;M^2/\mu^2_R,M^2/Q^2
)\}
\;\;, \nn
\end{align}
where the sum extends over the following set of flavour indices:
\begin{equation}
\{I\}= c, {\bar c}, i, j, a_1, b_1 \;.
\end{equation}
The exponent ${\cal G}_{\{I\}, \,N}$ of the universal form factor and
the process-dependent hard factor ${\cal H}_{ab, \,N}^{\{I\}, \,F}$
are
\begin{equation}
\label{calgcap}
{\cal G}_{\{I\}, \,N} = {\cal G}_{c} + {\cal G}_{i, \,N} + 
{\cal G}_{j, \,N} + {\cal G}_{ca_1, \,N} + {\cal G}_{{\bar c}b_1, \,N} \;\;,
\end{equation}
\begin{align}
\label{htildeflav}
{\cal H}_{ab, \,N}^{\{I\}, \,F}&\left(M, 
\as(\mu_R^2);M^2/\mu^2_R,M^2/\mu^2_F,M^2/Q^2
\right) = \sigma_{c{\bar c}, \,F}^{(0)}(\as(M^2),M) \;H_c^{F}(\as(M^2))
\nonumber \\
&\quad \times \;S_c(M,b_0/Q) \;\,{\widetilde C}_{ca_1, \,N}(\as(Q^2)) \;\,
{\widetilde C}_{{\bar c}b_1, \,N}(\as(Q^2))  \\
&\quad \times \;\left[ {\bom E}_{N}^{(i)} \;{\bom V}_{N}^{-1}(\as(Q^2))
\; {\bom U}_{N}(Q^2,\mu_F^2) \right]_{a_1 a}
\;
\left[ {\bom E}_{N}^{(j)} \;{\bom V}_{N}^{-1}(\as(Q^2)) 
\;{\bom U}_{N}(Q^2,\mu_F^2) \right]_{b_1 b}
\;\;. \nn
\end{align}
From Eqs.~(\ref{calgc}), (\ref{calgi}) and (\ref{calgca}) we see that
${\cal G}_{\{I\}, \,N}$ in Eq.~(\ref{calgcap})
has exactly the integral representation of 
Eq.~(\ref{gformfact}). The logarithmic expansion (see Eq.~(\ref{gexpan}))
of ${\cal G}_{c}$ and ${\cal G}_{i, \,N}$ starts at LL and NLL accuracy,
respectively. The term ${\cal G}_{ca, \,N}$ starts at NLL accuracy
in the flavour off-diagonal case $(c\neq a)$ and at NNLL accuracy
in the flavour diagonal case $(c=a)$. 
The hard function ${\cal H}_{ab, \,N}^{\{I\}, \,F}$ does not depend on the
impact parameter $b$. It can be
perturbatively expanded in powers of $\as(\mu_R^2)$ (with $\mu_R \sim M$),
since the various factors on the right-hand side
of Eq.~(\ref{htildeflav}) involve only scales $(M, Q, \mu_F)$ that are of the
order of the hard-scattering scale $M$.

We conclude this appendix with a comment on the solution (\ref{vuv1})
of the Altarelli--Parisi evolution equations and its relation with the 
resummation in Eq.~(\ref{gexpan})
of the logarithmic contributions to the impact-parameter form factor
$\exp \{{\cal G}_{\{I\}, \,N}(\as,L)\}$. 

The evolution operator ${\bom U}_{N}(b_0^2/b^2,Q^2)$ 
does not contribute to the LL function $g^{(1)}(\as L)$ in Eq.~(\ref{gexpan}).
It starts to contribute to the resummation at the level of the NLL 
function $g_N^{(2)}(\as L)$. Indeed, from Eqs.~(\ref{uloexp}) 
and (\ref{calgi}) we see that 
${\bom U}_{N}^{(\rm LO)}(\as(b_0^2/b^2),\as(Q^2))$, 
the solution of the evolution equations at the lowest-perturbative order,
contributes to the NLL terms $\as^n L^n$. The higher-order corrections to the
evolution equations are taken into account by the operator ${\bom V}_{N}(\as)$
in Eq.~(\ref{vuv1}). The role of these corrections can be examined 
by organizing them in classes of logarithmic
contributions $\as^k(\as L)^n$. Using the expansion in Eq.~(\ref{vexp}),
Eq.~(\ref{vuv1}) gives
\begin{equation}
\label{unll}
{\bom U}_{N}(b_0^2/b^2,Q^2) = 
{\bom U}_{N}^{(\rm LO)}(\as(b_0^2/b^2),\as(Q^2)) +
{\cal O}(\as^{n+2} L^{n+1}) \;\;, \quad (n \geq 0) \;,
\end{equation}
\begin{equation}
\label{unnll}
{\bom U}_{N}(b_0^2/b^2,Q^2) = 
{\bom U}_{N}^{(\rm NLO)}(\as(b_0^2/b^2),\as(Q^2)) +
{\cal O}(\as^{n+3} L^{n+1})
\;\;, \quad (n \geq 0) \;,
\end{equation}
where ${\bom U}_{N}^{(\rm NLO)}$ is the customary solution
[\ref{Furmanski:1981cw}, \ref{Vogt:2004ns}] of the evolution equations at NLO: 
\begin{align}
\label{unlo}
{\bom U}_{N}^{(\rm NLO)}(b_0^2/b^2,Q^2) &= 
{\bom U}_{N}^{(\rm LO)}(\as(b_0^2/b^2),\as(Q^2)) 
+ \frac{\as(b_0^2/b^2)}{\pi} \;
{\bom V}_{N}^{(1)} \;{\bom U}_{N}^{(\rm LO)}(\as(b_0^2/b^2),\as(Q^2)) \nn \\
&- \frac{\as(Q^2)}{\pi} \;{\bom U}_{N}^{(\rm LO)}(\as(b_0^2/b^2),\as(Q^2))
\; {\bom V}_{N}^{(1)} \;\;.
\end{align}
The terms denoted by ${\cal O}(\as^{n+2} L^{n+1})$ 
on the right-hand side of Eq.~(\ref{unll}) contribute at NNLL accuracy
(they are of the same logarithmic accuracy as those in the function 
$\as g_N^{(3)}(\as L)$ in Eq.~(\ref{gexpan})).
Analogously, the terms denoted by ${\cal O}(\as^{n+3} L^{n+1})$ 
on the right-hand side of Eq.~(\ref{unll}) contribute at NNNLL accuracy
(they are of the same logarithmic accuracy as those in the function 
$\as^2 g_N^{(4)}(\as L)$ in Eq.~(\ref{gexpan})).
Therefore, to resum the NLL (NNLL) contributions to the form factor
is sufficient to implement the solution of the evolution equations at the LO
(NLO). Note, however, that, 
to be consistent with the resummed logarithmic expansion, 
the scale dependence of the running couplings $\as(b_0^2/b^2)$ and $\as(Q^2)$ 
in Eq.~(\ref{unll}) (Eq.~(\ref{unnll}))
has to be evaluated at the NLO (NNLO).

\section{Appendix: Bessel transformation of logarithmic contributions}
\label{appb}

This appendix is devoted to the computation of the Bessel transformation
of logarithmic contributions.

We recall the definition of the
the functions ${\widetilde I}_n(q_T/Q)$ introduced in Eq.~(\ref{integrals}):
\begin{equation}
\label{integrals1}
{\widetilde I}_n(q_T/Q) = Q^2 \int_0^\infty db \;\f{b}{2} \,J_0(b q_T) 
\; \ln^n\left( \frac{Q^2b^2}{b_0^2}+1 \right) \;.
\end{equation}
These integrals are easily evaluated 
in terms of derivatives of the 
corresponding generating function ${\widetilde I}(x;\ep)$:
\begin{equation}
\label{win}
{\widetilde I}_n(x)= \lim_{\ep\to 0} 
\left(\frac{\partial}{\partial \ep}\right)^n {\widetilde I}(x;\ep) \;,
\end{equation}
where
\begin{equation}
\label{gfundef}
{\widetilde I}(x;\ep) \equiv \sum_{n=0}^{\infty} \frac{1}{n!} \;\ep^n \;
{\widetilde I}_n(x) \;\;.
\end{equation}
Inserting Eq.~(\ref{integrals1}) in the right-hand side of 
Eq.~(\ref{gfundef}}), we have
\begin{equation}
{\widetilde I}(x;\ep) = \int_0^\infty dt \;\f{t}{2} \,J_0(tx) 
\; \left( \frac{t^2}{b_0^2}+1 \right)^\ep \;\;,
\end{equation}
and this integral can be expressed [\ref{graryz}]
as follows in terms of $K_\nu(x)$, 
the modified Bessel function of imaginary argument (see Eq.~(\ref{interep})):
\begin{equation}
\label{wigf}
{\widetilde I}(x;\ep) =  - \left( \f{2}{b_0 x} \right)^{1+\ep} \;
\frac{\ep \,b_0^2}{2 \,\Gamma(1-\ep)} \;K_{1+\ep}(b_0 x)
\;\;.
\end{equation}
Inserting Eq.~(\ref{wigf}) in Eq.~(\ref{win}) and using the relation
\begin{equation}
\Gamma(1-\ep) = \exp \Big\{ \gamma_E \, \ep+
\sum_{k=2}^{\infty} \f{1}{k} \; \zeta_k \,\ep^k \Big\} \;\;,
\end{equation}
where $\zeta_n$ is the Riemann zeta-function $(\zeta_2=\pi^2/6=1.645\dots,
\zeta_3=1.202\dots)$, the integrals ${\widetilde I}_n(x)$ can  
straightforwardly be expressed
in terms of the derivatives,  $K_1^{(n)}(z)$, of the Bessel function
with respect to its index $\nu$:
\begin{equation}
K_1^{(n)}(z)\equiv \left[ \frac{\partial^n K_\nu(z)}{\partial 
\nu^n}\right]_{\nu=1} \;\;.
\end{equation}
These derivatives
have the following integral representation:
\begin{align}
\label{intereps}
K_1^{(2n)}(z)&=\int_0^\infty dt \; t^{2n} \;e^{-z \cosh t} \;\cosh t \;\;, \\
K_1^{(2n+1)}(z)&= \f{2n+1}{z} \int_0^\infty dt \; t^{2n} \; e^{-z \cosh t}
\;\;,
\end{align}
which can simply be obtained from Eq.~(\ref{interep}).

As discussed in Sect.~\ref{sec:matc}, the computation of the finite component 
of the $q_T$ distribution requires the evaluation of the functions
${\widetilde I}_n(x)$ when $x > 0$\footnote{The behaviour of 
${\widetilde I}_n(x)$ when $x=0$ is discussed at the end of this appendix.}.
In particular, the computation up to NLO (see Eq.~(\ref{resnlo})) requires
${\widetilde I}_n(x)$  with $n=1,2,3,4$; these functions are
\begin{equation}
\label{wi1}
{\widetilde I}_1(x)=-\frac{b_0}{x} K_1(b_0 x) \;,
\end{equation}
\begin{equation}
{\widetilde I}_2(x)= \frac{2b_0}{x} \left[ K_1(b_0 x) \;\ln x -
K_1^{(1)}(b_0 x) \right] \;,
\end{equation}
\begin{equation}
{\widetilde I}_3(x)= - \frac{3b_0}{x}\left[ 
K_1(b_0 x) \left( \ln^2 x - \zeta_2 \right) 
- 2 K^{(1)}_1(b_0 x) \ln x 
+  K_1^{(2)}(b_0 x) \right] \;,
\end{equation}
\begin{align}
\label{wi4}
{\widetilde I}_4(x)=& \,\frac{4b_0}{x}\Big[ 
K_1(b_0 x) \;(\ln^3 x -3 \zeta_2 \ln x + 2\zeta_3) 
- 3 K^{(1)}_1(b_0 x) \;(\ln^2 x- \zeta_2) \nn \\
&+3 K_1^{(2)}(b_0 x) \;\ln x  - K_1^{(3)}(b_0 x)\Big] \;.
\end{align}

The functions ${\widetilde I}_n(x)$ diverge when $x\to 0$.
To examine the divergent behaviour at small values of $x$,
we introduce the functions 
${\overline I}_n(x)$ and the corresponding generating function
${\overline I}(x;\ep)$:
\begin{equation}
{\widetilde I}_n(x) = {\overline I}_n(x) \left[ 1 + {\cal O}(x^2) \right]
\; ,
\end{equation}
\begin{equation}
{\widetilde I}(x;\ep) = {\overline I}(x;\ep) \left[ 1 + {\cal O}(x^2) \right]
\; .
\end{equation}
Using the small-$x$ behaviour of the Bessel function $K_{1+\ep}(x)$ 
[\ref{graryz}] and
performing the small-$x$ limit of Eq.~(\ref{wigf}), we get
\begin{equation}
\label{oigf}
{\overline I}(x;\ep) = - \,\ep \;D(\ep) \;\left( x^2 \right)^{-1-\ep} \; ,
\end{equation}
where
\begin{equation}
\label{dfun}
D(\ep) = \left( \f{2}{b_0} \right)^{\!2\ep} \;\f{\Gamma(1+\ep)}{\Gamma(1-\ep)}
= \exp \Big\{ -2 \sum_{k=1}^{\infty} \f{\zeta_{2k+1}}{2k+1} \;\ep^{2k+1} 
\Big\} \;\;.
\end{equation}

Note that the functions ${\overline I}_n(x)$ exactly correspond to 
the following Bessel transformations:
\begin{equation}
\label{integrals0}
{\overline I}_n(q_T/Q) = Q^2 \int_0^\infty db \;\f{b}{2} \,J_0(b q_T) 
\; \ln^n\left( \frac{Q^2b^2}{b_0^2} \right) \;\;,
\end{equation}
as can be checked by performing the limit $q_T \to 0$ of Eq.~(\ref{integrals1})
or by verifying that the generating function in Eq.~(\ref{oigf}) has the
following integral representation:
\begin{equation}
{\overline I}(x;\ep) = \f{1}{2} \;b_0^{- 2\ep} \;\int_0^\infty dt \;t^{1+2\ep} 
\,J_0(tx) \;\;.
\end{equation}
The relation between ${\overline I}_n(q_T/Q)$ and the small-$q_T$ limit
of ${\widetilde I}_n(q_T/Q)$ is not unexpected in view of the discussion
in Sect.~\ref{sec:rescross}. The integral in Eq.~(\ref{integrals1}) 
originates from Eq.~(\ref{integrals0}) after the replacement
$L=\ln (Q^2b^2/b_0^2) \to {\widetilde L}=\ln (1+ Q^2b^2/b_0^2)$
at the integrand level: when $q_T \to 0$, such a replacement has no effects 
on the singular behaviour at any logarithmic accuracy.

Though ${\overline I}_n(q_T/Q)$ and ${\widetilde I}_n(q_T/Q)$ coincide
when $q_T \to 0$, they behave quite differently at very large values of
$q_T$. When $x \to \infty$, from Eqs.~(\ref{wigf}) and (\ref{oigf})
we get
\begin{align}
{\widetilde I}_n(x) &= \left( -1 \right)^n \, \frac{n}{x} \,
{\sqrt {\f{\pi b_0}{2x} }} \, e^{-b_0 x} \;\ln^{n-1} \f{b_0x}{2}
\;\left[ 1 + {\cal O}\left(\f{1}{\ln x} \right) \right] \;\; , \\
{\overline I}_n(x) &= \left( -1 \right)^n \, \frac{2^{n-1}n}{x^2} \,
\;\ln^{n-1} x
\;\left[ 1 + {\cal O}\left(\f{1}{\ln x} \right) \right] \;\;.
\end{align}
Note, in particular, that ${\widetilde I}_n(x)$ is integrable over $x^2$
when $x \to \infty$, whereas ${\overline I}_n(x)$ it is not.

The function ${\overline I}_n(x)$ can easily be computed by performing the
$n$-th derivative of the generating function (\ref{oigf}) with respect to the
parameter $\ep$. To present the result, we first exclude the singular point
$x=0$ and consider only the region $x > 0$.
Since the generating function depends on $x$ only through
the factor $( x^2 )^{-1-\ep}$, $x^2 \;{\overline I}_n(x)$ is simply a 
polynomial of degree $n-1$ in the variable $\ln x^2$:
\begin{equation}
\label{inpol}
{\overline I}_n(x) = - \frac{1}{x^2} \;\sum_{k=0}^{n-1}  
\f{n!}{k! \,(n-k-1)!} \;
d_{k} \;\ln^{n-k-1} 
\f{1}{x^2}\;\;, \quad x > 0 \;\;,
\end{equation}
where the coefficients $d_n$ are obtained from Eq.~(\ref{dfun}):
\begin{equation}
d_n= \left[ \left( \f{d}{d\ep} \right)^n  D(\ep) \right]_{\ep=0} \;\;.
\end{equation}
The value of the first few coefficients is
\begin{equation}
d_0=1, \quad d_1=d_2=0, \quad d_3=-4 \zeta_3 , \quad d_4=0, 
\quad d_5=-48 \zeta_5 , \quad d_6=160 \zeta_3^2 \;\;.
\end{equation}
The result in Eq.~(\ref{inpol}) agrees with that in Ref.~[\ref{Kulesza:1999gm}],
where one can find the numerical values of $d_n$ with $n \leq 19$
($d_n=2^n {\overline b}_n(\infty)$, where ${\overline b}_n(\infty)$ are given 
in Table~1 of Ref.~[\ref{Kulesza:1999gm}]).
The small-$x$ limit of Eqs.~(\ref{wi1})--(\ref{wi4}) thus gives
\begin{equation}
{\overline I}_1(x) = - \frac{1}{x^2} \;, \quad
{\overline I}_2(x) = - \frac{2}{x^2} \ln \f{1}{x^2} \;, \quad
{\overline I}_3(x) = - \frac{3}{x^2} \ln^2 \f{1}{x^2} \;, \quad
{\overline I}_4(x) = - \frac{4}{x^2} \left( \ln^3 \f{1}{x^2} - 4 \zeta_3
\right) \;.
\end{equation}
Note that, since $d_1=d_2=d_4=0$, ${\overline I}_n(x)$ can be expressed in a
simple form to very high logarithmic accuracy. For example, 
we have: 
\begin{align}
{\overline I}_n(x) &= - \frac{n}{x^2} \left\{
\; \ln^{n-1} \f{1}{x^2} 
- \f{2}{3} \,\zeta_3 \,\f{(n-1)!}{(n-4)!} \;\ln^{n-4} \f{1}{x^2}
- \f{2}{5} \,\zeta_5 \,\f{(n-1)!}{(n-6)!} \;\ln^{n-6} \f{1}{x^2} 
\right. \nn \\
&+ \left. \f{2}{9} \,\zeta_3^2 \,\f{(n-1)!}{(n-7)!} \;\ln^{n-7} \f{1}{x^2}
+ {\cal O}\left( \ln^{n-8} \f{1}{x^2}\right)
\right\}\;\;,\quad  \quad x > 0 \;\;.
\end{align}

We now discuss how to deal with the region around the singular point $x=0$.
We first split the $x$ range in a large-$x$ $(x>x_0)$ and a small-$x$ 
$(x \leq x_0)$ region, where the parameter $x_0$ can be chosen arbitrarily.
Setting $x_0=1$, we have
\begin{equation}
\label{isplit}
{\overline I}_n(x) = {\overline I}_n(x) \;\Theta(x-1) + 
{\overline I}_n(x) \;\Theta(1-x) \;.
\end{equation}
In the large-$x$ region, which excludes the point $x=0$, ${\overline I}_n(x)$
is given by Eq.~(\ref{inpol}). In the small-$x$ region, to 
properly treat the singularity at $x=0$, we have to consider the generating
function in Eq.~(\ref{oigf}) and use the expansion
\begin{equation}
\label{plusdistex}
(x^2)^{-1-\ep} \;\Theta(1-x)=-\frac{1}{\ep} \;\delta(x^2)+ \left[ \frac{1}{x^2} \,
(x^2)^{-\ep} \right]_+ = -\frac{1}{\ep} \;\delta(x^2) +
\sum_{n=0}^{\infty} \frac{\ep^n}{n!} 
\left[ \frac{1}{x^2} \ln^n \frac{1}{x^2} \right]_+ \;\;,
\end{equation}
where the plus-distribution is customarily defined by its action onto
any function $h(x^2)$ that is finite at $x=0$:
\begin{equation}
\int_0^1 dx^2 \;h(x^2) \left[ \frac{1}{x^2} \ln^n \frac{1}{x^2} \right]_+
\equiv 
\int_0^1 dx^2 \;\f{h(x^2) - h(0)}{x^2} \;\ln^n \frac{1}{x^2} \;\;.
\end{equation}
Therefore the generalization of Eq.~(\ref{inpol}) to include the point $x=0$
is
\begin{equation}
\label{inpolx0}
{\overline I}_n(x) = d_{n} \;\delta(x^2) - \;\sum_{k=0}^{n-1}  
\f{n!}{k! \,(n-k-1)!} \;
d_{k} \;\left[ \,\frac{1}{x^2} \ln^{n-k-1} 
\f{1}{x^2} \right]_+  \;\;, \quad 0 \leq x \leq 1 \;\;.
\end{equation}

The procedure described in Eqs.~(\ref{isplit}) and (\ref{plusdistex})
can also be applied to properly define the integrals ${\widetilde I}_n(x)$
around the point $x=0$ in the small-$x$ region. Choosing $x_0=\infty$,
the final result is 
equivalent to start from ${\widetilde I}_n(x>0)$,
the expression of ${\widetilde I}_n(x)$
when $x \neq 0$ (for example, Eqs.~(\ref{wi1})--(\ref{wi4})), 
and then introduce a generalized plus-prescription
that acts in the entire range $0 \leq x < \infty$. Formally we can write
\begin{equation}
\label{winx0}
{\widetilde I}_n(x) = \left[ {\widetilde I}_n(x>0) \right]_{+\infty} \;\;,
\end{equation}
where the generalized plus-distribution is defined as
\begin{equation}
\int_0^\infty dx^2 \;h(x^2) \left[ {\widetilde I}_n(x>0) \right]_{+\infty}
\equiv 
\int_0^\infty dx^2 \;\left[ h(x^2) - h(0) \right] \;{\widetilde I}_n(x>0)
\;\;.
\end{equation}
The choice $x_0=\infty$ to define the plus-prescription in the case of
${\widetilde I}_n$ is feasible since ${\widetilde I}_n(x)$
(unlike  ${\overline I}_n(x)$) is integrable over $x^2$
when $x \to \infty$. This choice simplifies the definition of 
${\widetilde I}_n$ since the right-hand side of Eq.~(\ref{winx0})
(unlike Eq.~(\ref{inpolx0})) does not contain any contact term proportional to
$\delta(x^2)$. The contact term vanishes since the integrand factor
$\ln^n(1+Q^2b^2/b_0^2)$ in Eq.~(\ref{integrals1}) vanishes at $b=0$.
The vanishing of the contact term is thus ultimately related
to the unitarity constraint in Eqs.~(\ref{sigtotconst}) and (\ref{restot}).

\vspace*{0.5cm}

\noindent {\bf  Acknowledgements.} 
The work of G.B. was supported by a postdoctoral fellowship of the French 
ministry for education and research. The work of D.dF. was supported in part
by Fundaci\'on Antorchas, CONICET and UBACyT.
D.dF. wishes to thank the INFN for support and hospitality during his visit
at the Sezione di Firenze.

\section*{References}

\begin{enumerate}

\item \label{Bozzi:2003jy}
G.~Bozzi, S.~Catani, D.~de Florian and M.~Grazzini,
Phys.\ Lett.\ B {\bf 564} (2003) 65.

\item \label{Gunion:1989we}
For a review on Higgs physics in and beyond the Standard Model, see
J.~F.~Gunion, H.~E.~Haber, G.~L.~Kane and S.~Dawson,
{\it The Higgs Hunter's Guide} (Addison-Wesley, Reading, Mass., 1990);
M.~Carena and H.~E.~Haber,
Prog.\ Part.\ Nucl.\ Phys.\  {\bf 50} (2003) 63;
A.~Djouadi,
report LPT-ORSAY-05-17 [hep-ph/0503172],
report LPT-ORSAY-05-18 [hep-ph/0503173].

\item \label{Barate:2003sz}
R.~Barate {\it et al.}  [The LEP Collaborations and the LEP Working
Group for Higgs boson searches],
Phys.\ Lett.\ B {\bf 565} (2003) 61.

\item \label{Group:2004qh}
The LEP Collaborations, the LEP Electroweak Working Group and 
the SLD Electroweak and Heavy Flavour Groups, 
report LEPEWWG/2004--01 [hep-ex/0412015].

\item \label{Carena:2000yx}
M.~Carena {\it et al.},
{\it Report of the Tevatron Higgs working group},
hep-ph/0010338;
CDF and D0 Collaborations, {\it Results of the Tevatron Higgs
Sensitivity Study}, report FERMILAB--PUB--03/320-E.

\item \label{Yao:2004jv}
W.~M.~Yao  [CDF and D0 Collaborations],
report FERMILAB--CONF--04/307-E
[arXiv:hep-ex/0411053].

\item \label{atlascms}
CMS Coll., {\it Technical Proposal}, report CERN/LHCC/94-38 (1994);
ATLAS Coll., {\it ATLAS Detector and Physics Performance: Technical Design
Report}, Vol. 2, report CERN/LHCC/99-15 (1999).

\item \label{lhcupdate}
S.~Abdullin {\it et al.}, report CMS-NOTE-2003-033;
S.~Asai {\it et al.},
Eur.\ Phys.\ J.\ C {\bf 32S2} (2004) 19.

\item \label{Dawson:1991zj}
S.~Dawson,
Nucl.\ Phys.\ B {\bf 359} (1991) 283;
A.~Djouadi, M.~Spira and P.~M.~Zerwas,
Phys.\ Lett.\ B {\bf 264} (1991) 440.

\item \label{Spira:1995rr}
M.~Spira, A.~Djouadi, D.~Graudenz and P.~M.~Zerwas,
Nucl.\ Phys.\ B {\bf 453} (1995) 17.

\item \label{Harlander:2000mg}
R.~V.~Harlander,
Phys.\ Lett.\ B {\bf 492} (2000) 74;
V.~Ravindran, J.~Smith and W.~L.~van Neerven,
Nucl.\ Phys.\ B {\bf 704} (2005) 332.

\item \label{Catani:2001ic}
S.~Catani, D.~de Florian and M.~Grazzini,
JHEP {\bf 0105} (2001) 025,
JHEP {\bf 0201} (2002) 015.

\item \label{Harlander:2001is}
R.~V.~Harlander and W.~B.~Kilgore,
Phys.\ Rev.\ D {\bf 64} (2001) 013015.

\item \label{NNLOtotal}
R.~V.~Harlander and W.~B.~Kilgore,
Phys.\ Rev.\ Lett.\  {\bf 88} (2002) 201801;
C.~Anastasiou and K.~Melnikov,
Nucl.\ Phys.\ B {\bf 646} (2002) 220;
V.~Ravindran, J.~Smith and W.~L.~van Neerven,
Nucl.\ Phys.\ B {\bf 665} (2003) 325.

\item \label{Anastasiou:2004xq}
C.~Anastasiou, L.~J.~Dixon and K.~Melnikov,
Nucl.\ Phys.\ Proc.\ Suppl.\  {\bf 116} (2003) 193;
C.~Anastasiou, K.~Melnikov and F.~Petriello,
report SLAC-PUB-10673 [hep-ph/0409088].

\item \label{Kramer:1996iq}
M.~Kramer, E.~Laenen and M.~Spira,
Nucl.\ Phys.\ B {\bf 511} (1998) 523.

\item \label{Catani:2003zt}
S.~Catani, D.~de Florian, M.~Grazzini and P.~Nason,
JHEP {\bf 0307} (2003) 028.

\item \label{Ellis:1987xu}
R.~K.~Ellis, I.~Hinchliffe, M.~Soldate and J.~J.~van der Bij,
Nucl.\ Phys.\ B {\bf 297} (1988) 221;
U.~Baur and E.~W.~Glover,
Nucl.\ Phys.\ B {\bf 339} (1990) 38.

\item \label{DelDuca:2001fn}
V.~Del Duca, W.~Kilgore, C.~Oleari, C.~Schmidt and D.~Zeppenfeld,
Nucl.\ Phys.\ B {\bf 616} (2001) 367,
Phys.\ Rev.\ D {\bf 67} (2003) 073003.

\item \label{deFlorian:1999zd}
D.~de Florian, M.~Grazzini and Z.~Kunszt,
Phys.\ Rev.\ Lett.\  {\bf 82} (1999) 5209.

\item \label{Ravindran:2002dc}
V.~Ravindran, J.~Smith and W.~L.~Van Neerven,
Nucl.\ Phys.\ B {\bf 634} (2002) 247.

\item \label{Glosser:2002gm}
C.~J.~Glosser and C.~R.~Schmidt,
JHEP {\bf 0212} (2002) 016.

\item \label{Anastasiou:2005qj}
C.~Anastasiou, K.~Melnikov and F.~Petriello,
report UH-511-1066-05 [hep-ph/0501130].

\item \label{Smith:2005yq}
J.~Smith and W.~L.~van Neerven,
report YITP-SB-04-65 [hep-ph/0501098].

\item \label{Dokshitzer:hw}
Y.~L.~Dokshitzer, D.~Diakonov and S.~I.~Troian,
Phys.\ Rep.\  {\bf 58} (1980) 269.

\item \label{Parisi:1979se}
G.~Parisi and R.~Petronzio,
Nucl.\ Phys.\ B {\bf 154} (1979) 427.

\item \label{Curci:1979bg}
G.~Curci, M.~Greco and Y.~Srivastava,
Nucl.\ Phys.\ B {\bf 159} (1979) 451.

\item \label{Collins:1981uk}
J.~C.~Collins and D.~E.~Soper,
Nucl.\ Phys.\ B {\bf 193} (1981) 381
[Erratum-ibid.\ B {\bf 213} (1983) 545].

\item \label{Collins:va}
J.~C.~Collins and D.~E.~Soper,
Nucl.\ Phys.\ B {\bf 197} (1982) 446.

\item \label{Kodaira:1981nh}
J.~Kodaira and L.~Trentadue,
Phys.\ Lett.\ B {\bf 112} (1982) 66,
report SLAC-PUB-2934 (1982),
Phys.\ Lett.\ B {\bf 123} (1983) 335.

\item \label{Altarelli:1984pt}
G.~Altarelli, R.~K.~Ellis, M.~Greco and G.~Martinelli,
Nucl.\ Phys.\ B {\bf 246} (1984) 12.

\item \label{Collins:1984kg}
J.~C.~Collins, D.~E.~Soper and G.~Sterman,
Nucl.\ Phys.\ B {\bf 250} (1985) 199.

\item \label{Catani:2000vq}
S.~Catani, D.~de Florian and M.~Grazzini,
Nucl.\ Phys.\ B {\bf 596} (2001) 299.

\item \label{Catani:vd}
S.~Catani, E.~D'Emilio and L.~Trentadue,
Phys.\ Lett.\ B {\bf 211} (1988) 335.

\item \label{Kauffman:cx}
R.~P.~Kauffman,
Phys.\ Rev.\ D {\bf 45} (1992) 1512.

\item \label{deFlorian:2000pr}
D.~de Florian and M.~Grazzini,
Phys.\ Rev.\ Lett.\ {\bf 85} (2000) 4678,
Nucl.\ Phys.\ B {\bf 616} (2001) 247.


\item \label{Hinchliffe:1988ap}
I.~Hinchliffe and S.~F.~Novaes,
Phys.\ Rev.\ D {\bf 38} (1988) 3475.

\item \label{Kauffman:1991jt}
R.~P.~Kauffman,
Phys.\ Rev.\ D {\bf 44} (1991) 1415.

\item \label{Yuan:1991we}
C.~P.~Yuan,
Phys.\ Lett.\ B {\bf 283} (1992) 395.

\item \label{Balazs:2000wv}
C.~Balazs and C.~P.~Yuan,
Phys.\ Lett.\ B {\bf 478} (2000) 192.

\item \label{Balazs:2000sz}
C.~Balazs, J.~Huston and I.~Puljak,
Phys.\ Rev.\ D {\bf 63} (2001) 014021.

\item \label{Berger:2002ut}
E.~L.~Berger and J.~w.~Qiu,
Phys.\ Rev.\ D {\bf 67} (2003) 034026,
Phys.\ Rev.\ Lett.\  {\bf 91} (2003) 222003.

\item \label{Kulesza:2003wi}
A.~Kulesza and W.~J.~Stirling,
JHEP {\bf 0312} (2003) 056.

\item \label{Kulesza:2003wn}
A.~Kulesza, G.~Sterman and W.~Vogelsang,
Phys.\ Rev.\ D {\bf 69} (2004) 014012.

\item \label{Gawron:2003np}
A.~Gawron and J.~Kwiecinski,
Phys.\ Rev.\ D {\bf 70} (2004) 014003;
G.~Watt, A.~D.~Martin and M.~G.~Ryskin,
Phys.\ Rev.\ D {\bf 70} (2004) 014012
[Erratum-ibid.\ D {\bf 70} (2004) 079902].

\item \label{Lipatov:2005at}
A.~V.~Lipatov and N.~P.~Zotov,
hep-ph/0501172.


\item \label{PYTHIA}
T.~Sjostrand, P.~Eden, C.~Friberg, L.~Lonnblad, G.~Miu, 
S.~Mrenna and E.~Norrbin,
Comput.\ Phys.\ Commun.\  {\bf 135} (2001) 238;
T.~Sjostrand, L.~Lonnblad, S.~Mrenna and P.~Skands,
report LU-TP-03-38 [hep-ph/0308153].

\item \label{HERWIG}
G.~Marchesini, B.~R.~Webber, G.~Abbiendi, I.~G.~Knowles, M.~H.~Seymour 
and L.~Stanco,
Comput.\ Phys.\ Commun.\  {\bf 67} (1992) 465;
G.~Corcella {\it et al.},
JHEP {\bf 0101} (2001) 010,
report CERN-TH-2001-173 [hep-ph/0107071].

\item \label{MCatNLO}
S.~Frixione and B.~R.~Webber,
JHEP {\bf 0206} (2002) 029;
S.~Frixione, P.~Nason and B.~R.~Webber,
JHEP {\bf 0308} (2003) 007;
S.~Frixione and B.~R.~Webber,
report CERN-TH-2003-207 [hep-ph/0309186].

\item \label{Corcella:2004fr}
G.~Corcella and S.~Moretti,
Phys.\ Lett.\ B {\bf 590} (2004) 249.

\item \label{Balazs:2004rd}
C.~Balazs, M.~Grazzini, J.~Huston, A.~Kulesza and I.~Puljak,
hep-ph/0403052,
published in
M.~Dobbs {\it et al.},
hep-ph/0403100, p.~51, Proceedings of the Les Houches 2003
Workshop on {\em Physics at TeV Colliders}.


\item \label{bozzi}
G.Bozzi, PhD Thesis, University of Florence, April 2004.

\item \label{Collins:gx}
J.~C.~Collins, D.~E.~Soper and G.~Sterman,
in {\it Perturbative Quantum Chromodynamics},
ed. A.H.~Mueller (World Scientific, Singapore, 1989), p.~1. 

\item \label{Catani:1991kz}
S.~Catani, G.~Turnock, B.~R.~Webber and L.~Trentadue,
Phys.\ Lett.\ B {\bf 263} (1991) 491.

\item \label{Catani:1992ua}
S.~Catani, L.~Trentadue, G.~Turnock and B.~R.~Webber,
Nucl.\ Phys.\ B {\bf 407} (1993) 3.

\item \label{shapedis}
M.~Dasgupta and G.~P.~Salam,
Eur.\ Phys.\ J.\ C {\bf 24} (2002) 213,
JHEP {\bf 0208} (2002) 032.

\item \label{Banfi:2004nk}
A.~Banfi, G.~P.~Salam and G.~Zanderighi,
JHEP {\bf 0408} (2004) 062.

\item \label{threshold}
G.~Sterman,
Nucl.\ Phys.\ B {\bf 281} (1987) 310;
S.~Catani and L.~Trentadue,
Nucl.\ Phys.\ B {\bf 327} (1989) 323,
Nucl.\ Phys.\ B {\bf 353} (1991) 183.

\item \label{Catani:1996yz}
S.~Catani, M.~L.~Mangano, P.~Nason and L.~Trentadue,
Nucl.\ Phys.\ B {\bf 478} (1996) 273.

\item \label{deFlorian:2004mp}
D.~de Florian and M.~Grazzini,
Nucl.\ Phys.\ B {\bf 704} (2005) 387.

\item \label{beta2}
O.~V.~Tarasov, A.~A.~Vladimirov and A.~Y.~Zharkov,
Phys.\ Lett.\ B {\bf 93} (1980) 429;
S.~A.~Larin and J.~A.~Vermaseren,
Phys.\ Lett.\ B {\bf 303} (1993) 334.

\item \label{Laenen:2000de}
E.~Laenen, G.~Sterman and W.~Vogelsang,
Phys.\ Rev.\ Lett.\  {\bf 84} (2000) 4296;
A.~Kulesza, G.~Sterman and W.~Vogelsang,
Phys.\ Rev.\ D {\bf 66} (2002) 014011.

\item \label{Qiu:2000ga}
J.~w.~Qiu and X.~f.~Zhang,
Phys.\ Rev.\ Lett.\  {\bf 86} (2001) 2724,
Phys.\ Rev.\ D {\bf 63} (2001) 114011.

\item \label{book}
R.~K.~Ellis, W.~J.~Stirling and B.~R.~Webber,
{\it QCD and collider physics},
(Cambridge University Press, Cambridge, 1996) and references therein.

\item \label{Vogt:2000ci}
A.~Vogt,
Phys.\ Lett.\ B {\bf 497} (2001) 228;
C.~F.~Berger,
Phys.\ Rev.\ D {\bf 66} (2002) 116002.

\item \label{mvv}
S.~Moch, J.~A.~M.~Vermaseren and A.~Vogt,
Nucl.\ Phys.\ B {\bf 688} (2004) 101, 
Nucl.\ Phys.\ B {\bf 691} (2004) 129.

\item \label{Davies:1984hs}
  C.~T.~H.~Davies and W.~J.~Stirling,
  Nucl.\ Phys.\ B {\bf 244} (1984) 337.

\item \label{Chetyrkin:1997iv}
  K.~G.~Chetyrkin, B.~A.~Kniehl and M.~Steinhauser,
  Phys.\ Rev.\ Lett.\  {\bf 79} (1997) 353.

\item \label{code}
{\tt http://arturo.fi.infn.it/grazzini/codes.html}

\item \label{Martin:2004ir}
  A.~D.~Martin, R.~G.~Roberts, W.~J.~Stirling and R.~S.~Thorne,
  Phys.\ Lett.\ B {\bf 604} (2004) 61.

\item \label{Alekhin:2002fv}
S.~Alekhin,
Phys.\ Rev.\ D {\bf 68} (2003) 014002, talk presented at EPS05, {\it HEP2005
Europhysics Conference}, Lisboa, July 2005.

\item \label{Pumplin:2002vw}
  J.~Pumplin, D.~R.~Stump, J.~Huston, H.~L.~Lai, P.~Nadolsky and W.~K.~Tung,
  JHEP {\bf 0207} (2002) 012.

\item \label{Gluck:1998xa}
M.~Gluck, E.~Reya and A.~Vogt,
Eur.\ Phys.\ J.\ C {\bf 5} (1998) 461

\item \label{DSW}
C.~T.~H.~Davies, B.~R.~Webber and W.~J.~Stirling,
Nucl.\ Phys.\ B {\bf 256} (1985) 413.

\item \label{LY}
G.~A.~Ladinsky and C.~P.~Yuan,
Phys.\ Rev.\ D {\bf 50} (1994) 4239.


\item \label{BLNY}
F.~Landry, R.~Brock, P.~M.~Nadolsky and C.~P.~Yuan,
Phys.\ Rev.\ D {\bf 67} (2003) 073016.

\item \label{Konychev:2005iy}
A.~V.~Konychev and P.~M.~Nadolsky,
report ANL-HEP-PR-05-52 [hep-ph/0506225].

\item \label{Assamagan:2004mu}
G.~Bozzi, S.~Catani, D.~de Florian and M.~Grazzini,
in  K.~A.~Assamagan {\it et al.}  [Higgs Working Group Collaboration],
hep-ph/0406152, p.~14,
proceedings of the Les Houches 2003
Workshop on {\em Physics at TeV Colliders}.

\item \label{Frixione:1998dw}
S.~Frixione, P.~Nason and G.~Ridolfi,
Nucl.\ Phys.\ B {\bf 542} (1999) 311.


\item \label{Furmanski:1981cw}
W.~Furmanski and R.~Petronzio,
Z.\ Phys.\ C {\bf 11} (1982) 293.

\item \label{Vogt:2004ns}
A.~Vogt,
report NIKHEF-04-011 [hep-ph/0408244].

\item \label{graryz}
I.S.~Gradshteyn and I.M.~Ryzhik, 
{\it Tables of Integrals, Series and Products} 
(Academic Press, San Diego, 2000).

\item \label{Kulesza:1999gm}
A.~Kulesza and W.~J.~Stirling,
Nucl.\ Phys.\ B {\bf 555} (1999) 279.

\end{enumerate}

\end{document}